# Atomic-scale Nucleation and Growth Pathways of Complex Plate-like Precipitates in Aluminum Alloys

Junyuan Bai[1], Gaowu Qin[1,2,3]*, Xueyong Pang[1,3], Zhihao Zhao[1,3]

[1]*Key Laboratory for Anisotropy and Texture of Materials (Ministry of Education), School of Materials Science and Engineering, Northeastern University, Shenyang 110819, China*

[2]*Institute for Strategic Materials and Components, Shenyang University of Chemical Technology, Shenyang 110142, China*

[3]*Research Center for Metal Wires, Northeastern University, Shenyang 110819, China*

**Abstract:**

Aluminum alloys, the most widely utilized lightweight structural materials, predominantly depend on coherent complex-structured nanoplates to enhance their mechanical properties. Despite several decades of research, the atomic-scale nucleation and growth pathways for these complex-structured nanoplates remain elusive, as probing and simulating atomistic events like solid nucleation is prohibitively challenging. Here, using theoretical calculations and focus on three representative complex-structured nanoplates—$T_1$, Ω, and $\eta'/\eta_2$—in commercial Al alloys, we explicitly demonstrate their associated structural transitions follow an "*interlayer-sliding+shuffling*" mode. Specifically, partial dislocations complete the interlayer-sliding stage, while atomic shuffling occurs upon forming the unstable basic structural transformation unit (BSTU) of the nanoplates. By identifying these BSTUs, we propose structural evolution pathways for these nanoplates within the Al matrix, which align well with experimental observations and enable the evaluation of critical nuclei. These findings provide long-sought mechanistic details into how coherent nanoplates nucleate and grow, facilitating the rational design of higher-performance Al alloys and other structural materials.

*Corresponding author.

Email: qingw@smm.neu.edu.cn (G.W. Qin)

## Introduction

Aluminum, the most abundant metallic element on Earth, is inherently soft. Nevertheless, the discovery of duralumin in 1906 opened a new era for extensive applications of age-hardening Al alloys across industries. By precipitating numerous dense and coherent nanoprecipitates within the matrix during aging treatment, the mechanical properties of Al alloys can be significantly improved[1–3]. Among various nanoprecipitates discovered, coherent $\{111\}_{Al}$-oriented plate-like precipitates are particularly important as they often form interleaved three-dimensional networks, effectively impeding dislocation motion and even enhancing the creep resistance of Al alloys[4–6]. For instance, well-known $\{111\}_{Al}$-oriented $T_1$, Ω, and $\eta'/\eta_2$ nanoplates are typically recognized as key strengthening phases in their respective alloys, as illustrated in Table 1. Moreover, the recently discovered V-Sc$(Al_2Cu)_4$ nanoplate[7], which forms in situ within the Ω nanoplates, can even raise the serving temperature of Al alloys to an unprecedented ~400°C. Unlike the metastable $\{100\}_{Al}$-oriented θ′-$Al_2Cu$ nanoplate, these $\{111\}_{Al}$-oriented complex-structured nanoplates are generally stable phases, offering substantial promise for enhancing the strength and creep resistance of Al alloys. Despite extensive research spanning several decades, the atomic-scale nucleation and growth mechanisms of these nanoplates remain poorly understood, which is undoubtedly indispensable for further developing high-strength and heat-resistant Al alloys.

At the atomic scale, tracking nucleation and growth pathways in solids is always challenging. Experimental measurements often lack the necessary spatiotemporal resolution to capture atomistic events, while theoretical simulations may struggle due to inaccurate interatomic potentials or an inability to simulate rare events like solid nucleation within accessible timescales[8–10]. According to the crystal structures of parent and product phases, in-situ precipitation of coherent nanoprecipitates in alloys can be classified into three types: Type-I (simple→$simple_{identical}$, such as face-centered cubic(fcc)→$L1_2$), Type-II (simple→$simple_{distinct}$, such as fcc→hexagonal close-packed (hcp)), and Type-III (simple→complex, such as hcp→topologically close-packed (TCP)). As indicated in Table 1, apart from γ′-$AlAg_2$ precipitation (fcc→hcp), the formation of coherent nanoplates in Al alloys generally involves simple→complex transformations. Actually, early investigations into plate-like precipitates began with the γ′ phase, leading to the proposal of a terrace-ledge growth model[11,12], where (a/6)<$11\bar{2}$> (where a is the lattice constant) Shockley partial dislocations at the ledge's edge were thought to accomplish the fcc→hcp transformation. Despite subsequent observations of partial dislocations in precipitates like $T_1$[13–15], Ω[16], and θ′[17], their specific roles in simple→complex transformations remain elusive. Consequently, there is a lack of comprehensive physical models describing the atomic-scale nucleation and growth of these nanoplates, particularly regarding the conditions triggering structural transformations and the essential kinetic processes governing plate thickening.

For simple→complex transitions in Al alloys, our recent study[18] on hcp→TCP transitions in Mg alloys provides valuable insights. Utilizing density functional theory (DFT) calculations to sample solute clusters, we identified an unstable 3-layer hcp-ordering, which spontaneously collapses into a TCP (icosahedron) structure upon structural relaxation, serving as the basic structural transformation unit (BSTU) that governs the entire hcp→TCP transformation. Apart from atomic diffusion, only shuffle-based displacements are involved during TCP formation. Although technological constraints impede modeling the complete precipitation kinetic process[19] using techniques such as kinetic monte carlo or molecular dynamics (MD) simulations, the identification of BSTU offers a workaround for deducing the structural evolution pathway of TCP plates within the hcp matrix, enabling us to understand how precipitates nucleate and grow. Comparatively, TCP precipitation within the fcc-Al matrix is more intricate than in the hcp-Mg alloys.

This study employs DFT and MD approaches to explore the atomic-scale nucleation and growth pathways of complex nanoprecipitates in Al alloys, with three representative $\{111\}_{Al}$-oriented nanoplates ($T_1$, Ω, and $\eta'/\eta_2$) as model systems. We demonstrate that the simple→complex diffusive transitions in Al alloys typically follow an "*Interlayer-sliding+shuffling*" mode, in which partial dislocations complete the interlayer-sliding stage, followed by atomic shuffling upon BSTU formation. The structural evolution pathways for $T_1$, Ω, and $\eta'/\eta_2$ nanoplates were established by identifying their BSTUs, and these pathways align well with experimental observations. Furthermore, these pathways enable a quantitative assessment of the critical nucleus and provide explanations for various intricate precipitation behaviors. This study offers long-sought mechanistic insights into the atomic-scale nucleation and growth of complex nanoplates in Al alloys or other fcc-based alloys, and establishes a solid theoretical foundation for rationally designing alloy compositions and enhancing the thermal stability of nanoprecipitates.

**Table 1**. Orientation relationships and structural transformation types of $\{111\}_{Al}$-oriented nanoplates in Al alloys.

| Phases | Alloy Systems | Habit Plane | Orientation Relationships | Structural Transformation Types |
|---|---|---|---|---|
| $\gamma'$-$AlAg_2$ | Al-Ag | $\{111\}_{Al}$ | $(0001)_{\gamma'}//(111)_{Al}$, $[11\bar{2}0]_{\gamma'}//[1\bar{1}0]_{Al}$[12] | fcc→hcp, Type-II |
| $T_1$- $Al_6Cu_4Li_3$[20,21] | Al-Cu-Li | $\{111\}_{Al}$ | $(0001)_{T_1}//(111)_{Al}$, $[10\bar{1}0]_{T_1}//[1\bar{1}0]_{Al}$[15] | fcc→TCP, Type-III |
| Ω-$Al_2Cu$ | Al-Cu-Mg-Ag | $\{111\}_{Al}$ | $(001)_{\Omega}//(111)_{Al}$, $[010]_{\Omega}//[10\bar{1}]_{Al}$[16] | fcc→*C*16, Type-III |
| V-Sc($Al_2Cu$)$_4$[7] | Al-Cu-Mg-Ag-Sc | $\{111\}_{Al}$ | $(001)_V//(111)_{Al}$, $[010]_V//[10\bar{1}]_{Al}$[7] | fcc→$C16$→$D2_b$, Type-III |
| $\eta'$-$Mg_6Zn_{13}$** /$\eta_2$-$MgZn_2$ | Al-Mg-Zn-Cu | $\{111\}_{Al}$ | $(0001)_{\eta'/\eta_2}//(111)_{Al}$, $[10\bar{1}0]_{\eta'/\eta_2}//[1\bar{1}0]_{Al}$[22] | fcc→TCP, Type-III |

** represents that is was proposed in the present study.

**Interlayer-sliding via partial dislocations**

In the hcp→TCP transformation[18], we found that TCP formation occurs exclusively through atomic shuffling, without requiring any dislocations, as the similarity in stacking sequence between TCP structures (-*TKTKTK*-, where *T* represents a Triangular lattice net and *K* denotes a Kagomé lattice net) and the hcp lattice (-*ABABAB*-). Although TCP nanoplates like hcp-structured $T_1$ and $\eta_2$-$MgZn_2$ also form in Al alloys, the fcc lattice of Al does not favor their direct in-situ precipitation solely via shuffle-based displacements. Hence, partial dislocations, capable of modifying the local stacking sequence of lattice, may play a key role in fcc→TCP transformations. The main text uses the $T_1$ phase as an example to illustrate the "*interlayer-sliding*+*shuffling*" mode of simple→complex transformations in Al alloys, with detailed results for Ω and $\eta'/\eta_2$ available in the Supplementary Information.

As the thinnest structure discovered in Al-Cu-Li alloy, depicted in Fig. 1a, the $T_1$ nanoplate often maintains a single-unit-cell height, analogous to the $\gamma''$ phases[23,24] in Mg-RE (rare earth)-Zn series alloys. The $T_1$ nanoplate possesses a symmetrical 5-layer structure consisting of $U_T|K|\omega_T|K|U_T$ lattice nets. Detailed introductions for these lattice nets are presented in Fig. 1b. The thinnest Ω nanoplate observed is isostructural to this 5-layer $T_1$ nanoplate[16,25]. Although the $T_1$ phase was previously assigned the chemical formula $Al_2CuLi$[13,14], this cannot be definitively confirmed without $T_1$ thickening, as the stacking periodicity for the $T_1$ nanoplate along $<111>_{Al}$ direction is disrupted by the outward Al atoms (marked by red-edged ellipses) on the outermost $U_T$ lattice nets. Unlike the conservative transformation of Laves formation in the hcp lattice[18], $T_1$ formation requires additional atoms due to the higher atomic density of $\omega_T$ compared to other types of lattice nets ($\rho_{\omega_T} > \rho_{U_T} = \rho_K = \rho_T$). Furthermore, Yang et al.[15] recently confirmed (a/6)<11$\bar{2}$> partial dislocations at the edges of (3, 5)-planes within the $T_1$ nanoplate (see Fig. 2a). However, pure Al typically exhibits high stacking fault energy, making partial dislocations hard to occur[26].

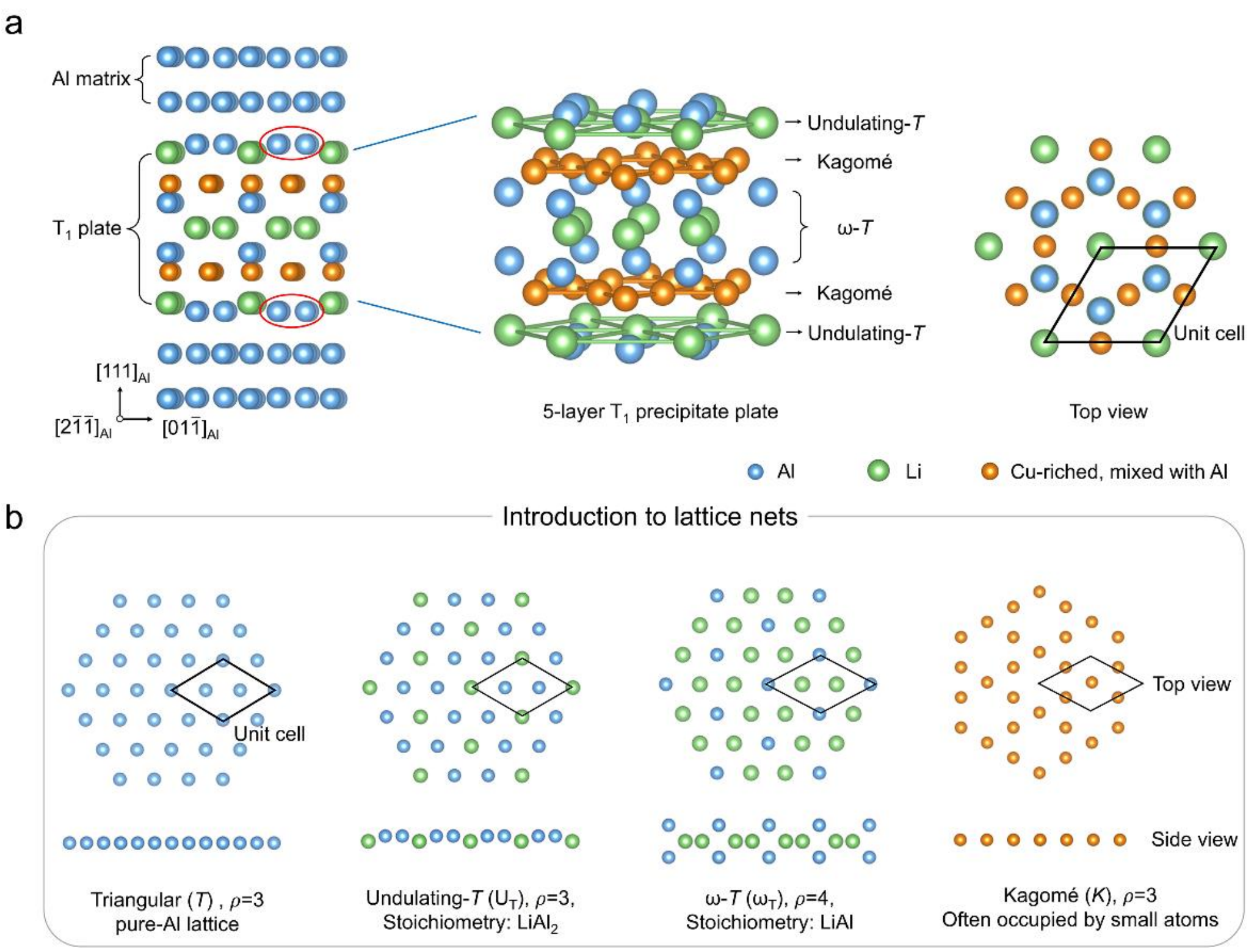

**Figure 1**. **Schematic illustration of the atomic structure of experimentally identified $T_1$ precipitate plate. a** Atomic structural model of the 5-layer single-unit-cell height $T_1$ nanoplate; blue and green balls represent Al and Li atoms; while yellow balls denote the enrichment of Cu, mixed with Al. **b**. An introduction to the Triangular and Kagomé lattice nets present in the $T_1$ nanoplate. The $\omega_T$ is typically observed in $\mu$-$A_6B_7$ type (where A denotes large atoms and B represents small atoms) TCP phases, having an atomic density ($\rho(\omega_T)$ =4) that exceeds that of other lattice nets.

Given that experiments[13,27] have observed a direct transformation of the $T_1$ nanoplate from the (Cu, Li)-rich Guinier-Preston (GP) zone, the high solute concentrations in the GP zone may significantly impact the lattice's stacking fault energy ($\gamma_{sf}$). The effect of Cu was examined first. According to experimental observations[27], Fig. 2b constructs a Cu-GP zone configuration, where Cu fully occupies the (2, 4)-planes, and the remainder consists of the Al lattice. DFT calculations of the generalized stacking fault (GSF) energy curves for the (3, 5)-planes along $[11\bar{2}]_{Al}$ (see Fig. 2c) show $\gamma_{sf}^{(3)\text{-plane}}$= -0.075 J/m$^2$ and $\gamma_{sf}^{(5)\text{-plane}}$= 0.171 J/m$^2$. This suggests that $\gamma_{sf}^{(3)\text{-plane}}$ can be substantially reduced to a negative value due to its adjacent Cu (2, 4)-planes, while $\gamma_{sf}^{(5)\text{-plane}}$ remains positive. If we further consider the effect of Li, the inset in Fig. 2d presents a (Li, Cu)-GP zone model with (1, 5)-planes and (3)-plane occupied by one-third and two-thirds Li, respectively. The results show that both $\gamma_{sf}^{(3)\text{-plane}}$ (-0.034 J/m$^2$) and $\gamma_{sf}^{(5)\text{-plane}}$ (-0.008 J/m$^2$) achieve negative values with the introduction of Li, indicating a notable reduction in both $\gamma_{sf}^{(3)\text{-plane}}$ and $\gamma_{sf}^{(5)\text{-plane}}$ under the combined action of Cu and Li. Actually, there are fluctuations in the GP zone's composition during aging[27], which cannot be considered in the current DFT calculations. Nonetheless, these calculations still offer a qualitative understanding of whether different solutes affect $\gamma_{sf}$ by examining specific, idealized compositions, such as the (Li, Cu)-GP model shown in Fig. 2d.

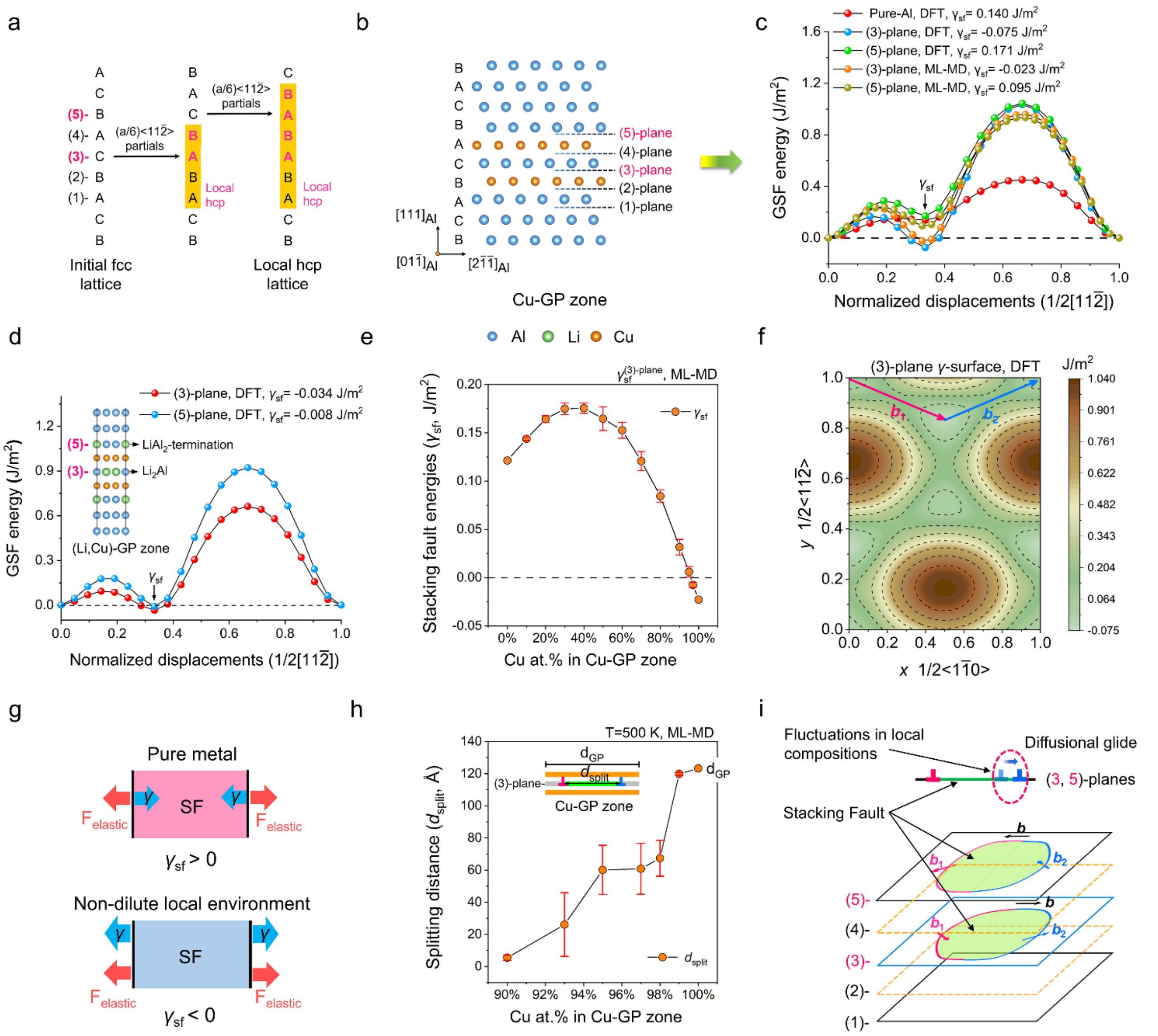

**Figure 2. Interlayer-sliding stage in structural transformation. a**, Illustration of the structural transformation from initial fcc lattice to local hcp stacking sequence via (a/6) <11$\bar{2}$> Shockley partial dislocations. **b**, Structural model of Cu-GP zone with (2, 4)-planes fully occupied by Cu atom; blue, green, and yellow balls represent Al, Li, and Cu atoms, respectively. **c**, GSF energy curves along the 1/2[11$\bar{2}$] direction in the (3, 5)-planes of the Cu-GP zone according to DFT and ML-MD calculations. **d**, GSF energy curves along the [11$\bar{2}$] direction in the (3, 5)-planes of the (Li, Cu)-GP zone based on DFT calculations. Inset shows a structural model of the (Li, Cu)-GP zone, with (2, 4)-planes fully occupied by Cu atoms, (1, 5)-planes by one-third of Li, and (3)-plane by two-thirds of Li. **e**, Variation of ML-MD measured $\gamma_{sf}$ of the (3)-plane with different concentrations of Cu in the Cu-GP zone. Error bars of standard deviation (SD) indicate $\gamma_{sf}$ from 10 independent sampling configurations. **f**, GSF energy ($\gamma$) surface for the (3)-plane in Cu-GP zone. **g**, Schematic illustration of various forces acting on partial dislocations during dissociation. **h**, Variation of ML-MD measured equilibrium splitting distance $d_{split}$ of partial dislocations at 500 K with Cu at.% in the Cu-GP zone. Error bars of SD indicate $d_{split}$ from 5 independent models. **i**, Schematic illustration of configuration and migration mechanism for partial dislocations on the (3, 5)-planes.

In Fig. 2e, we further utilized a machine learning (ML) Al-Cu potential[28] to evaluate the impact of solute content on $\gamma_{sf}$, given this potential yields a reasonable, quantitative agreement with $\gamma_{sf}^{(3,\ 5)\text{-planes}}$ values derived from our DFT calculations (Fig. 2c). Since the absence of ML Al-Cu-Li potentials, we focused exclusively on evaluating the impact of varying Cu content within the (2, 4)-planes on $\gamma_{sf}^{(3)\text{-plane}}$, with the insights gained should also be applicable to the scenarios involving Li. Detailed calculation procedures are provided in the Methods section. The results reveal an evident trend: $\gamma_{sf}^{(3)\text{-plane}}$ initially increases, then sharply decreases as the Cu content increases, eventually reaching negative values when the concentration approaches ~96 at.%. This demonstrates the effect of high-content Cu within the GP zone in reducing $\gamma_{sf}$. Additionally, the γ-surface of the (3)-plane in Fig. 2f, derived from the Cu-GP model (Fig. 2b), indicates that (a/2)<$1\bar{1}0$> perfect dislocation still energetically favors dissociation into two (a/6)<$11\bar{2}$> partials. Therefore, it is concluded that the formation of a (Li, Cu)-rich GP zone creates a favorable chemical environment for the generation of local hcp lattices (that is, stacking fault) through dislocations dissociation at the (3, 5)-planes.

Prior experimental researchers[13,14,29] have suggested that the migration of (a/6)<$11\bar{2}$> partials at the (3, 5)-planes follows a diffusional glide mechanism driven by Cu and Li diffusion. According to dislocation theory[30], the negative $\gamma_{sf}$ calculated here, unlike the typical positive $\gamma_{sf}$ in pure metals, means that the force γ acting on partial dislocations keeps in the same direction as the elastic force ($F_{elastic}$), as illustrated in Fig. 2g. Consequently, if Shockley partials occur at the (3, 5)-planes (the source of dislocations will be discussed later), they should accumulate at the GP zone's edge under the action of γ and $F_{elastic}$, and migrate outward as GP zone expands. However, one should note that the abovementioned negative $\gamma_{sf}$ is derived from the scenario of high-content (~100 at.%) compositions within the GP zone, which does not contain the impact of compositional fluctuations. Therefore, analyzing dislocation migration is difficult due to the force γ varies with compositional fluctuations.

Given the difficulties in directly observing the migration process of dislocations, which involves long-term dynamic evolution that exceeds the time span achieved by classical MD techniques, we adopted an indirect approach considering composition fluctuations to probe the migration mechanism in Fig. 2h. This involved measuring the equilibrium width of an (a/2)<$1\bar{1}0$> edge dislocation splitting ($d_{split}$ at 500K) at the (3)-plane under varying Cu concentrations (which equivalent to considering composition fluctuations) in the Cu-GP zone through ML-MD simulations. Our measurements indicate that $d_{split}$ is generally smaller than the width of the GP zone ($d_{GP}$) for Cu concentrations below 98 at.%; while as the Cu contents approach 100 at.%, the $d_{split}$ gradually increases until it equals the $d_{GP}$. Moreover, Fig. S1 also reveals that variations in Cu content in regions distant from partial dislocations do not affect their $d_{split}$ after the passage of partial dislocations. This indicates that the migration of partial dislocations primarily depends on their surrounding solute distribution. As local composition fluctuations facilitate the dispersion of solutes around the dislocations at a specific moment, dislocations can migrate further and gradually reach the edge of the GP zone, as schematically depicted in Fig. 2i. Hence, while the migration of partial dislocations follows a diffusional glide mechanism, it is driven by the local composition fluctuations in their vicinity.

Notably, according to the elastic theory of dislocations[30], (a/6)<11$\bar{2}$> partials at the (3, 5)-planes cannot be dislocation loops because the glide dislocation loops are inherently unstable due to attraction between opposite segments of the loop. Our ML-MD simulations also corroborate this instability in Fig. S2. Thus, these partial dislocations should keep a structure resembling a loop, as depicted in Fig. 2i, formed by the connection of two dissociated (a/6)< 11$\bar{2}$ > partials ($b_1$ and $b_2$). Moreover, Supplementary Section 1 indicates that the high-content Mg, Cu, and Ag within the GP zone of Ω precipitate effectively lower the lattice's $\gamma_{sf}$ to negative values, facilitating stacking fault formation. In contrast, the chemical composition of Mg and Zn in the GP zone fails to reduce the lattice's $\gamma_{sf}$ for $\eta'/\eta_2$ nanoplates. However, $\gamma_{sf}$ is much decreased upon the introduction of Cu into the GP zone, indicating that Cu is crucial in facilitating $\eta'/\eta_2$ precipitation. This is one of the key reasons for adding Cu into 7xxx series Al alloys.

**Atomic shuffling upon BSTU formation**

During simple→complex transformations, complex products often occur in the parent matrix with a multi-layered form that diverges from conventional layer-by-layer perspectives. An unstable cluster enabling the direct formation of this multi-layered structure is termed the BSTU of simple→complex transformations, which is crucial for revealing the structural evolution of complex precipitates. For instance, in Mg alloys, a 3-layer unstable hcp-ordering was identified as a BSTU (termed as $M_R$-type) responsible for forming Laves ($AB_2$) and Laves-like ($AB_3$, $A_2B_7$, and $AB_5$) nanoplates[18], as detailed in Supplementary Section 4. Importantly, the unstable BSTU can be directly captured via structural relaxation in DFT calculations, which establishes a foundation for using DFT calculations to explore different BSTUs under appropriate modelings. Moreover, our prior study[18] demonstrates this spontaneous structural transition (an energy reduction process) caused by unstable BSTU formation originates from the internal-coordinate instability rather than dynamic instability in other common solid-state diffusionless-displacive phase transitions[31].

Likewise, uncovering the formation pathway of the 5-layer $T_1$ nanoplate relies on identifying its BSTU. Following the passage of (a/6)<11$\bar{2}$> partials along the (3, 5)-planes, the resultant local hcp lattice in the (1-5)-planes provides a favorable structural environment for subsequent TCP formations.

Since the formation of the $T_1$ nanoplate's $\omega_T$ requires additional atoms, Al atoms are placed directly onto the tetrahedral interstitial sites (Tet, pink balls) of the model shown in Fig. 3a. This placement directly produces the $T_1$ nanoplate upon relaxation. Despite the model lacking matrix portions, the resultant $T_1$ formation implies that atomic hopping into Tet sites may trigger structural transitions. Hence, the role of atomic hopping events needs to be carefully examined. Specifically, there are two modes of atomic hopping into the Tet sites: direct hopping involving a single atom (see Fig. 3b) and concerted hopping involving multiple atoms (see Fig. 3c). Hopping atoms can originate from either the (2, 4)-planes or the (3)-plane. However, direct hopping is ineffective because Al, Cu, and Li atoms cannot stably occupy Tet sites and return to their original positions after relaxation, as depicted in Fig. 3b. Consequently, only the concerted hopping mode is considered. Based on the source of hopping atoms, concerted hopping events can be divided into Ep1-ABC (from the (2, 4)-planes) and Ep2-AB (from the (3)-plane) sub-modes, encompassing six distinct hopping events, as summarized in Fig. 3d.

Using a large-sized model incorporating the lateral matrix portion (more details in Methods), we observed a sign of structural transformation around the Tet site following a single Ep1-AlAlCu hopping event, clearly visible in the top-view of (3, 4, 5)-planes depicted in Fig. 3e. Subsequent analysis revealed that the Ep1-AlAlCu event (Supplementary Movie 1) contributes mainly to the $T_1$ formation compared to the other five hopping events. During Ep1-AlAlCu event, the hopping of Al from the (2)-plane into the Tet site pushes the Al above it on the (3)-plane to move outward; meanwhile, the dissociative displacement ($\xi_y$, indicated by red arrows) of these two Al atoms also drives collective displacement ($\xi_x$, marked by black arrows) of Cu atoms on the (2, 4)-planes, forming two openings, which attract Li from the (1, 5)-planes to move towards them via $\xi_y$. Consequently, the atomic hopping event into the Tet sites triggers atomic shuffling, and a continuous occurrence of Ep1-AlAlCu events is expected to lead to the $T_1$ nanoplate formation through $\xi_x$ and $\xi_y$ gradually. Specifically, the (2, 4)-planes transition into the *K* net via $\xi_x$, while the (3)-plane and (1, 5)-planes convert into the $\omega_T$ and $U_T$, respectively, through $\xi_y$ to relax high strains due to the presence of large Li atoms. Furthermore, by mapping the energy (($E(\xi_x, \xi_y)$-$E(1, 1)$) landscape of structural transformation relative to normalized $\xi_x$ and $\xi_y$ in Fig. 3f, we found that only synchronous action of $\xi_x$ and $\xi_y$ leads to energy decline along the Minimum Energy Path (MEP). This indicates that $\xi_x$ and $\xi_y$ displacements are interdependent during structural transformation, and the five planes involved represent the minimum number of layers necessary to form the $\omega_T$. Therefore, we designate this 5-layer unstable structure as $M_T$-type BSTU for $T_1$ formation.

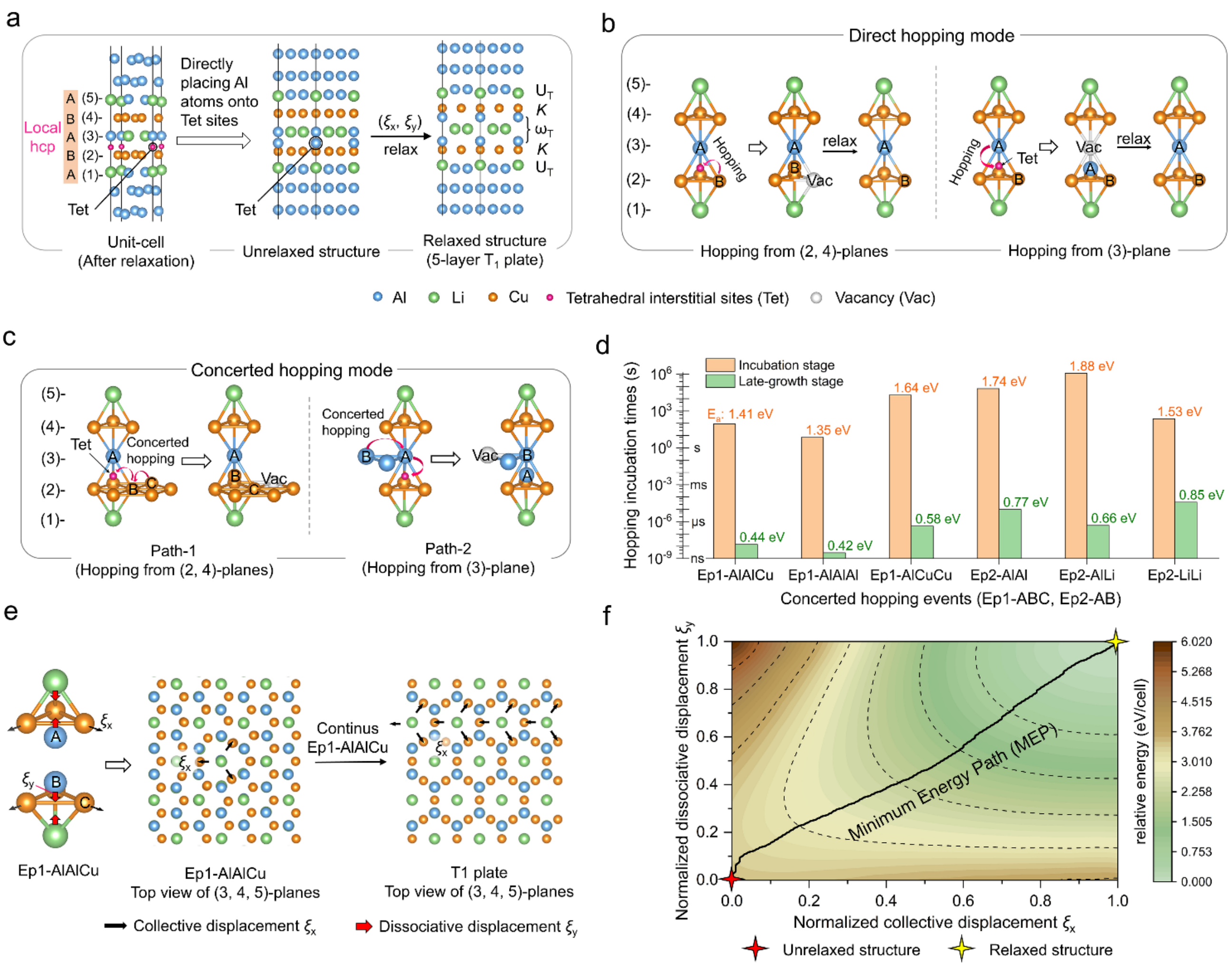

**Figure 3. Atomic shuffling triggered by concerted hopping events. a**, Schematic illustration depicts the direct formation of a 5-layer $T_1$ nanoplate, achieved by placing Al atoms into tetrahedral interstitial (Tet) sites using a model without containing a lateral matrix portion; blue, green, yellow, pink, and white balls represent Al, Li, Cu atoms, Tet sites, and Vacancy, respectively. **b**, Illustration showing how a single atom hops into the Tet sites in the direct hopping mode. **c,** Schematic depicting concerted hopping mode involving multiple atoms. **d**, The hopping incubation times for six concerted hopping events, along with their activation energies ($E_a$) marked at the top of the bar, during incubation and late-growth stages. **e**, Detailed results following a single Ep1-AlAlCu event (Supplementary Movie 1), showing a slight structural transformation around the A-B atomic column in a large-sized model containing a lateral matrix portion. Black and red arrows represent collective displacement $\xi_x$ and dissociative displacement $\xi_y$, respectively. **f**. Variation of energy during the local-hcp→TCP structural transformation with normalized $\xi_x$ and normalized $\xi_y$, with the unit of eV/Cell and the final relaxed structure's energy being the reference. To reduce computation cost, we employed the model shown in Fig. 3a to map the energy, yielding equivalent results to a large-sized model.

According to Vineyard's harmonic transition state theory[32], the hopping incubation times for six concerted hopping events during the incubation and late-growth stages are evaluated in Fig. 3c (more details in Methods), with corresponding activation energies ($E_a$) indicated in the panel. During both stages, Ep1-AlAlCu and Ep1-AlAlAl display lower $E_a$ values and shorter hopping incubation times than the other four events, suggesting that the Al-Al pair kinetically favors the formation of $\omega_T$, leading to a LiAl stoichiometry for $\omega_T$ ultimately. Moreover, the subsequent section reveals that the $T_1$ phase also thermodynamically supports this stoichiometry. Although Ep1-AlAlAl has slightly lower $E_a$ values than Ep1-AlAlCu, we suggest that Ep1-AlAlCu contributes mainly to the $T_1$ nanoplate formation due to the lower probability of AlAlAl combination in Cu-rich (2, 4)-planes. Notably, during the incubation stage, Ep1-AlAlCu events occur on a scale of seconds, consistent with the general understanding of solid nucleation as a rare event[9] requiring a longer incubation time. Afterward, the much-reduced $E_a$ values in the late-growth stage indicate a rapid $T_1$ nanoplate lengthening. This evaluation not only qualitatively depicts the dynamic characteristics of $T_1$ formation across different stages but also validates the feasibility of concerted hopping triggering structural transformation from a kinetic perspective.

**Structural evolution pathway of plate-like precipitates guided by BSTU**

Despite the $T_1$ phase being the stable precipitate in the Al-Cu-Li system, there remains controversy[4,20,21] regarding its specific phase due to the thin thickness of $T_1$ nanoplates and uncertainties in their composition during aging. In Supplementary Section 3, through constructing Al-Cu-Li convex hull diagrams at aging temperature (400-500 K) to examine phase stability, we demonstrate that the $T_1$ nanoplate should be the thermodynamically stable $Al_6Cu_4Li_3$ (see Fig. S9a) Laves-like phase. However, due to kinetic constraints, the realistic $T_1$ nanoplates are hard to reach this $Al_6Cu_4Li_3$ structure and more closely correspond to the metastable $Al_4Cu_6Li_3$ phase (see Fig. S9d). Unlike typical Laves phases, the $Al_4Cu_6Li_3$ ($Al_6Cu_4Li_3$) Laves-like phase contains $\omega_T$ and adopts -$Pl_T|K|\omega_T|K$- periodic stacking unit, where the Planar-$T$ ($Pl_T$) possesses a $LiAl_2$ composition and a planar structure in side view.

Consistent with the initial 5-layer $T_1$ nanoplate formation, the subsequent $T_1$ thickening/lengthening still follows an "*interlayer-sliding+shuffling*" mode. During the interlayer-sliding stage (Fig. 4a), a local hcp environment within the (6-9)-planes needs to be established through the formation of (Li, Cu)-rich GP zone within these planes, which facilitates the passage of (a/6)$<11\bar{2}>$ partials along the (7, 9)-planes. The presence of this (Li, Cu)-rich GP zone has been experimentally verified[27]. In the subsequent shuffling stage, a 9-layer $T_1$ nanoplate with a $U_T|K|\omega_T|K|Pl_T|K|\omega_T|K|U_T$ structure can be achieved by forming new unstable $M_T$-type BSTU along the outermost $U_T$, according to Laves-like plate thickening in Mg alloys[18] proceed via $U_T \rightarrow Pl_T$ interfacial transition (see Fig. S7). Notably, the $U_T(LiAl_2) \rightarrow Pl_T(LiAl_2)$ transition during thickening is essential for eliminating the stacking periodicity constraints imposed by the original $U_T$ lattice, thereby enabling $T_1$ nanoplate thickening/lengthening. Meanwhile, our simulated high-angle annular dark-field scanning transmission electron microscopy (HAADF-STEM, Fig. 4a) from <11$\bar{2}$>$_{Al}$ and <1$\bar{1}$0>$_{Al}$ projections for this 9-layer configuration closely match experimental observations (Figs. 4b), strongly supporting this thickening process. Thus, as schematically illustrated in Fig. 4c, the structural evolution of $T_1$ nanoplates within the Al matrix should follow a rule of 5-layer→9-layer→13-layer→···→ (5+4*n*)-layer (where *n* represents the number of thickening events). This evolution rule effectively interprets the regular thickness pattern observed experimentally (Figs. 4b and S11b) in thickened-$T_1$ nanoplates, demonstrating its predictability.

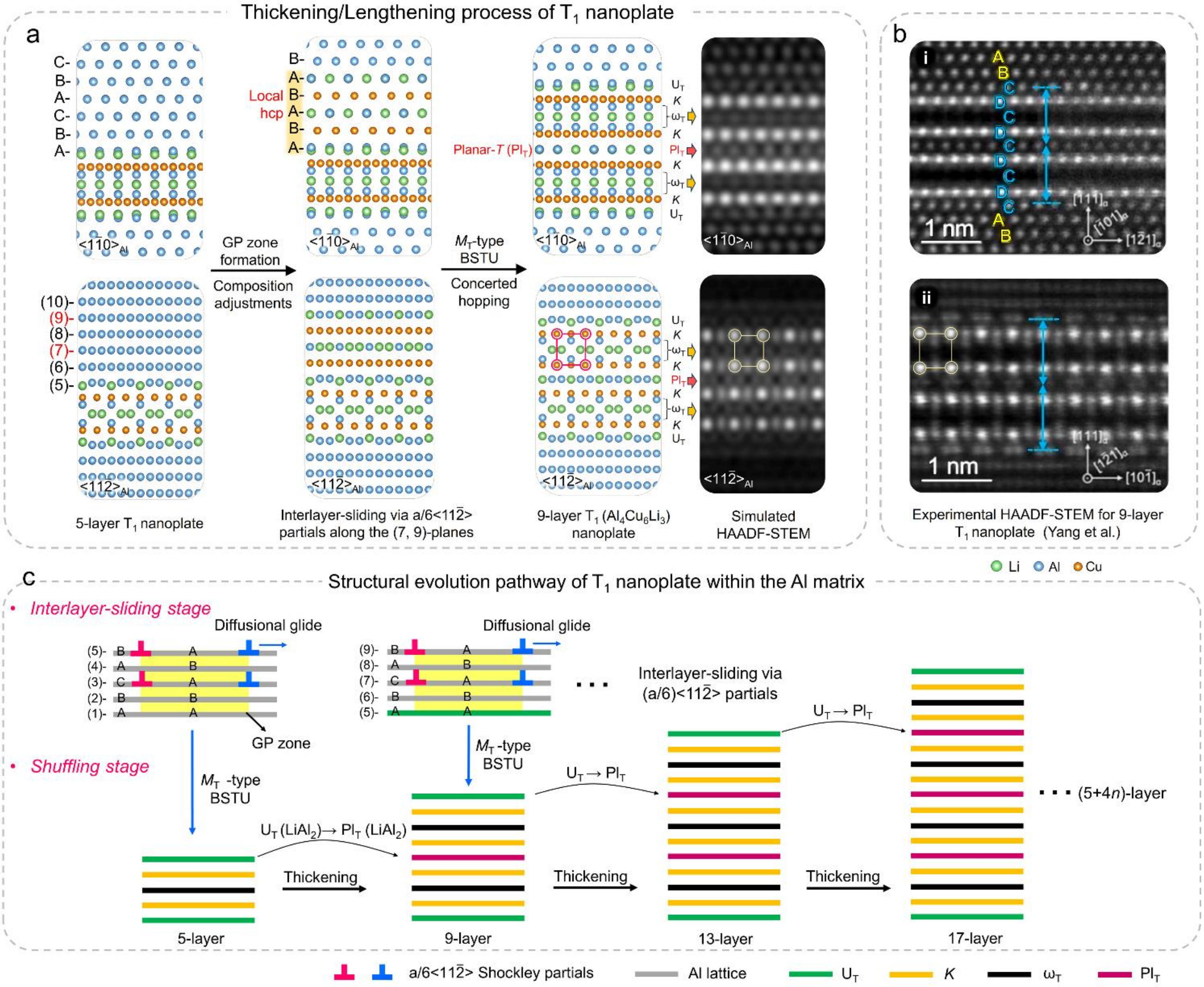

**Figure 4. Structural evolution pathway of $T_1$ nanoplate within the Al matrix.** **a**, Schematic depiction of the thickening/lengthening process from the 5-layer $T_1$ nanoplate to the 9-layer $U_T|K|\omega_T|K|Pl_T|K|\omega_T|K|U_T$ structure sequentially through interlayer-sliding and shuffling stages. The Planar-*T* lattice nets in this phase are abbreviated as $Pl_T$, and blue, green, and yellow balls represent Al, Li, and Cu atoms, respectively. **b**, Experimental HAADF-STEM images of the 9-layer thickened $T_1$ nanoplate, reported by Yang et al.[15] recently. **c**, Structural evolution pathway of $T_1$ nanoplate within the Al matrix.

Apart from the 5-layer $M_T$-type BSTU, the 3-layer $M_R$-type BSTU can also operate within the Al matrix when a local hcp environment is created by (a/6)<11$\bar{2}$> partials. For instance, Supplementary

Section 3 shows that an 11-layer defective $T_1$ nanoplate forms through the participation of both $M_R$-type and $M_T$-type BSTUs. Supplementary Section 4 further reveals that the 7-layer $\eta'$ nanoplate with a $U_T|K|\omega_T|K|Pu_T|K|U_T$ structure (where the Puckered-*T* ($Pu_T$) has a $Mg_2Zn$ composition and displays a puckered structure in side view) forms via $M_T$-type and $M_R$-type BSTUs. However, after $\eta_2$ transforms in situ from the $\eta'$ nanoplate (7-layer $\eta'$→ 9-layer $\eta_2$), the subsequent $\eta_2$-$MgZn_2$ formation exclusively involves $M_R$-type BSTU and utilizes $U_T$→$Pu_T$ transitions to thicken the $\eta_2$ nanoplates. The structural evolution of $\eta_2$-$MgZn_2$ within the Al matrix follows a rule of 9-layer→11-layer→13-layer→⋯, which is consistent with experimental observations[22] that $\eta_2$ forms only with certain discrete thickness. Therefore, upon establishing a local hcp environment, various TCP precipitates can be generated using these BSTUs through different assembly strategies.

Unlike TCP nanoplates discussed above, Supplementary Section 5 reveals that the Ω nanoplate possesses a unique $U_T|K|$distorted θ-$Al_2Cu$ (termed $\theta_d$)$|K|U_T$ sandwich structure, where the $U_T|K$ layers always serve as the shell during precipitation. Despite $\theta_d$-$Al_2Cu$ adopting a distorted *C*16 structure, it can be viewed as a TCP-like structure, consisting of alternating $U_T$-like and *K*-like lattice nets (see Fig. S19), which results in a conservative transition during the thickening of Ω nanoplates from their initial 5-layer $U_T|K|\omega_T|K|U_T$ form. We found the shell portion, $U_T|K$, plays a structural template role in transformation, kinetically inducing the coherent formation of $\theta_d$-$Al_2Cu$ along the $\{111\}_{Al}$ planes. The formation mechanism of the 5-layer Ω nanoplate is consistent with that of the 5-layer $T_1$ nanoplate due to their identical structures. However, during the subsequent Ω thickening, besides the (a/6)$<11\bar{2}>$ partials, the interlayer-sliding stage also involves the (a/4)$<1\bar{1}0>$ partials. Notably, the migration of these (a/4)$<1\bar{1}0>$ partials can trigger atomic shuffling across four {111} planes to thicken/lengthen Ω nanoplates (see Fig. S21). We designate this 4-layer unstable structure as $M_\Omega$-type BSTU, which governs the thickening and lengthening of thickened-Ω nanoplates. Overall, the Al→Ω transformations still follow an "*interlayer-sliding*+*shuffling*" mode, with the thickness of Ω nanoplates given by $N$=5+2$n$ (Fig. S22), where $N$ is the total number of layers and $n$ specifies the number of thickening steps. For further details, refer to Supplementary Section 5.

In reality, $T_1$[33], Ω[34], and $\eta'/\eta_2$[35,36] nanoplates frequently display intricate phenomena such as off-stoichiometry (mainly containing excess solvent Al) or polymorphism, which are also ubiquitous in TCP nanoplates within Mg alloys[37,38]. Our prior study[18] on hcp→TCP transitions demonstrates that the off-stoichiometry arises from the nonclassical nucleation behaviors during structural transitions; additionally, this off-stoichiometry in precipitates may further induce polymorphism through the interplay between thermodynamics and kinetics. Unlike classical nucleation theory (CNT), the nonclassical nucleation mechanism allows a composition in the nucleus that differs from the target precipitate phase or nucleation via intermediates. Supplementary Section 6 reveals that the formation of $T_1$, Ω, and $\eta'/\eta_2$ also displays an evident nonclassical nucleation behavior. Despite having a fully non-uniform composition in their nuclei, structural transitions can still proceed. Therefore, the observed off-stoichiometry and polymorphism in these three nanoplates are attributed to their nonclassical nucleation behaviors. Moreover, due to this nonclassical nucleation nature and associated kinetic constraints, the actual nanoplates are often highly metastable, such as the aforementioned $T_1$ nanoplates, and require a long-term dynamic evolution to achieve equilibrium.

Classical solid-state phase transformation theory[11,12] typically attributes the growth of plate-like precipitates to partial dislocations at ledge edges, which has been extensively validated in simple→$\text{simple}_{\text{distinct}}$ transformations[12]. This study elucidates the precise roles of partial dislocations for simple→complex transformations: they can not only create local stacking faults that facilitate subsequent atomic shuffling but may also directly trigger atomic shuffling in specific transitions. Our findings indicate that the partial dislocations should occur on alternating planes to sustain precipitate growth. However, the origin of such high-density partial dislocations remains unclear. In Supplementary Section 7, inspired by a diffusive nucleation mechanism[39] for dislocations at metal surfaces observed recently, we found a strong tendency for the quenched-in excessive vacancies to segregate into Cu-rich atomic planes, which may facilitate dislocation nucleation at their adjacent *T*-forming atomic planes due to increased misfit strain. Thus, this vacancy-induced nucleation mechanism is expected to dominate during precipitation, while those matrix-inherent or pre-deformation-induced dislocations may contribute occasionally to initial nucleation.

As described above, a spontaneous structural transformation occurs when the nucleus evolves into an unstable BSTU at a certain moment; however, according to CNT, only nuclei with a size *R* (*n,* representing the number of atoms in the nuclei) equal to or greater than the critical radius ($R(n) \geq R^*(n^*)$) can grow stably, while nuclei with $R(n) < R^*(n^*)$ will dissipate. Based on the obtained structural evolution pathways, we employed CNT to evaluate the $R^*$ and critical nucleation barriers ($\Delta G^*$) for the homogeneous nucleation of Ω and $\eta'$ nanoplates in Supplementary Section 8. The results contribute to clarifying the formation mechanisms of unique microstructures and resolving related debates. Specifically, for Ω nucleation, the 5-layer Ω nucleus has the lowest $\Delta G^*$($\Delta G^*_{\text{5-layer}}$=667 $k_B$T, where $k_B$ is the Boltzmann constant and T=500 K) and $R^*(n^*)$($R^*_{\text{5-layer}}\approx$1 nm, $n^*_{\text{5-layer}}$=58 Cu atoms) compared to the 7-layer Ω nucleus, indicating both energetic and kinetic favorability for the 5-layer Ω nucleus in nucleation. This explains why the thinnest observable Ω nanoplate in experiments is the 5-layer $U_T|K|\omega_T|K|U_T$ structure. For $\eta'$ nucleation, we found that the 7-layer $U_T|K|\omega_T|K|Pu_T|K|U_T$ $\eta'$ nuclei are kinetically favored as the critical nucleus over the 5-layer $U_T|K|\omega_T|K|U_T$ $\eta'$ nucleus (see Fig. S27).

Given that $\eta'$ nanoplates typically occur in a 7-layer form[22,25,35] rather than a 5-layer one like $T_1$ and $\Omega$ nanoplates, it can be concluded that kinetics dominate $\eta'$ nucleation.

During early precipitation, the nucleus rim typically maintains coherence with the surrounding matrix. However, as nanoplates thicken, preserving coherence along the rim interface becomes increasingly difficult due to the formation of TCP or $\Omega$ nanoplates producing contraction strain perpendicular to the broad interface[15,16,22,34]. This readily leads to Frank-type misfit-compensating dislocations occurring at the rim interface, releasing interfacial elastic strain. Such phenomena have been extensively observed at the edges of thicker $\eta_2$-$MgZn_2$[22,35,36] and $\Omega$[34] nanoplates. As nanoplates continuously thicken, the rim interface will eventually become incoherent, causing a significant reduction in mechanical performance. Hence, it is crucial to prevent the thickening/coarsening of nanoplates during high-temperature service. This work provides mechanistic insights into stabilizing nanoprecipitates in Al alloys. For instance, a promising strategy is to hinder dislocation nucleation. Our other study[40] indicates that the exceptional heat resistance in V-Sc($Al_2Cu$)$_4$ nanoplates (mentioned above) arises from their formation within $\Omega$ nanoplates, effectively blocking the nucleation of (a/4) $<1\bar{1}0>$ partials by modifying the $K$ lattice structure in the $U_T|K$ shell. Besides this approach, other strategies, such as interfacial solute segregation[41] or the development of core/shell structure[42], require further exploration in the future.

**Conclusions:**

The atomistic-scale nucleation and growth mechanisms of complex nanoplates in Al alloys have not been established due to difficulties in capturing atomistic events experimentally and covering rare events within conventional theoretical simulation timescales. Here, we circumvent these difficulties by employing reasonable modeling in DFT calculations to identify the BSTU of fcc→complex diffusive transformations. Using three representative $\{111\}_{Al}$-oriented complex nanoplates ($T_1$, $\Omega$, and $\eta'/\eta_2$) as model systems, their formation pathways were systematically explored. Our results show that the fcc→complex transformations generally follow an "*interlayer-sliding*+*shuffling*" mode, in which the entire structural transformation occurs sequentially through interlayer sliding provided by partial dislocations, and atomic shuffling upon BSTU formation. We demonstrate that structural transformations require compositional variations, both of which are dynamically coupled during precipitation, and the formation of the GP zone, enriched with specific components, creates a favorable chemical environment for nucleation, dissociation, and migration of dislocations. Furthermore, we show how different BSTUs can be employed to deduce the structural evolution pathway of nanoplates within the fcc matrix, enabling the prediction of critical nuclei and the interpretation of intricate experimental phenomena. These insights are expected to establish a universal framework for comprehending and studying nucleation and growth processes of in-situ precipitation of nanoprecipitates in alloys, contributing to the design of higher-performance Al alloys and other fcc-based materials, such as Ni-based superalloys, Cu alloys, and austenitic stainless steel.

## Methods

### First-principles calculations

First-principles calculations based on density functional theory (DFT) were performed using the Vienna ab initio simulation package (VASP)[43,44] with Blochl's projector augmented wave (PAW) potential method[45]. The exchange-correlation energy functional was described with the generalized gradient approximation (GGA) as parameterized by Perdew-Burke-Ernzerhof (PBE)[46]. A 520 eV plane wave cutoff was adopted with the convergence criteria for energy and the atomic force set as $10^{-6}$ eV and $10^{-2}$ eV/Å, respectively. Partial occupancies were determined by using the first-order Methfessel-Paxton method with a smearing width of 0.2 eV[47]. A Monkhorst-Pack $k$-mesh with a spacing of 0.25 $Å^{-1}$ (corresponding to a 10×10×10 $k$-point grid for an Al unit cell) between $k$ points was employed, ensuring energy convergence to better than 1 meV/atom. Structures were visualized using the VESTA[48] software.

The GSF energies were calculated using the slab-vacuum supercell method. A 15 Å vacuum layer was added along the $z$-axis direction to separate adjacent slabs. The GSF energy is defined as: $\gamma=[E(\delta_a, \delta_b)-E(0,0)]/A$, where $E(\delta_a, \delta_b)$ is the energy of the supercell with the upper half part shifted by $(\delta_a, \delta_b)$ relative to the down half part along the slip plane; $E(0, 0)$ is the energy of the system without shifting; A is the area of the slip plane. During relaxations, all atoms except those just above and below the fault plane were allowed to fully relax. This relaxation scheme better reflects realistic scenarios[49] and facilitates the capture of the structural transitions during the Ω precipitate thickening process.

To determine the position of the saddle point (i.e., activation energy $E_a$) and the associated minimum energy pathway during atomic hopping, calculations were performed employing the climbing image nudged elastic band (CI-NEB) method[50,51] with 3 interpolated images. According to Vineyard's harmonic transition state theory[32], the hopping incubation time for a diffusion event can be written as: $\tau = v_0^{-1}\exp\left[\frac{\Delta H_m}{k_B T}\right]$, where $k_B$ denotes the Boltzmann constant, $\Delta H_m$ represents the migration enthalpy, and $v_0$ is the attempt frequency. Here, the hopping incubation time $\tau$ was evaluated at T=500 K. For a system comprising $N$ atoms, $v_0$ is given by: $v_0=\prod_{l=1}^{3N-3} v_l / \prod_{k=1}^{3N-4} \tilde{v}_k$, where the frequencies $v_l$ signifies the 3$N$-3 normal-mode vibrational frequencies of the initial configuration, and $\tilde{v}_k$ are the 3$N$-4 non-imaginary normal-mode vibrational frequencies at the saddle-point configuration. A detailed assessment of attempt frequency $v_0$ for various hopping events is provided in Supplementary Section 2. Moreover, since prior work[52] has demonstrated that anharmonic effects resulting from volume expansion are negligible for aluminum, we employ the harmonic approximation in this study, equating $E_a$ to $\Delta H_m$. Notably, it is currently infeasible to compute the exact $E_a$ for every possible hopping across all local atomic configurations during the late-growth stages. Therefore, the results for the late-growth stages can only be used for qualitative comparison with those of the incubation stage.

The energy landscape of local-hcp→$T_1$ transformation was obtained by interpolating 9×9 intermediate configurations between initial unrelaxed ($\xi_x$=0, $\xi_y$=0) and finally relaxed structures ($\xi_x$=1, $\xi_y$=1), with the internal $T_1$-associated atoms fixed and minimizing energy with respect to all degrees of freedom. Considering the substantial size of models incorporating the lateral matrix, we adopt a model excluding the lateral matrix to plot the energy landscape, as this type of model can replicate the same relaxation results as the former model without consuming high computation costs. The minimum energy path (MEP) in the 2D energy landscape was identified using the multidimensional lowest energy (MULE) code[53].

In this work, we employed models that rigorously incorporated the matrix (the 3D environment) to perform DFT calculations of the precipitate plate formation process, with obtained results that have ruled out the influence of periodic boundary conditions in the models. The unit-cell structure of the fcc-Al lattice was constructed based on their $x$-axis, $y$-axis, and $z$-axis parallel to the $[1\bar{1}2]$, $[211]$, and $[\bar{1}11]$, respectively. Based on this unit-cell structure, GSF calculations utilized 1×1×5 models. When calculating activation energies $E_a$ for various concerted hopping events, a local hcp-lattice needs to be created in the large-sized model. However, introducing this local hcp-lattice into the fcc-model also creates a surface in the model, leading to discontinuities in the electronic density and significantly slowing down the convergence of electronic self-consistency[49]. To improve computational efficiency, we constructed hcp-Al models (unit-cell structure: $x$-axis: $[10\bar{1}0]_{hcp}$, $y$-axis: $[01\bar{1}0]_{hcp}$, and $z$-axis: $[0001]_{hcp}$) to evaluate $E_a$ values, with 600-atom (5×5×4) models utilized for the incubation stage and 870-atom (6×6×4) models for the late-growth stage. A Γ-centered 1×1×1 $k$-point mesh was adopted for these calculations. In Supplementary Section 2, we demonstrate that these hcp-Al models can quantitatively replicate the $E_a$ values for Cu, Al, and Li diffusions compared to the vacuum-containing models with hybrid hcp and fcc lattices, indicating the validity of the hcp-Al models.

**Molecular dynamics simulations**

MD simulations were conducted using the Large-scale Atomic/Molecular Massively Parallel Simulator (LAMMPS)[54]. We utilized neural-network machine learning Al-Mg-Zn-Cu potential, recently developed by Marchand et al.[28], for these simulations through the n2p2[55] interface. In contrast to traditional interatomic potentials such as embedded-atom or modified embedded-atom (EAM and MEAM, respectively), ML potentials are an emerging method that can create computationally efficient atomistic potentials achieving near-DFT accuracy, enabling modeling of many complex metallurgical phenomena. Through examining the accuracy of this ML potential in calculating GSF energies across various component GP zones, we found that this potential can quantitatively replicate near-DFT accuracy in describing Al-Cu (Fig. 2c) and Al-Zn-Cu (Fig. S3) interactions, while accuracy was lacking for Al-Mg-Cu interaction. Therefore, this ML potential was primarily employed to explore dislocation dissociation and migration behaviors within the Cu-GP zone.

Simulation cells containing dislocations at the (3)-plane of Cu-GP zone are oriented along $x$=[110], $y$=[$\bar{1}$11], and $z$=[1$\bar{1}$2] directions, respectively, where the Cu-GP zone was created by randomly distributing Cu on (2, 4)-planes to a given concentration. The cell size was chosen as 17.02×14.02×2.48 nm in $x$, $y$, and $z$ directions and consists of 35690 atoms. Using the Atomsk[56] package, a full edge dislocation was introduced at the (3)-plane, displacing half of the upper simulation cell with the shortest lattice vector (that is, Burgers vector) relative to the lower cell. A Poisson ratio of 0.33 for Al was employed, and an isotropic displacement-field solution was applied to create (a/2)<110> type dislocation on the (3)-plane, with the dislocation line parallel to the $z$-direction. Periodic boundary conditions were utilized in the glide plane of the edge dislocation ($x$ and $z$ directions), while shrink-wrapped boundary conditions were applied out-of-plane ($y$ direction). The dislocation cells were first relaxed to their equilibrium positions using energy minimization. Then, they were thermalized at 500 K for 40 ps using a Nose-Hoover thermostat under a constant temperature and volume (NVT) ensemble. This simulation time was chosen to ensure convergence of the measured $d_{split}$ values of two split (a/6)<11$\bar{2}$> Shockley partials, and the difference between the last two $d_{split}$ values was less than 2%. The $d_{split}$ values at each composition were measured from five independent models using the dislocation analysis (DXA) unity in the Ovito[57] Python package and were averaged over the final 30 ps intervals. In GSF calculations, a 6-atom fcc unit cell with its $x$, $y$, and $z$ axes oriented along the [1$\bar{1}$2], [110], and [$\bar{1}$11] directions, respectively, was employed to construct calculation cells (in total 2798 atoms) using a 20×1×19 scheme. The bulk reference cells were first constructed and minimized along the three axes directions, where the GP zone was created by randomly distributing specific solute atoms on the (2, 4)-planes according to the given concentration. Subsequently, a translation vector was applied to the perfect cell to obtain the stacking fault cell structure, and minimization was performed with x, y positions fixed. Finally, the zero-temperature $\gamma_{sf}$ was computed by using the difference in energy of the two cells divided by the stacking fault area.

**Simulation of HAADF-STEM images**

The HAADF-STEM images were simulated with the code of QSTEM[58], and the corresponding parameters used for simulations are: a spherical aberration coefficient $C_3$= 1 mm, defocus values $\Delta f$= = -60 nm beam convergence angle $\alpha$= 15 mrad, annular dark-field detector range is from 70 to 200 mrad.

## Data availability

The authors declare that all data supporting the findings of this study are available within the paper and Supplementary Information files. Additional images are available from the corresponding authors upon request.

## Acknowledgements

This research is supported by the Fundamental Research Funds for the Central Universities of China (N2102011, N2007011, N160208001), 111 Project (B20029).

## Author contributions

G. Qin and J. Bai conceived the original idea and designed the work. J. Bai conducted the simulations with help from X. Pang and Z. Zhao. J. Bai wrote the paper. G. Qin supervised the project.

# Supplementary Information for
# Atomic-scale Nucleation and Growth Pathway of Complex Plate-like Precipitates in Aluminum Alloys

Junyuan Bai[1], Gaowu Qin[1,2,3]*, Xueyong Pang[1,3], Zhihao Zhao[1,3]

[1]*Key Laboratory for Anisotropy and Texture of Materials (Ministry of Education), School of Materials Science and Engineering, Northeastern University, Shenyang 110819, China*

[2]*Institute for Strategic Materials and Components, Shenyang University of Chemical Technology, Shenyang 110142, China*

[3]*Research Center for Metal Wires, Northeastern University, Shenyang 110819, China*

## Contents:

## 1. Supplementary Section 1: GSF calculations for the Ω and $\eta'/\eta_2$ nanoplates

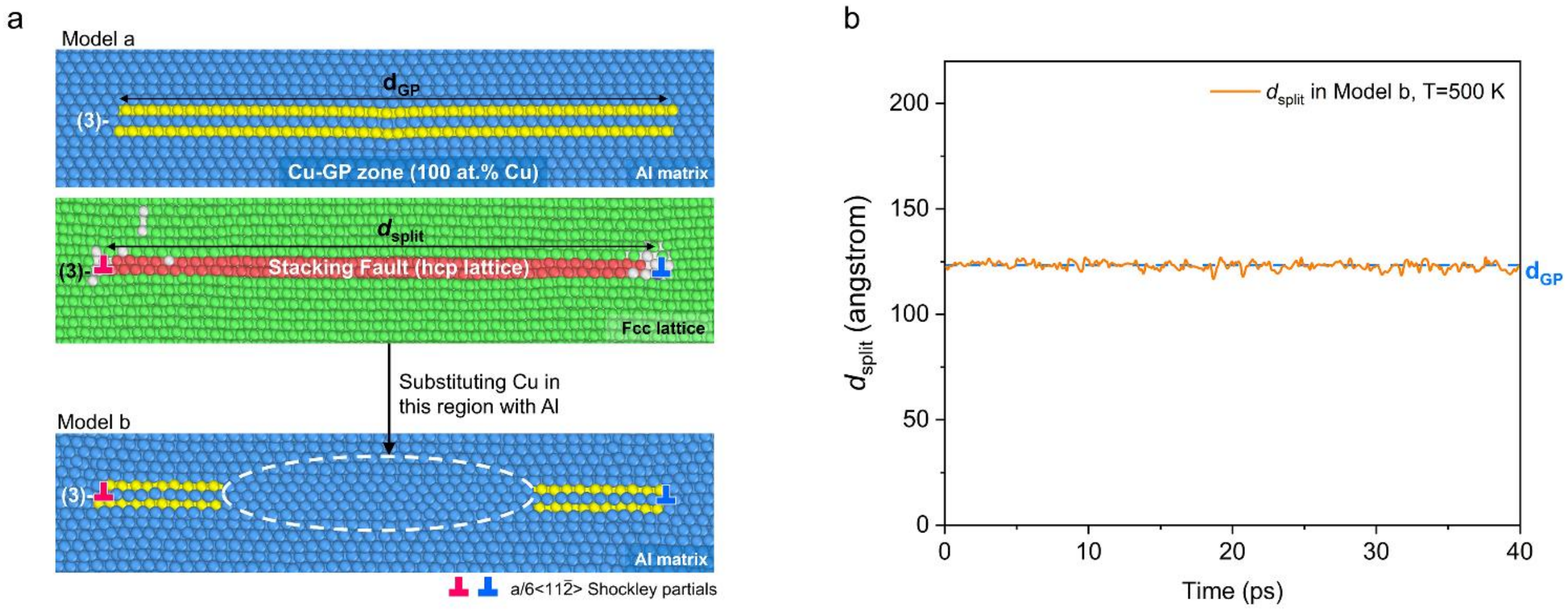

**Figure S1. a**. Splitting of the (a/2)<1$\bar{1}$0> edge dislocation at the (3)-plane in the Cu-GP zone, with Model a showing the scenario with 100 at.% Cu in the Cu-GP zone, while Model b depicts Cu substitution with Al in the middle region (indicated by a white dashed ellipse). The parameters $d_{GP}$ and $d_{split}$ refer to the width of the GP zone and the two dissociated (a/6)<11$\bar{2}$> Shockley partials. Cu and Al atoms are represented by yellow and blue balls, respectively, while hcp and fcc lattices are denoted by red and green balls, respectively. **b**. The evolution of the average split distance $d_{split}$ in Model b as a function of time, showing nearly constant values equal to the $d_{GP}$ at T=500 K.

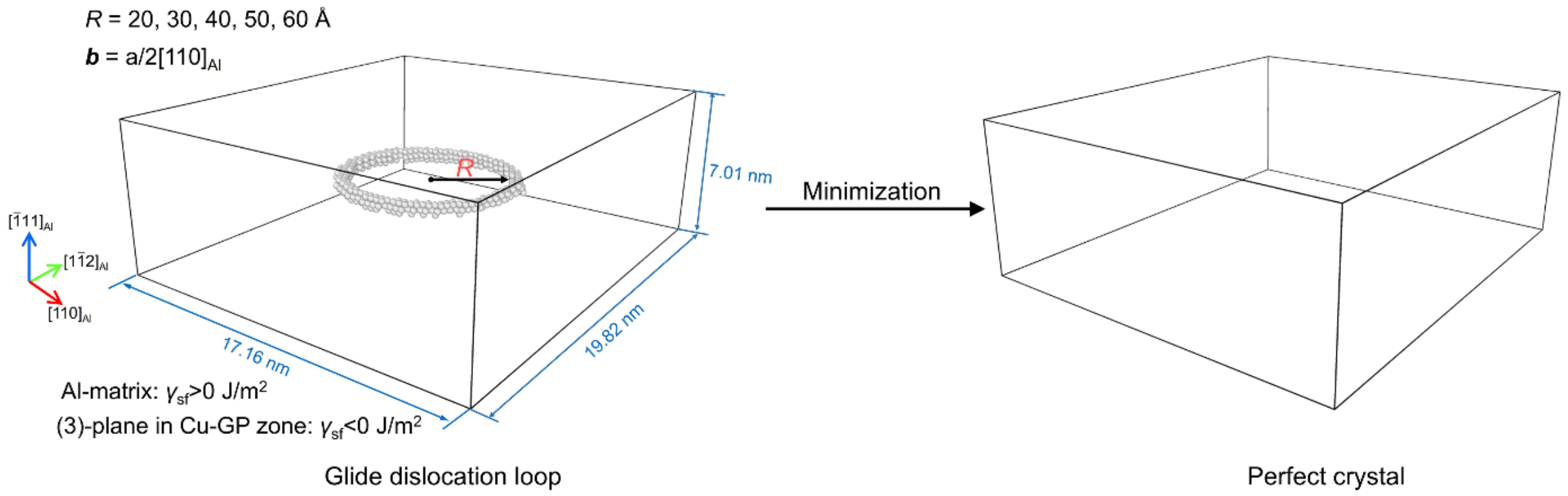

**Figure S2.** Schematic illustration showing the minimization process of a glide dislocation loop (***b***=(a/2)[110]$_{Al}$ ) with different radii ($R$=20, 30, 40, 50, 60 Å) in two scenarios: when $\gamma_{sf} > 0$ in pure Al matrix and when $\gamma_{sf} < 0$ at the (3)-plane in Cu-GP zone. In the absence of stress, both cases demonstrate the instability of the glide dislocation loop, changing into a perfect lattice structure during minimization. The models were constructed using the Atomsk[1] software.

Ω precipitates were initially observed in high Cu: Mg ratio Al-Cu-Mg ternary alloys, albeit at low densities. The trace addition of Ag, however, greatly facilitates the formation of thin Ω nanoplates along $\{111\}_{Al}$ matrix planes, significantly enhancing yield strength and creep resistance at temperatures up to 150-220 °C[2–4]. This makes Al-Cu-Mg-Ag alloys particularly suitable for aerospace and defense applications. The most accepted structure for Ω precipitate is orthorhombic (space group: *Fmmm*), with an orientation relationship of $(001)_{\Omega}//(111)_{Al}$ and $[010]_{\Omega}//[10\bar{1}]_{Al}$ relative to the Al matrix. Since the little difference between the orthorhombic Ω structure and the equilibrium tetragonal θ-$Al_2Cu$ *C*16 phase (space group: *I4/mcm*) in binary Al-Cu alloys, prior researchers[5,6] viewed the Ω as a distorted θ or a variant of equilibrium θ phase. A recent theoretical study[3] has confirmed this perspective, indicating that the experimentally observed orthorhombic symmetry of Ω results from elastic distortion imposed by the surrounding α-Al matrix. Thus, while the $Al_2Cu$ *C*16 phase is not a TCP structure, the introduction of Mg and Ag into Al-Cu alloys modifies the precipitation sequence, enabling the coherent precipitation of the pristine, incoherent θ-$Al_2Cu$ phase in a distorted state within the matrix. This study designates this distorted θ as $\theta_d$ to distinguish it from the ideal θ-$Al_2Cu$ *C*16 structure. An intriguing feature of the Ω nanoplates is Mg and Ag always occupy their outermost layers, with no presence in the interior. However, the exact impact of Mg and Ag on Ω formation remains elusive. Their roles will be discussed in Supplementary Section 5.

The matured/thickened Ω nanoplates, as detailed in Supplementary Section 5, have a sandwiched structure comprising $U_T|K|\theta_d$-$Al_2Cu|K|U_T$ lattice nets, with the outermost $U_T(MgAg_2)|K$(Cu-riched) lattice nets always serving as the shell. Similar to the $T_1$ nanoplate, numerous experiments[7–9] have demonstrated that the thinnest Ω nanoplates also possess a 5-layer $U_T|K|\omega_T|K|U_T$ structure, from which the matured Ω plates evolve. Also, Yang et al.[9] recently confirmed the presence of (a/6)<11$\bar{2}$> partial dislocations along the edges of (3, 5)-planes within these 5-layer Ω nanoplates. Although Yang et al.[9] suggested this 5-layer structure as a precursor to the matured Ω nanoplates and designated it as the Ω′ phase, our findings presented in Supplementary Section 8 indicate that, from both energetic and kinetic perspectives, the coherent precipitation of θ-$Al_2Cu$ along $\{111\}_{Al}$ planes requires forming this 5-layer structure first. Hence, this study still recognizes this 5-layer structure as the Ω phase.

For Ω nanoplate nucleation, previous experimental studies[10,11] have identified that Ω nanoplates evolve continuously from co-clusters or GP zones containing Mg, Ag, and Cu aligned along $\{111\}_{Al}$ matrix planes. In Fig. S3a-c, we evaluated the effects of different chemical compositions in GP zones on stacking fault energies ($\gamma_{sf}$), with the inset displaying the structural model of GP zones and detailed $\gamma_{sf}^{(3,\,5)\text{-planes}}$ values summarized in Table. S1. Specifically, concerning Ag, a high Ag content within the Ag-GP zone (Fig. S3a) can substantially reduces $\gamma_{sf}^{(3)\text{-plane}}$(-0.027 J/m$^2$) to a negative value, while $\gamma_{sf}^{(5)\text{-plane}}$(0.129 J/m$^2$) remains positive, indicating that Ag behaves similarly to Cu. Furthermore, the (Mg, Cu, Ag)-GP zone, characterized by $MgAg_2$ layers as terminations, $Mg_2Al$ layer on the (3)-plane, and Cu on the (2, 4)-planes (inset, Fig. S3b), shows both $\gamma_{sf}^{(3)\text{-plane}}$(-0.075 J/m$^2$) and $\gamma_{sf}^{(5)\text{-plane}}$(-0.010 J/m$^2$) reaching negative values under the combined action of Mg, Cu, and Ag. Notably, the separation between the (Mg, Ag) and Cu within the GP zone has been demonstrated in prior atom probe tomography (APT) study[10], primarily attributed to the positive mixing enthalpy between Ag and Cu. Moreover, the (Mg, Cu)-GP zone with $MgAl_2$ terminations maintains a positive $\gamma_{sf}^{(5)\text{-plane}}$ (0.050 J/m$^2$), implying an active role of Ag in inducing dislocation dissociations at (5)-plane (i.e., facilitate Ω nucleation). Figure. S3c reveals ML-MD calculated $\gamma_{sf}^{(3)\text{-plane}}$ (0.068 J/m$^2$) for (Mg, Cu)-GP zone lacks the negative characteristic obtained from DFT calculations, indicating the limitations of ML potential in describing the intricate interactions between Al, Mg, and Cu. Overall, analogous to (Li, Cu)-rich GP zones observed in the $T_1$ nanoplate, these DFT calculations qualitatively indicate that (Mg, Cu, Ag)-rich GP zones of Ω nanoplate also create a favorable chemical environment for inducing dislocations dissociation at the (3, 5)-planes.

The {111}-oriented $\eta'/\eta_2$ nanoplates are crucial for strengthening 7xxx Al-Zn-Mg-(Cu) series alloys. The commonly accepted precipitation sequence in these alloys is as follows[12–15]: super-saturated solid solution (SSSS)→GP zones→$\eta'$→$\eta_2$, where the equilibrium $\eta_2$-$MgZn_2$ Laves nanoplates evolve progressively from solute clusters, through GP zones, and the precursor $\eta'$ phase, rather than precipitating directly. Besides this $\eta_2$-$MgZn_2$ nanoplate, there are also many other $MgZn_2$ precipitates with diverse morphologies and orientation relationships found in the 7xxx series alloys[16], which are not the focus of this study. Moreover, adding trace Cu is common in 7xxx series alloys, as it remarkably facilitates the nucleation of $\eta'$ nanoplates and enhances the alloys' strength. According to previous experimental observations[12–14], GP zones typically consist of seven $\{111\}_{Al}$ atomic planes, with their composition/structure continuously evolving until they transform into a 7-layer $\eta'$ structure comprising $U_T|K|\omega_T|K|Pu_T|K|U_T$ lattice nets (see Fig. S15).

Unlike $T_1$ and Ω nanoplates, the presence of partial dislocations at the edge of GP zones or $\eta'/\eta_2$ nanoplates has not been experimentally reported, requiring further investigation. Nevertheless, given that hcp-structured $\eta'/\eta_2$ nanoplates precipitate coherently from the fcc-Al matrix, we suggest that (a/6)<11$\bar{2}$> partials also participate in $\eta'/\eta_2$ nucleation, and the possibility for the presence of these partial dislocations was examined below. Using a 5-layer GP zone model (Fig. S3d-f), this section aims to elucidate how Cu segregation within GP zones facilitates dislocation dissociation and migration. Further insights into why $\eta_2$ requires $\eta'$ as its precursor phase and why GP-zone/$\eta'$ occurs with a 7-layer form in the matrix, unlike the 5-layer structure of $T_1$ and Ω nanoplates, will be provided in Supplementary Sections 4 and 8, respectively.

Figure S3d and e show that both $\gamma_{sf}^{(3)\text{-plane}}$ and $\gamma_{sf}^{(5)\text{-plane}}$ values for Zn-GP and (Mg, Zn)-GP zones remain positive, indicating that high Zn content or the co-action of (Mg, Zn) cannot effectively reduce $\gamma_{sf}$ to create a favorable chemical environment for dislocation dissociation at the (3, 5)-planes. Given the observed Cu enrichment in GP zones[13,14], it is inferred that Cu may contribute mainly to dislocation dissociations. Since Fig. S3f and Table. S1 reveal that the ML potential produces acceptable $\gamma_{sf}^{(3,\,5)\text{-planes}}$ values compared to DFT results, we utilize this ML potential to further assess the impact of Cu content in the Zn-GP zone on $\gamma_{sf}^{(3)\text{-plane}}$, as depicted in Fig. S4a. The results indicate that $\gamma_{sf}^{(3)\text{-plane}}$ initially increases slightly, then sharply decreases as Cu content increases, eventually reaching negative values. Hence, linking the diffusional glide mechanism for partial dislocations demonstrated in the main text, it is proposed that Cu is crucial for $\eta'/\eta_2$ nucleation by segregating within the GP zone, and the migration of partial dislocations at the GP zone edge relies on Cu fluctuations around their vicinity (see Fig. S4b), without requiring complete substitution of Zn.

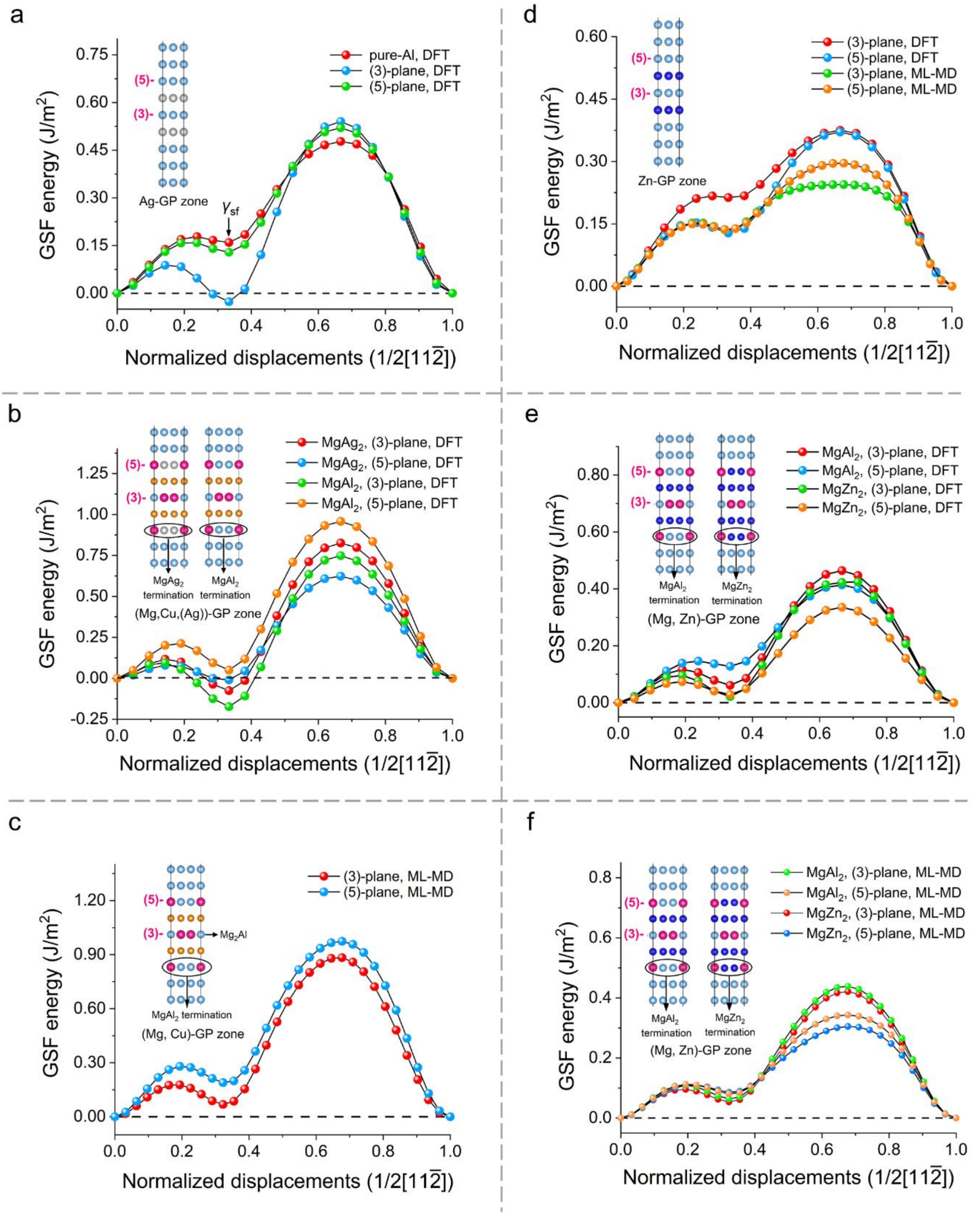

**Figure S3.** GSF energy curves along the [11$\bar{2}$] direction in the (3, 5)-planes of the (Mg, Cu, (Ag))-GP (**a-c**) and (Mg, Zn)-GP (**d-f**) zones, computed using DFT and ML-MD methods. The inset shows the atomic structure of the GP zone model, where light-blue, grey, pink, yellow, and dark-blue balls correspond to Al, Ag, Mg, Cu, and Zn atoms, respectively. The detailed $\gamma_{sf}$ values for (3, 5)-planes in various GP zones are summarized in Table S1.

**Table S1.** Summary of $\gamma_{sf}$ values for (3, 5)-planes in different GP zones.

| Different GP zones | Termination | Planes | $\gamma_{sf}$ (DFT, J/m$^2$) | $\gamma_{sf}$ (ML-MD , J/m$^2$) |
|---|---|---|---|---|
| Ag-GP zone | — | (3)-plane | -0.027 | — |
| | — | (5)-plane | 0.129 | — |
| (Mg, Cu, (Ag))-GP zone | $MgAg_2$ | (3)-plane | -0.075 | — |
| | $MgAg_2$ | (5)-plane | -0.010 | — |
| | $MgAl_2$ | (3)-plane | -0.173 | 0.068 |
| | $MgAl_2$ | (5)-plane | 0.050 | 0.189 |
| Zn-GP zone | — | (3)-plane | 0.213 | 0.136 |
| | — | (5)-plane | 0.128 | 0.136 |
| (Mg, Zn)-GP zone | $MgAl_2$ | (3)-plane | 0.061 | 0.064 |
| | $MgAl_2$ | (5)-plane | 0.128 | 0.081 |
| | $MgZn_2$ | (3)-plane | 0.021 | 0.054 |
| | $MgZn_2$ | (5)-plane | 0.028 | 0.086 |

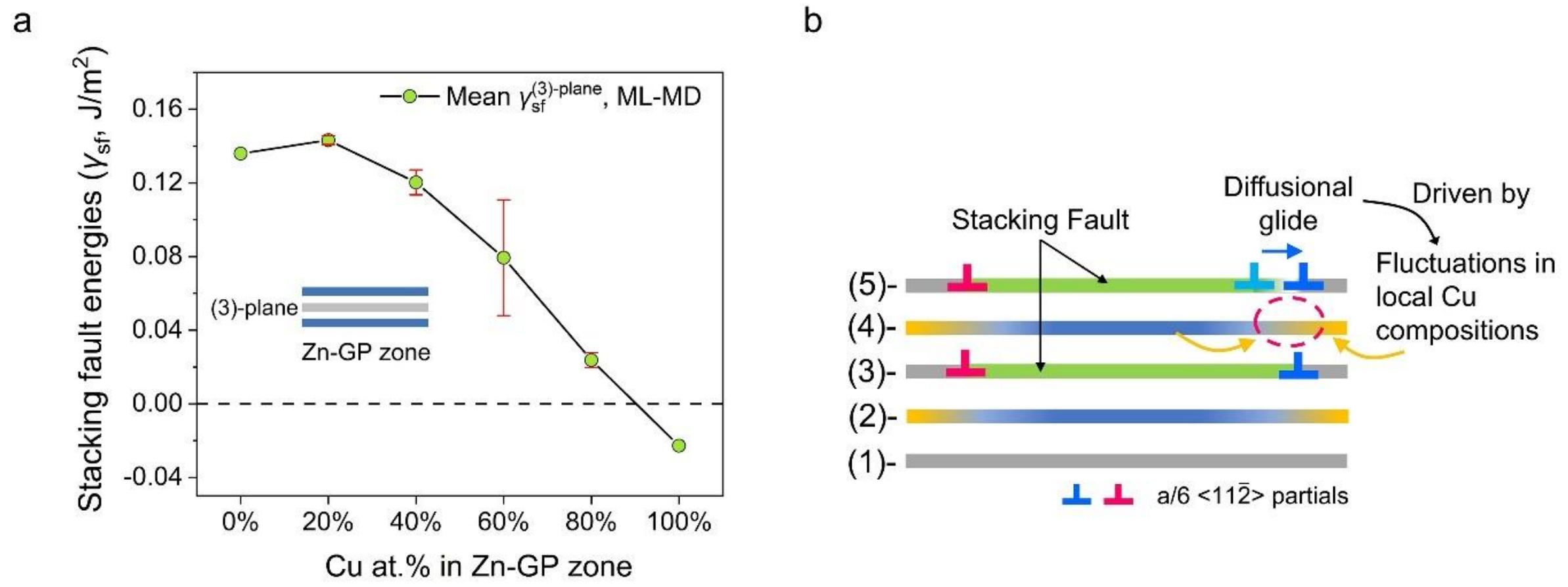

**Figure S4. a.** Variation of ML-MD measured $\gamma_{sf}$ of the (3)-plane with varying Cu concentrations in the Zn-GP zone. Error bars represent the standard deviation (SD) of $\gamma_{sf}$ from 10 independent sampling configurations. **b**. Schematic illustration of the diffusional glide mechanism for partial dislocations on the (3, 5)-planes, driven by fluctuations in local Cu compositions.

## 2. Supplementary Section 2: Diffusion attempt frequency of various concerted hopping events

Figure S5 presents the atomic structures of hcp-Al and vacuum-containing models. These two models were employed to evaluate the migration barriers of different solutes across various atomic planes, as detailed in Table S2. The computed migration barriers exhibit negligible variations between the two models, indicating that the hcp-Al model accurately reproduces these barriers. This allows for precise quantitative assessments of activation energies for concerted hopping events via hcp-Al models.

To circumvent the high computational costs associated with directly calculating vibrational frequencies in the large 600-atom and 870-atom models, which are used for evaluating migration barriers of concerted hopping events, we approximated the attempt frequencies for six concerted hopping events using the diffusion attempt frequency of solutes within the Al matrix. Figure S6 illustrates two possible vacancy-mediated diffusion jumps within and between $\{111\}_{Al}$ planes, each with corresponding diffusion attempt frequencies $v_{-}$ and $v_{|}$. Table S3 presents the calculated $v_{-}$ and $v_{|}$ values for Al, Cu, Li, Mg, and Zn in the Al matrix, derived from a 35-atom fcc-Al model with a single vacancy. These values are converged after validation using a larger 95-atom model. Additionally, $v_{-}$ and $v_{|}$ values are in the same order of magnitude as the lattice vibration frequency[17,18], and the $v_{-}^{Li} > v_{-}^{Al} > v_{-}^{Cu} > v_{-}^{Zn}$ ($v_{|}^{Li} > v_{|}^{Al} > v_{|}^{Cu} > v_{|}^{Zn}$) relations display that they scale inversely with their respective mass, further affirming their reliability. Since concerted hopping events involve multiple atoms, the attempt frequency $v_0$ for the entire event should be based on the lowest among them. For instance, as $v_0$ for different hopping events presented in Table S4, the attempt frequency $v_0$ for the Ep1-AlAlCu event should equal $v_{-}^{Cu}$ (1.79 THz).

**Table S2.** Migration barriers ($E_{X}^{(3,\ 4)\text{-planes}}$) of various solutes X across different atomic planes in the hcp-Al and Vacuum-containing models. Values are in eV.

| Solutes | Hcp-Al Model (eV) | Vacuum-containing Model (eV) | Deviation (%) |
|---|---|---|---|
| $E_{Al}^{(3)\text{-plane}}$ | 0.524 | 0.509 | +2.95 |
| $E_{Cu}^{(4)\text{-plane}}$ | 0.305 | 0.307 | -0.65 |
| $E_{Li}^{(3)\text{-plane}}$ | 0.456 | 0.431 | +5.80 |
| $E_{Mg}^{(3)\text{-plane}}$ | 0.395 | 0.415 | -4.82 |
| $E_{Zn}^{(4)\text{-plane}}$ | 0.112 | 0.117 | -4.27 |

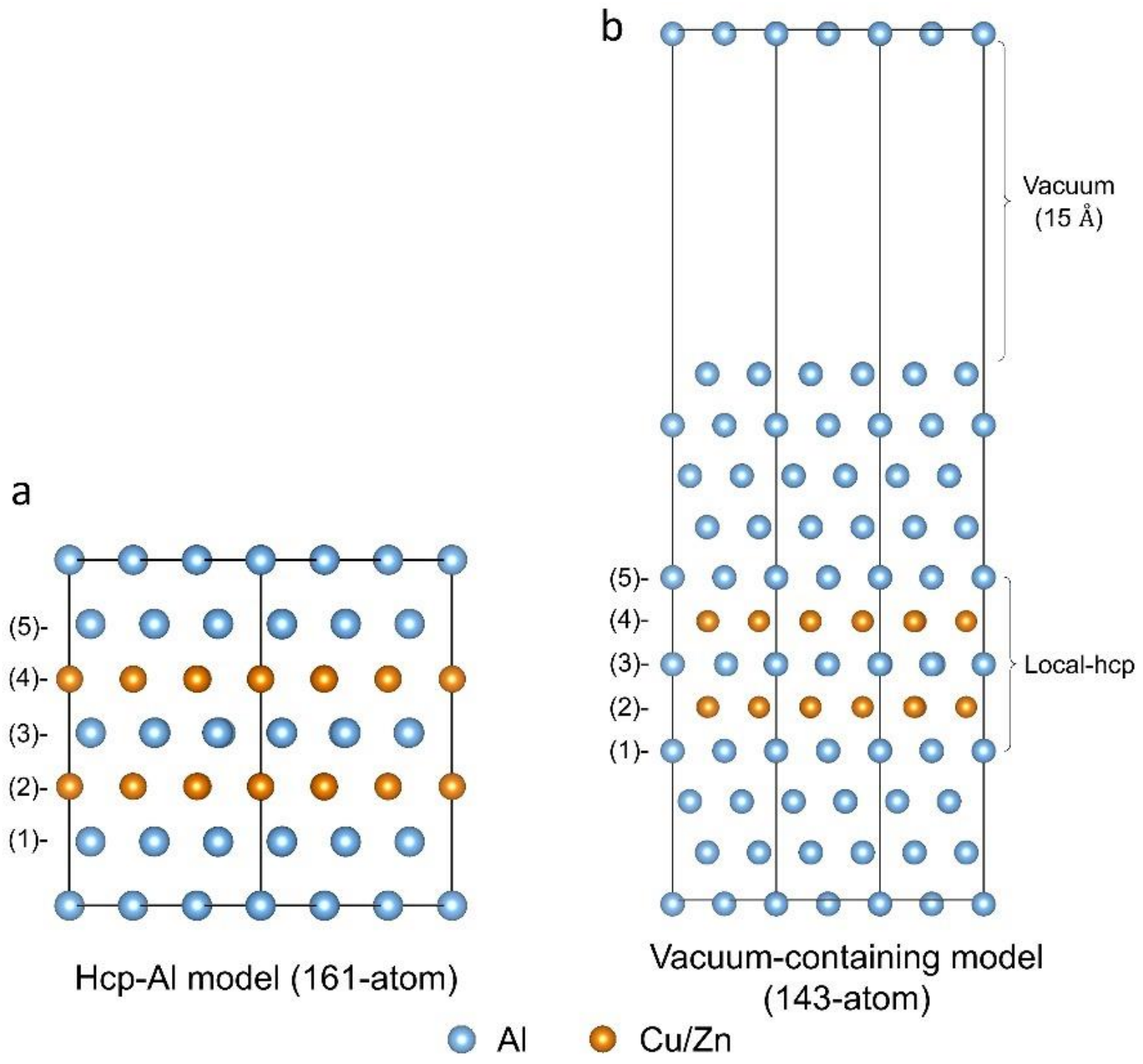

**Figure S5.** Atomic structure of the hcp-Al model (a) and vacuum-containing model (b), with which (2, 4)-planes are occupied by Cu/Zn atoms. The light-blue and yellow balls represent Al and Cu/Zn atoms, respectively.

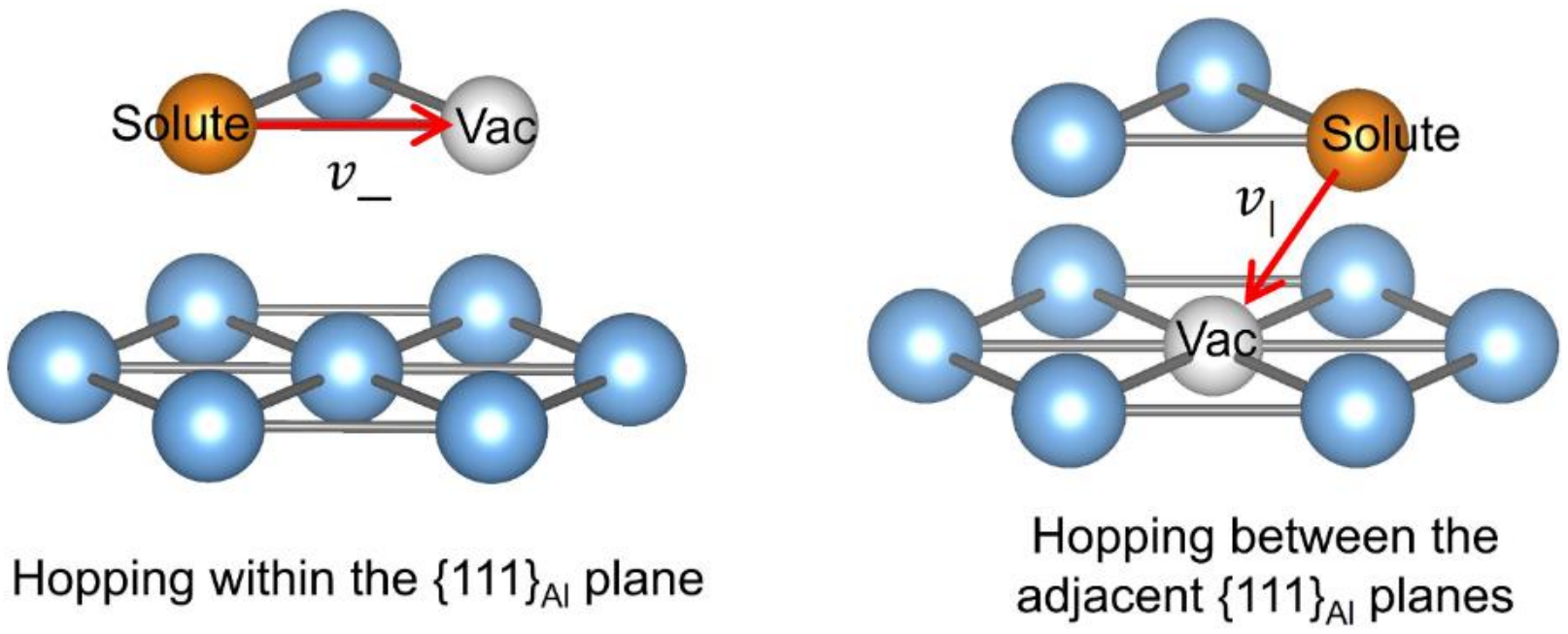

**Figure S6.** Schematic illustration of the two possible vacancy-mediated diffusion jumps in an fcc lattice for the solute hopping within the $\{111\}_{Al}$ plane and between the adjacent $\{111\}_{Al}$ planes. $v_{-}$ and $v_{|}$ are the corresponding diffusion attempt frequencies.

**Table S3**. Calculated attempt frequency of different solutes in the Al matrix. Values are in THz.

| Diffusion attempt frequency | Al | Cu | Li | Mg | Zn |
|---|---|---|---|---|---|
| Hopping within the plane $v_{-}$ (THz) | 5.41 | 1.79 | 12.95 | 13.67 | 1.63 |
| Hopping between the adjacent planes $v_{\|}$ (THz) | 8.25 | 1.71 | 10.40 | 9.48 | 1.15 |

**Table S4.** Diffusion attempt frequency $v_0$ for various concerted hopping events.

| | Concerted hopping | Events | A | B | C | $v_0$ (THz) |
|---|---|---|---|---|---|---|
| $T_1$ | Path-1 | Ep1-AlAlCu | $v_{\|}^{Al}$ | $v_{-}^{Al}$ | $v_{-}^{Cu}$ | $v_{-}^{Cu}$ (1.79) |
| | | Ep1-AlAlAl | $v_{\|}^{Al}$ | $v_{-}^{Al}$ | $v_{-}^{Al}$ | $v_{-}^{Al}$ (5.41) |
| | | Ep1-AlCuCu | $v_{\|}^{Al}$ | $v_{-}^{Cu}$ | $v_{-}^{Cu}$ | $v_{-}^{Cu}$ (1.79) |
| | Path-2 | Ep2-AlAl | $v_{\|}^{Al}$ | $v_{-}^{Al}$ | — | $v_{-}^{Al}$ (5.41) |
| | | Ep2-AlLi | $v_{\|}^{Al}$ | $v_{-}^{Li}$ | — | $v_{\|}^{Al}$ (8.25) |
| | | Ep2-LiLi | $v_{\|}^{Li}$ | $v_{-}^{Li}$ | — | $v_{\|}^{Li}$ (10.40) |
| $\eta'$ | Path-1 | Ep1-AlAlZn | $v_{\|}^{Al}$ | $v_{-}^{Al}$ | $v_{-}^{Zn}$ | $v_{-}^{Zn}$ (1.63) |
| | | Ep1-AlAlAl | $v_{\|}^{Al}$ | $v_{-}^{Al}$ | $v_{-}^{Al}$ | $v_{-}^{Al}$ (5.41) |
| | | Ep1-AlZnZn | $v_{\|}^{Al}$ | $v_{-}^{Zn}$ | $v_{-}^{Zn}$ | $v_{-}^{Zn}$ (1.63) |
| | Path-2 | Ep2-AlAl | $v_{\|}^{Al}$ | $v_{-}^{Al}$ | — | $v_{-}^{Al}$ (5.41) |
| | | Ep2-AlMg | $v_{\|}^{Al}$ | $v_{-}^{Mg}$ | — | $v_{\|}^{Al}$ (8.25) |
| | | Ep2-MgMg | $v_{\|}^{Mg}$ | $v_{-}^{Mg}$ | — | $v_{\|}^{Mg}$ (9.48) |

## 3. Supplementary Section 3: Thermodynamic origin of the $T_1$ nanoplate

**Methods:**

This section employs convex hull analysis to evaluate the phase stability of the $T_1$ nanoplate at both 0 K and high temperatures. A comprehensive overview of this method can be found in ref.[19]. For a compound to be thermodynamically stable, it must not only be lower in energy than all other phases at the same composition but also be lower in energy than all linear combinations of phases. The collection of all ground states in a given phase space constitutes the convex hull, with phases located above the convex hull considered unstable and those on the hull recognized as stable phases.

The Al-Cu-Li DFT-calculated convex hull diagrams at 0 K and high temperatures were constructed using the *pymatgen* code[20], with structure data sourced from the Materials Project[21] and Open Quantum Material Database (OQMD)[22,23] databases, alongside manually incorporated $Al_6Cu_4Li_3$ structures from previous studies[24,25]. At 0 K, the ternary convex hull considers only internal energy, while the high-temperature convex hulls further account for vibrational entropy contributions. Typically, the entropy associated with harmonic atomic vibrations at high temperatures is proportional to the logarithmic moment of the phonon densities of states (DOS), $g(\omega)$[26]:

$$S_{vib}=k_B\int_0^\infty\left[1+ln\left(\frac{k_BT}{\hbar\omega}\right)+\ldots\right]g(\omega)d\omega \tag{S-1}$$

Where $k_B$ is the Boltzmann constant, $\hbar$ the reduced Planck's constant, and $\omega$ the phonon frequencies. Higher-order terms in Eq. (S-1) are negligibly small for temperatures higher than the characteristic Debye temperature $\theta_D$ of atomic vibrations. Because typical metals $\theta_D$ is well below room temperature, Eq. (S-1) can be truncated after the first two terms and thus the contribution of $S_{vib}$ to the free energy can be obtained by $E_v = -TS_{vib}$. For phonon dispersion and $g(\omega)$ calculations, the harmonic interatomic force constants were obtained by density functional perturbation theory (DFPT) using the supercell approach, which calculates the dynamical matrix through the linear response of electron density[27]. Then, phonon dispersion and $g(\omega)$ were computed using the *Phonopy*[28] code with the obtained harmonic interatomic force constants as input.

The second-order elastic constants play a crucial role in governing materials's mechanical and dynamical properties, particularly their stability and stiffness[29]. These elastic constants are calculated using the stress-strain method via the *VASPKIT* code[30] for predicted phases, and then their mechanical stability is further evaluated. Born-Oppenheimer molecular dynamics is used for the *ab initio* molecular dynamics (AIMD) calculations such that the electronic and ionic subsystems are fully decoupled. The energy tolerance for the electronic relaxation is set as $1\times10^{-5}$ eV. All the simulations were performed in a canonical NVT ensemble (i.e., keeping constant atom number, volume, and temperature) with temperature (500 K) controlled by the Nosé thermostat[31,32]. The total number of time steps is 5000 with each time step corresponding to 2 fs.

**Results and discussions:**

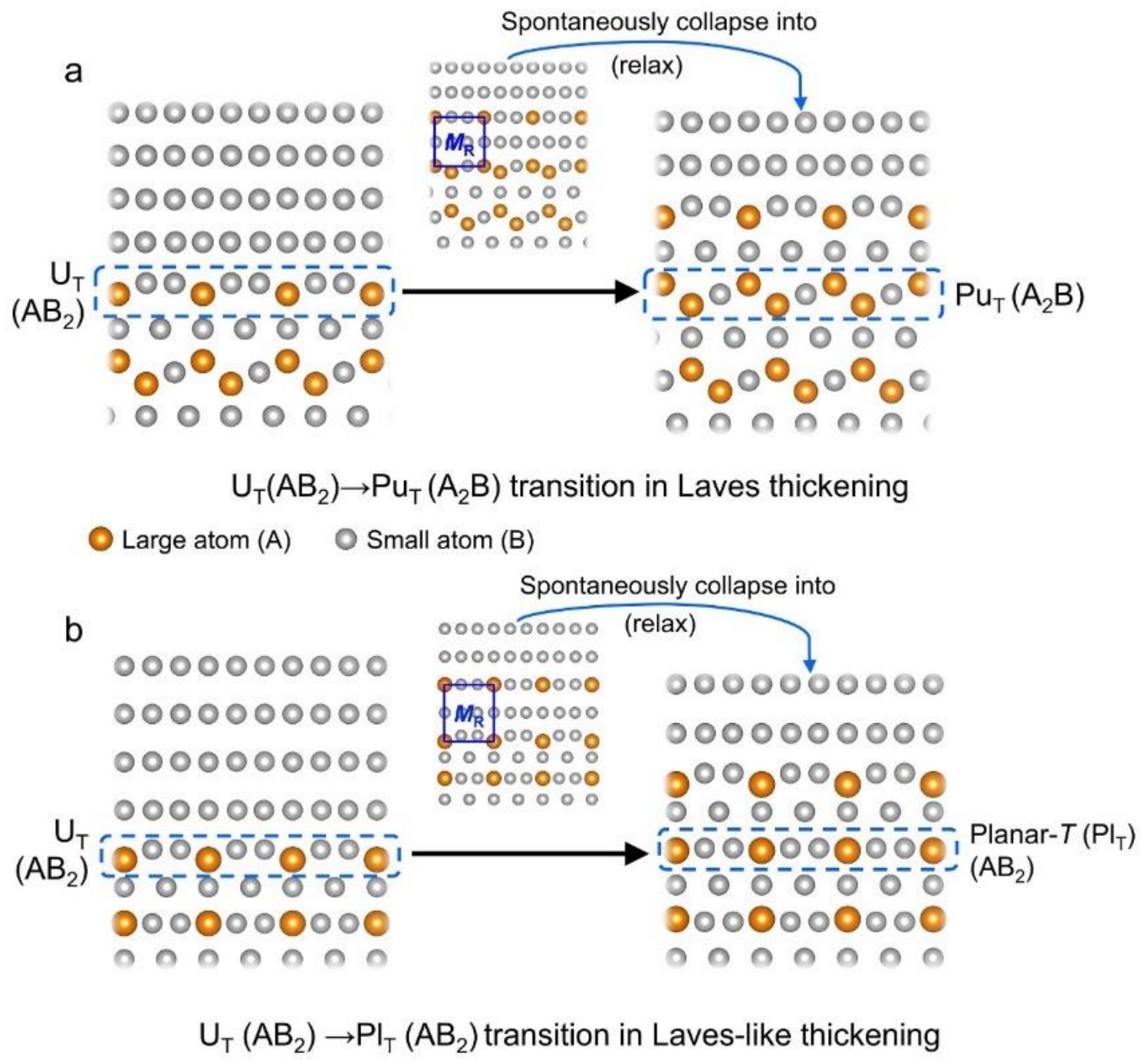

**Figure S7.** Schematic illustrating $U_T(AB_2)$→$Pu_T(A_2B)$ and $U_T(AB_2)$→$Pl_T(AB_2)$ transitions in Laves and Laves-like phases thickening via forming $M_R$-type BSTU on the broad terrace (i.e., $U_T$ lattice net) of nanoplates. Yellow and white balls represent large (A) and small (B) atoms, respectively.

According to our prior work[33] on TCP thickening in Mg alloys (see Fig. S7), while both Laves and Laves-like ($A_2B_7$, $AB_3$, and $AB_5$ types with Planar-*T* ($Pl_T$) lattice nets) phases require the formation of new $M_R$-type BSTUs along their outermost $U_T$ lattice, the former thickens via the $U_T(AB_2)$→ $Pu_T(A_2B)$ transition and the latter thickens through the $U_T(AB_2)$→$Pl_T(AB_2)$ transition without changing $U_T$'s composition. This study reveals that both transitions can occur during the thickening of $T_1$ nanoplates, with the $U_T$→$Pl_T$ transition predominantly driving $T_1$ nanoplate growth. Due to the extremely thin thickness of the $T_1$ nanoplate, early experimental studies[34–37] on its crystal structures and stoichiometry have been controversial. Although aberration-corrected high-angle annular dark-field scanning transmission electron microscopy (HAADF-STEM) has confirmed the 5-layer $U_T|K|\omega_T|K|U_T$ structure of the thinnest $T_1$ nanoplates, the exact stoichiometry, particularly the Cu: Al ratio within the *K* lattice nets, remains unclear. To resolve these disputes, Kim et al.[24], Wang et al.[25], and Liu et al.[38] have successively employed cluster expansion (CE) methods to search for a thermodynamically stable structure for the $T_1$ phase, proposing the $Al_6Cu_4Li_3$ structure at 0 K. However, our reconstructed 0 K Al-Cu-Li convex hull (see Fig. S8a) shows that $Al_6Cu_4Li_3$ does not form a tie-line with Al, indicating it is not in equilibrium with Al. This contradicts the two-phase region between $T_1$ and Al solid solution observed in the experimental phase diagram[39].

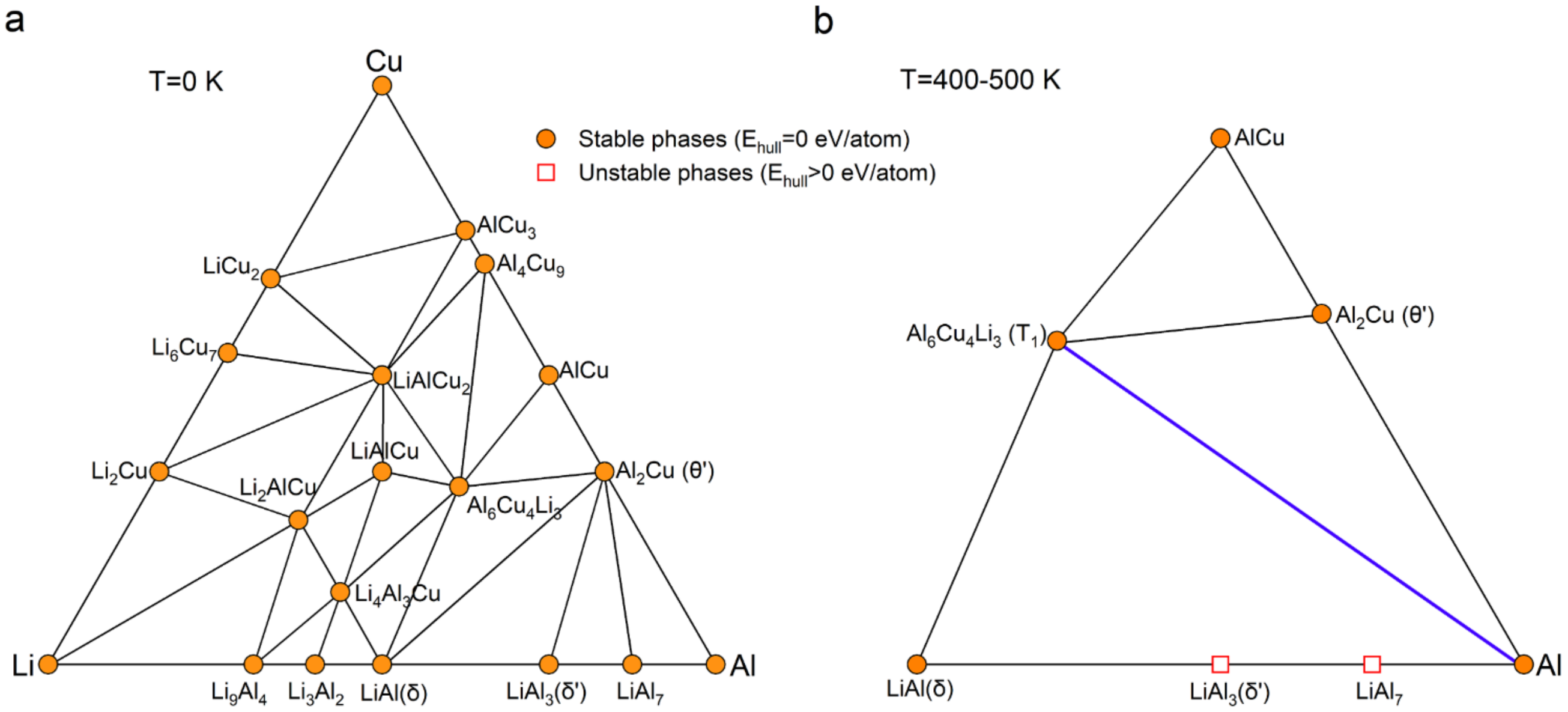

**Figure S8**. Ternary Al-Cu-Li convex hull at T=0 K (a) and T=400-500 K (b).

The 0 K convex hull diagram considers only the internal energy; however, vibration entropy ($S_{vib}$) at elevated temperatures may significantly affect phase stability, such as the θ-$Al_2Cu$ phase is stabilized by $S_{vib}$ compared to the θ′-$Al_2Cu$ phase[40]. This study further incorporates $S_{vib}$ to evaluate phase stability at aging temperatures (400-500K). Specifically, we calculated $S_{vib}$ for phases in the Al-rich corner of the Al-Cu-Li convex hull, including δ-LiAl, δ′-$LiAl_3$, $LiAl_7$, $Al_6Cu_4Li_3$, AlCu, θ-$Al_2Cu$ (*C*16 structure), and θ′-$Al_2Cu$ (*C*1 structure), to construct the convex hull at 400-500K (see Fig. S8b). Our results indicate that the $Al_6Cu_4Li_3$ phase forms a tie-line (marked in blue) with Al, as the $LiAl_3$(δ′) and $LiAl_7$ phases, stable at 0 K, become unstable within this temperature range. This is consistent with experimental observations that $LiAl_3$(δ′) is typically a metastable phase[41]. Consequently, the stable $T_1$ nanoplate should be the $Al_6Cu_4Li_3$ phase thermodynamically. The dynamical, structural, and mechanical stability of the $Al_6Cu_4Li_3$ phase is confirmed in Fig. S9(b-c) and Table. S5.

**Table S5**. Computed elastic constants for $Al_6Cu_4Li_3$ and $Al_4Cu_6Li_3$ phases. Values are in GPa.

| Phase | C11 | C12 | C13 | C22 | C23 | C33 | C44 | C55 | C66 | Mechanical Stability |
|---|---|---|---|---|---|---|---|---|---|---|
| $Al_6Cu_4Li_3$ | 167.95 | 49.41 | 40.08 | 172.85 | 36.18 | 127.88 | 39.17 | 33.63 | 62.13 | Stable |
| $Al_4Cu_6Li_3$ | 171.28 | 60.82 | 33.23 | — | — | 141.23 | 23.44 | — | — | Stable |

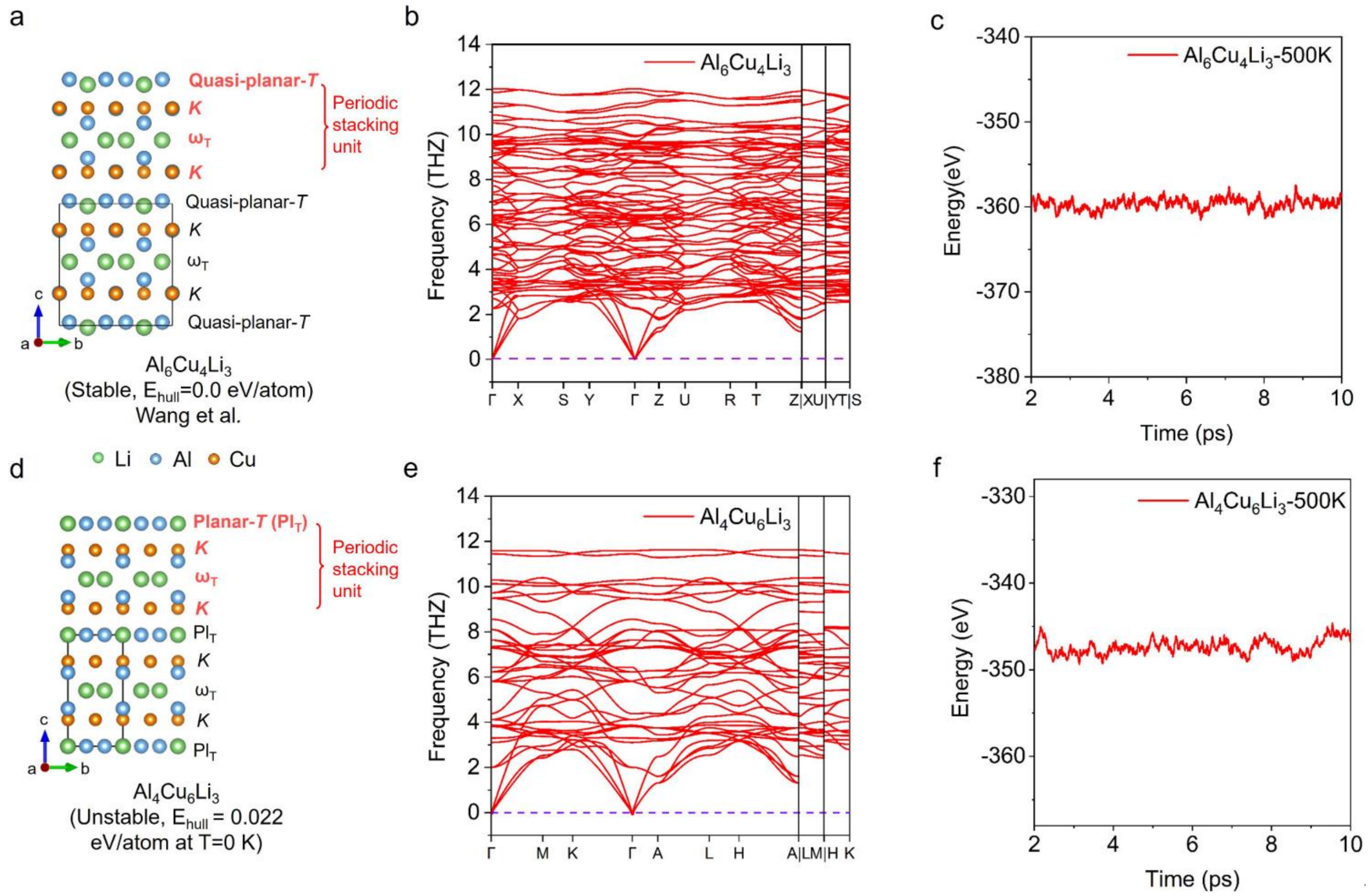

**Fig. S9**. The atomic structures, phonon dispersion curves, and system's energy variation as a function of time at T=500 K for the $Al_6Cu_4Li_3$ (a-c) and $Al_4Cu_6Li_3$ (d-f) phases.

Nevertheless, since each atomic column in the *K* lattice of $Al_6Cu_4Li_3$ contains identical Cu content along the $<11\bar{2}>_{Al}$ direction (see Fig. S10), the simulated HAADF-STEM image of $Al_6Cu_4Li_3$ cannot show an "O-unit" feature. This contrasts with the distinct "O-unit" structure observed experimentally in the $T_1$ nanoplate (see Figs. 4b and S11b). Notably, the interlayer-sliding stage requires a substantial Cu content within the *K*-forming planes to facilitate the nucleation and migration of partial dislocations; meanwhile, Supplementary Section 6 reveals that $T_1$ nucleation follows a nonclassical behavior that does not require the same composition as the equilibrium $Al_6Cu_4Li_3$ phase. As a result of kinetic constraints, the experimentally observed $T_1$ nanoplates are typically metastable and require long-term compositional adjustments, primarily involving the absorption of Al into the *K* lattices, to evolve into the final $Al_6Cu_4Li_3$ structure. Consequently, we suggest these observed $T_1$ nanoplates as a metastable $Al_4Cu_6Li_3$ phase containing a pure Cu *K* lattice (see Fig. S9d). The dynamical, structural, and mechanical stability of the $Al_4Cu_6Li_3$ phase is corroborated in Fig. S9(e-f) and Table. S5.

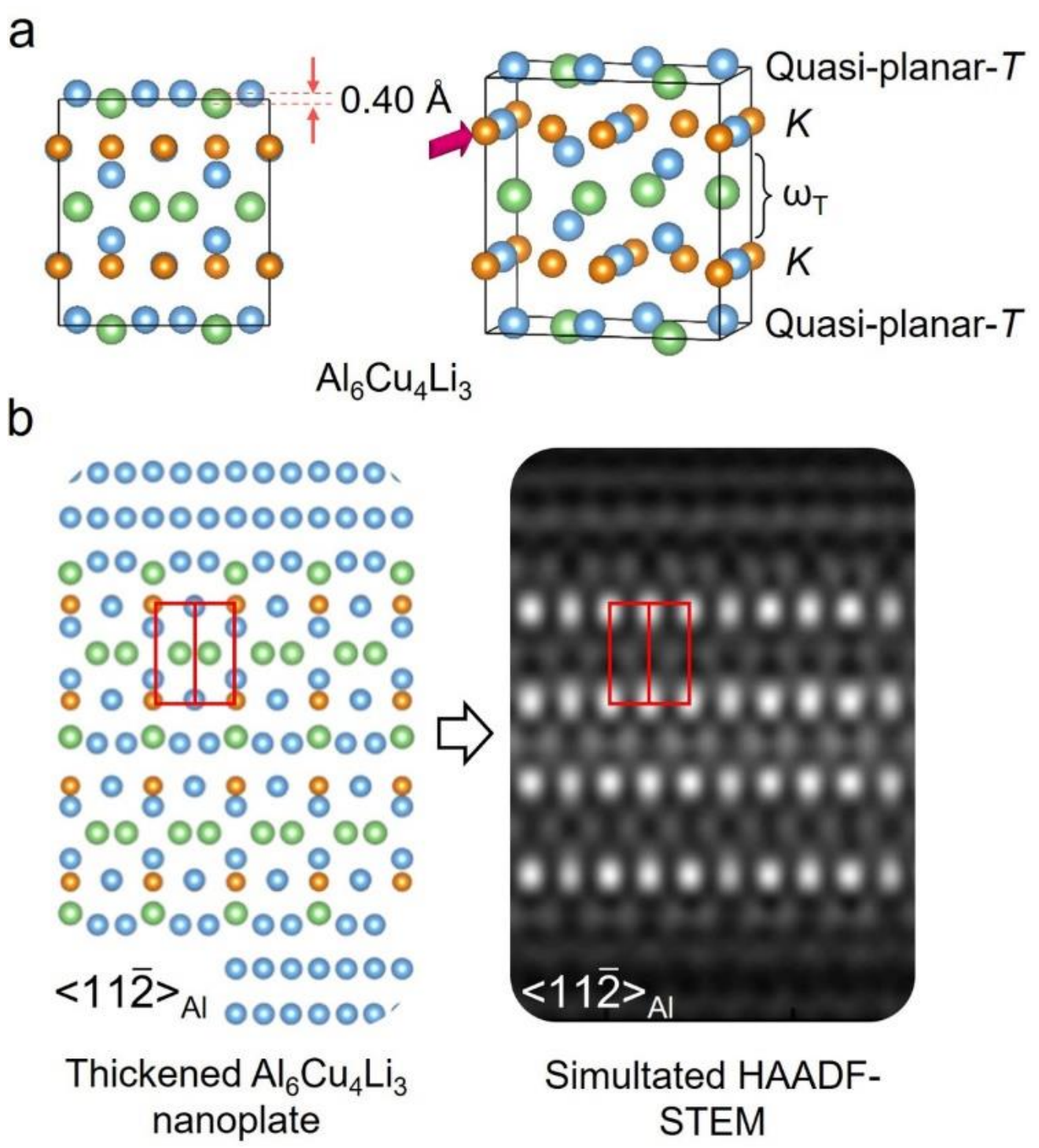

**Figure S10. a.** Atomic structure of the unit cell for the ternary $Al_6Cu_4Li_3$ phase, as suggested by Wang et al.[25]. **b**. The structural model for a thickened $Al_6Cu_4Li_3$ nanoplate and the corresponding simulated HAADF-STEM image. Li, Cu, and Al atoms are denoted by light-green, yellow, and light-blue balls, respectively.

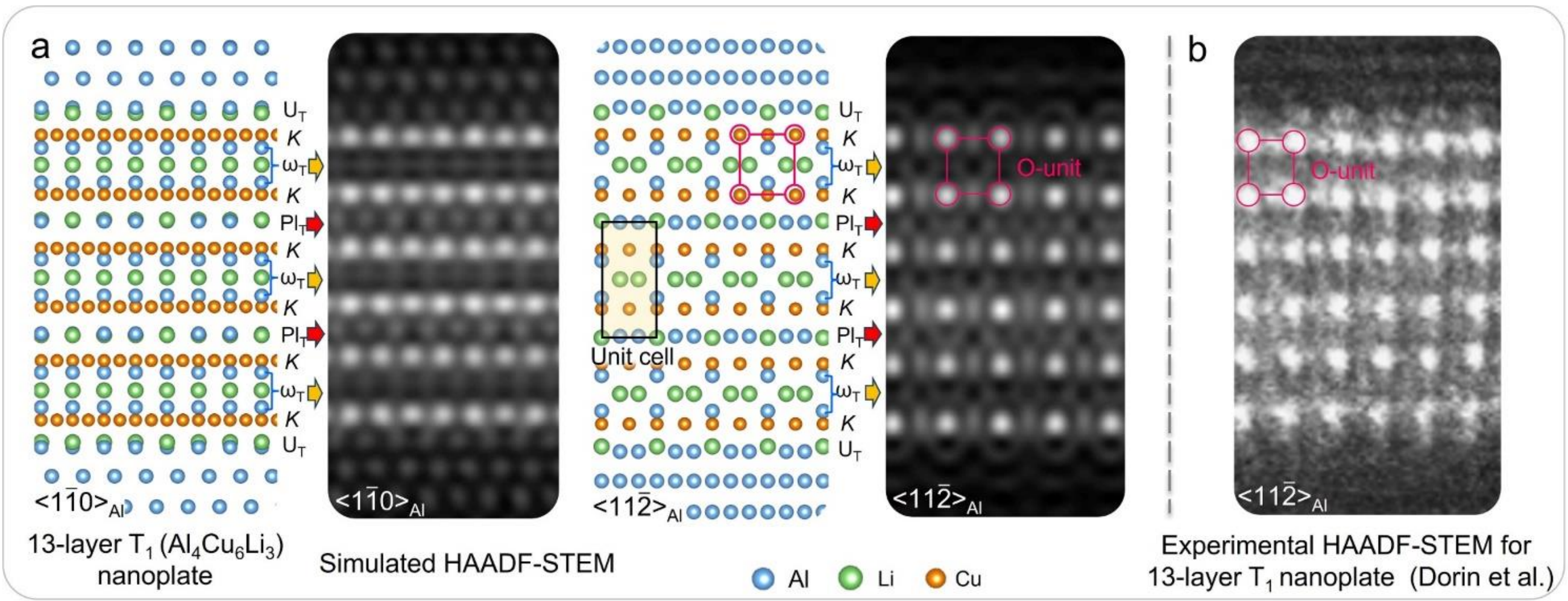

**Figure S11. a.** Simulated HAADF-STEM images of a 13-layer $T_1$ ($Al_4Cu_6Li_3$) nanoplate from $<1\bar{1}0>_{Al}$ and $<11\bar{2}>_{Al}$ observations. Light-blue, light-green, and yellow balls represent Al, Li, and Cu atoms, respectively. **b**. Experimental HAADF-STEM image of a 13-layer $T_1$ ($Al_4Cu_6Li_3$) nanoplate from $<11\bar{2}>_{Al}$ projection, as reported by Dorin et al.[42].

Typically, as detailed in the main text, the structural evolution of $T_1$ nanoplates within the Al matrix is expected to follow a rule of 5-layer→9-layer→13-layer→···→ (5+4*n*)-layer (where *n* represents the number of thickening times). However, deviations from this expected rule, such as the 11-layer and 10-layer defective $T_1$ nanoplates shown in Fig. S12a and Fig. S13a, respectively, may frequently occur. Here, we mainly explain the formation mechanisms of these 11-layer and 10-layer defective $T_1$ nanoplates, by which insights obtained can be a reference for resolving other types of defective $T_1$ nanoplates. Specifically, the 11-layer $T_1$ nanoplate, previously reported by Häusler et al.[43], is formed from a 5-layer $T_1$ structure through sequential $M_R$-type and $M_T$-type BSTU formations after a local hcp-lattice has been created, as shown in Fig. S12d. However, since Häusler et al.[43] only provide a $<1\bar{1}0>_{Al}$ projection of the 11-layer $T_1$ nanoplate, it is unclear whether the $U_T \rightarrow Pu_T$ or $U_T \rightarrow Pl_T$ transition occurred during the second-step thickening. Consequently, both Type-I and Type-II configurations may be possible for this 11-layer nanoplate. This also indicates that discerning the difference between the $<1\bar{1}0>_{Al}$ projection of $Pu_T$ and $Pl_T$ is difficult, making the $<11\bar{2}>_{Al}$ projection essential experimentally. Regarding the 10-layer $T_1$ nanoplate (Fig. S13) reported by Gao et al.[44], we found that this structure arises when the $M_T$-type BSTU forms along the (6-10)-planes instead of the (5)-plane (i.e., $U_T$), leading to the formation of two 5-layer $T_1$ nanoplates without gaps between them. We emphasize that all the transitions above have been confirmed in large-sized models containing the surrounding matrix. Moreover, such defective $T_1$ nanoplates are often metastable, like the Type-II 11-layer $T_1$ configurations, which can be considered as the metastable $Al_2Cu_{11}Li_6$ ($E_{hull}$=0.017 eV/atom at 0K) phase. The reasons behind the prevalence of these defective nanoplates are elucidated in Supplementary Section 6.

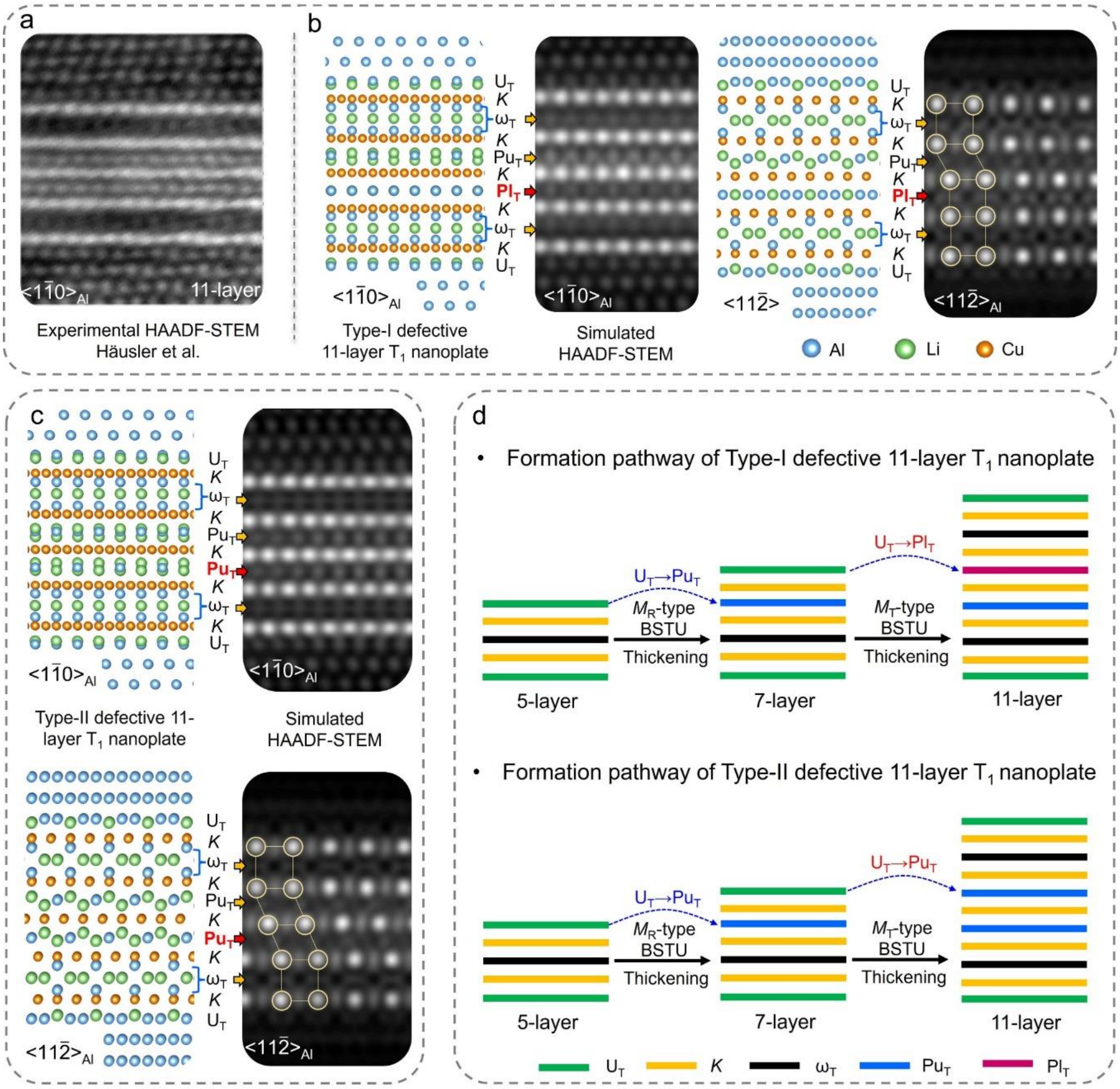

**Figure S12**. **a**. Experimental HAADF-STEM image of a defective 11-layer $T_1$ nanoplate from the $<1\bar{1}0>_{Al}$ projection, reported by Häusler et al.[43]. **b-c**. Simulated HAADF-STEM images of the defective 11-layer $T_1$ nanoplate based on two configurations: Type-I and Type-II. **d**. Formation pathways for Type-I and II defective 11-layer $T_1$ nanoplates.

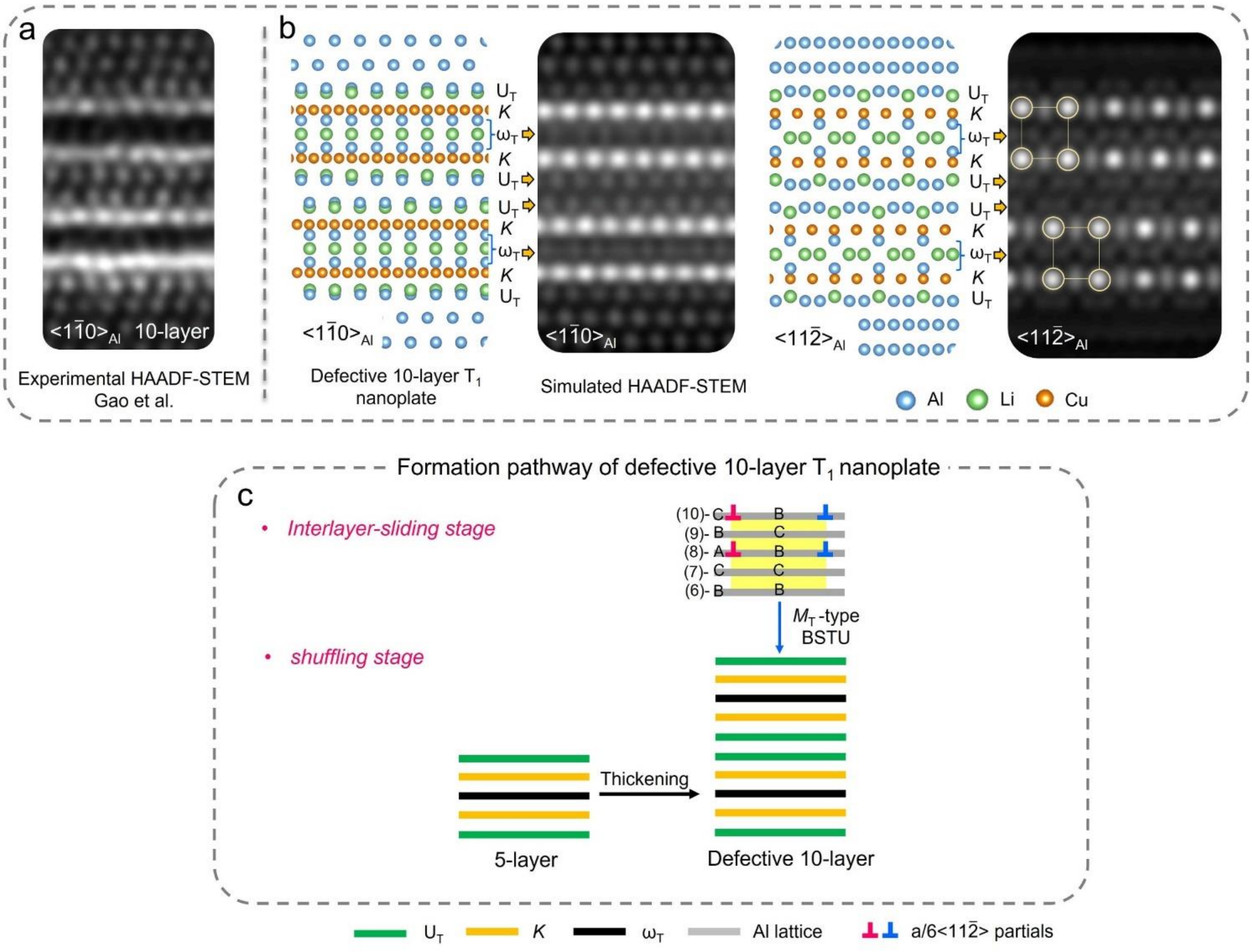

**Figure S13. a.** Experimental HAADF-STEM image of a defective 10-layer $T_1$ nanoplate from the <1$\bar{1}$0>$_{Al}$ projection, reported by Gao et al.[44]. **b**. Simulated HAADF-STEM images of the defective 10-layer $T_1$ nanoplate. **c**. Formation pathway for defective 10-layer $T_1$ nanoplate.

## 4. Supplementary Section 4: Structural evolution pathway of the $\eta'/\eta_2$ nanoplates within the Al matrix

Analogous to 7xxx series Al alloys, hcp-based Mg alloys can also in situ precipitate $MgZn_2$ Laves nanoplates along their $\{0001\}_{Mg}$ basal planes during aging. Our recent study[33] has identified a 3-layer unstable $M_R$-type hcp-ordering (see Fig. S14a) that acts as the BSTU governing the entire hcp→Laves transitions, resulting in Laves formation starting from a 3-layer $U_T|K|U_T$ ($M$-type) structure and thickening two {0001} atomic planes each time, as illustrated in Fig. S14b. Notably, the thickening of these Laves nanoplates depends on changing the outermost $U_T$ into a $Pu_T$ (i.e., $U_T \rightarrow Pu_T$) through forming new $M_R$-type BSTU based on these $U_T$ lattice nets, without requiring any dislocations. Moreover, based on this pathway, we evaluated that the critical nucleus for $MgZn_2$ precipitation should possess a 7-layer $U_T|K|Pu_T|K|Pu_T|K|U_T$ structure[33], which might be the thinnest $MgZn_2$ nanoplate observable in experiments. Smaller nuclei, such as 3-layer or 5-layer nuclei, are expected to dissipate due to thermal fluctuations and thus cannot be clearly observed.

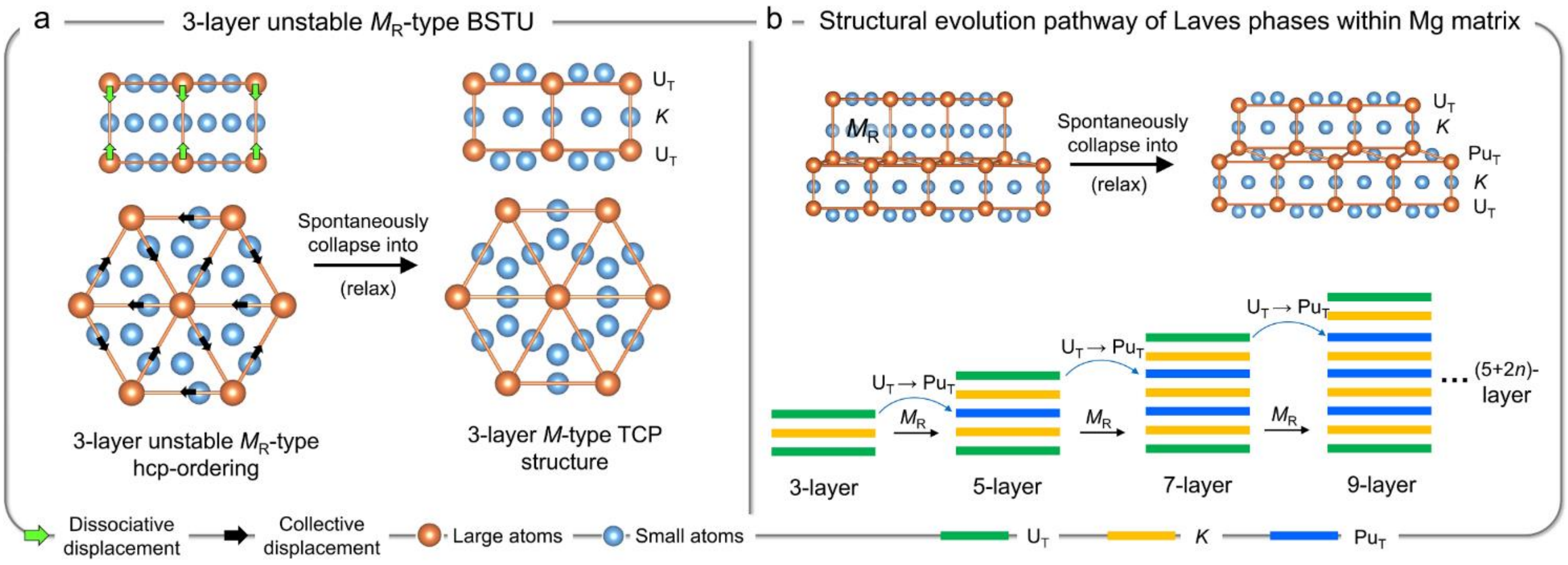

**Figure S14. a**. Schematic illustrating the spontaneous transformation of a 3-layer unstable $M_R$-type hcp-ordering into a 3-layer $M$-type TCP structure composed of $U_T|K|U_T$ lattice nets. Large and small atoms are represented by yellow and light-blue balls, respectively. **b**. Schematic illustrating the structural evolution pathway of the Laves phases within the hcp-base Mg matrix via $M_R$-type BSTU[33].

Owing to the fcc matrix of Al alloys, $MgZn_2$-$\eta_2$ nanoplates in 7xxx Al-Zn-Mg-(Cu) series alloys exhibit a unique precipitation pathway compared to hcp-based Mg alloys. As introduced in Supplementary Section 1, equilibrium $MgZn_2$-$\eta_2$ nanoplates evolve sequentially from 7-layer GP zones and precursor $\eta'$ phase rather than precipitating directly. Additionally, the segregation or partitioning of Cu into the GP zones can effectively induce the nucleation and migration of $(a/6)\langle 11\bar{2}\rangle$ partial dislocations to create local hcp lattices, facilitating $\eta'/\eta_2$ formation. Figure S15a shows the $\eta'$ nanoplate's 7-layer $U_T|K|\omega_T|K|Pu_T|K|U_T$ structure, which is also interpreted by researchers[12–14] as comprising “O” and “R” subunits based on Zn columns.

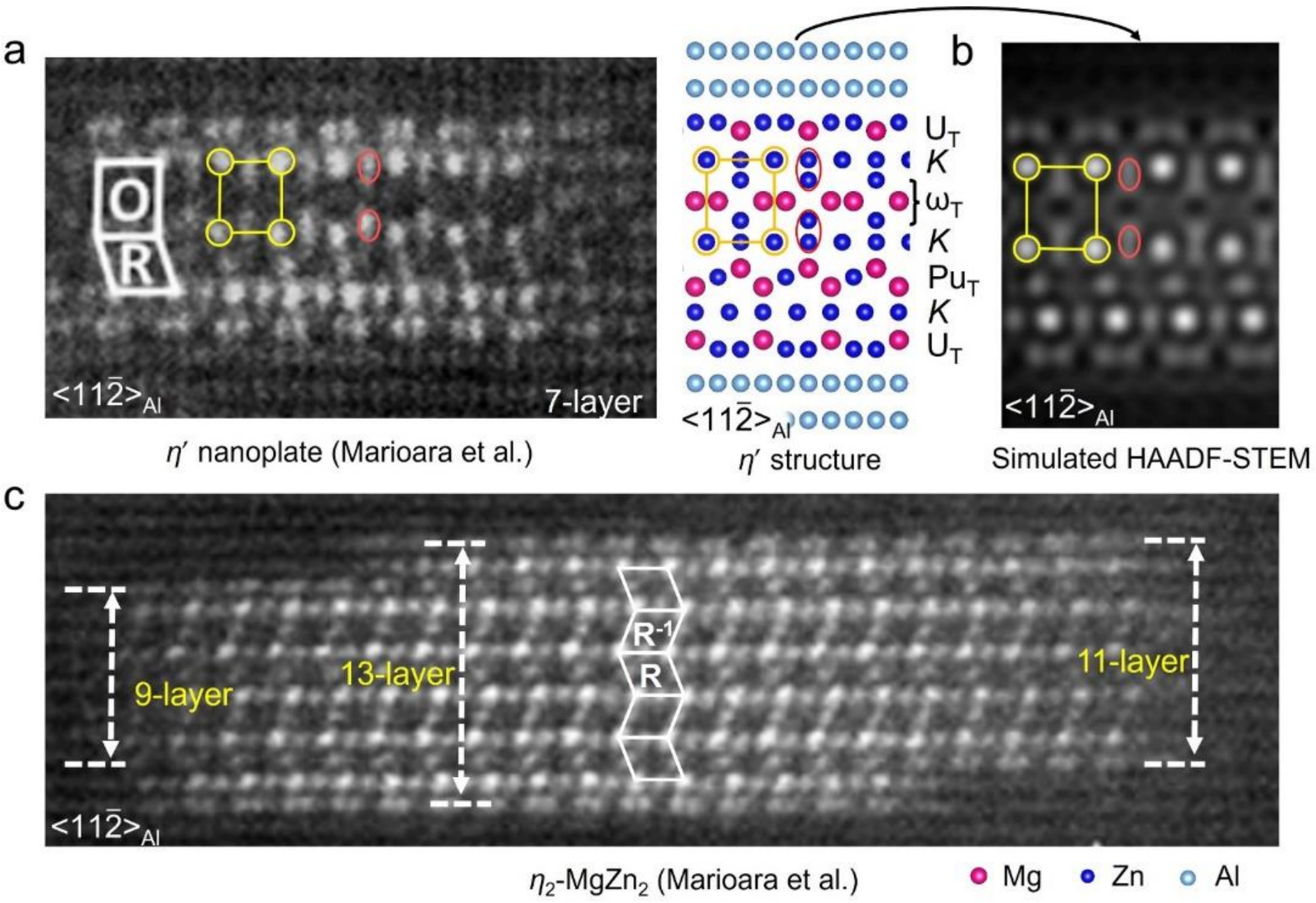

**Figure S15**. **a**. The left panel shows the experimental observation of a 7-layer ($U_T|K|\omega_T|K|Pu_T|K|U_T$) $\eta'$ nanoplate, while the right panel displays the corresponding atomic structure. Mg, Zn, and Al atoms are represented by pink, dark-blue, and light-blue balls, respectively. **b**. A HADDF-STEM image for the $\eta_2$-$MgZn_2$ nanoplate showing various thicknesses. These images are reported by Marioara et al.[12].

Unlike the 5-layer $T_1$ and Ω nanoplates, where Al-Al pairs establish their $\omega_T$, previous experimental studies[12,45] suggest that the $\omega_T$ in $\eta'$ plates consists of Zn-Zn pairs. Our evaluation (Fig. S16a) of activation energies ($E_a$) for six concerted hopping events indicates that Ep1-AlZnZn predominantly drives structural transitions during the incubation and late-growth stages due to its lower $E_a$ values and high Zn content in the (2, 4)-planes. This kinetically facilitates the formation of Zn-rich $\omega_T$. Additionally, the simulated HAADF-STEM image in Fig. S16b reproduces the characteristically bright contrast (highlighted by pink-edged ovals) shown in Fig. S16a, further supporting Zn enrichment in the $\omega_T$ lattice. In Fig. S16b, by changing the outermost $U_T$ of $\eta'$ nanoplate into a $Pu_T$ lattice and stacking the remaining periodic lattice nets, the $\eta'$ nanoplate can be identified as metastable $Mg_6Zn_{13}$ phase ($E_{hull}$=0.030 eV/atom) with a periodic stacking unit of -$Pu_T|K|\omega_T|K|Pu_T|K$-. Although $Mg_6Al_2Zn_{11}$ (where Al-Al pairs establish its $\omega_T$) has a slightly lower $E_{hull}$ (0.022 eV/atom), $Mg_6Zn_{13}$ is kinetically favored. Thus, $\eta'$ is considered to be the $Mg_6Zn_{13}$ phase.

Once a local hcp lattice is created through the glide of (a/6)<11$\bar{2}$> partial dislocations along (3, 5, 7)-planes, as schematically shown in Fig. S17a, a 7-layer $\eta'$ nanoplate can form via 5-layer $M_T$-type and 3-layer $M_R$-type BSTUs within a 7-layer GP zone. Specifically, $M_T$-type BSTU produces a 5-layer $U_T|K|\omega_T|K|U_T$ structure, and $M_R$-type BSTU forming on its outermost $U_T$ lattice can lead to $\eta'$ nanoplate formation. Furthermore, in Supplementary Section 8, the 7-layer $\eta'$-structured nucleus is determined to be kinetically preferred as the critical nucleus over the 5-layer $U_T|K|\omega_T|K|U_T$ nucleus. Therefore, during aging, GP zones occur within the matrix in a 7-layer configuration, and we emphasize that both types of BSTUs form synchronously within these GP zones without a specific order.

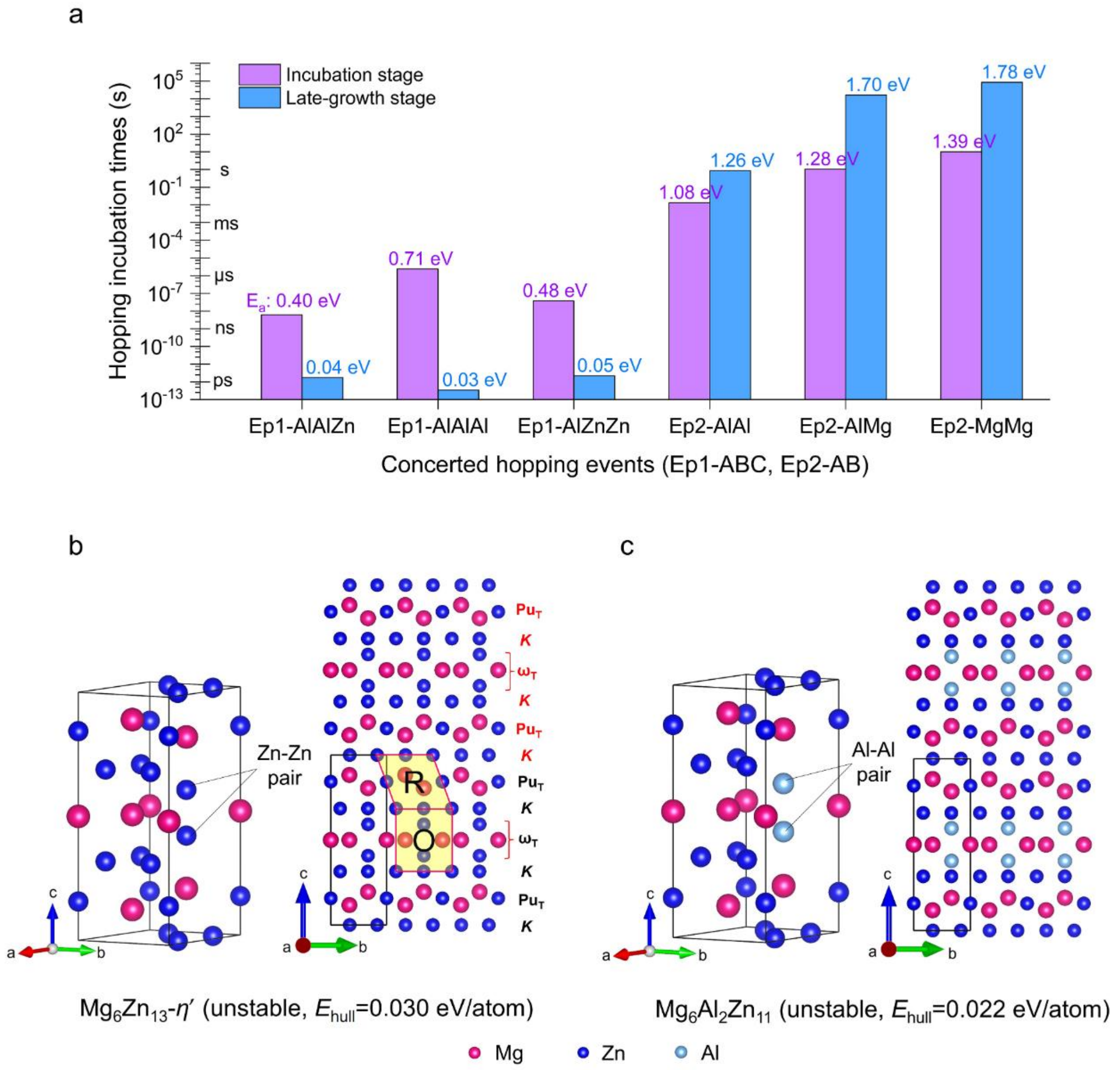

**Figure S16**. **a**. The hopping incubation time for six concerted hopping events, along with their activation energies ($E_a$) marked at the top of the bar, during both incubation and late-growth stages. **b-c**. Atomic structures of $Mg_6Zn_3$-$\eta'$ and $Mg_6Al_2Zn_{11}$ unstable phases with detailed lattice nets marked on the right side of the panel. Mg, Zn, and Al atoms are denoted by pink, dark-blue, and light-blue balls, respectively.

Subsequently, equilibrium $MgZn_2$-$\eta_2$ nanoplates evolve from 7-layer $\eta'$ nanoplates. Both lengthening and thickening of $MgZn_2$-$\eta_2$ nanoplates primarily depend on forming 3-layer $M_R$-type BSTUs along their original outermost $U_T$ lattice following the local hcp lattice creation. For instance, in Fig. S17a, a 9-layer $\eta_2$ structure can thicken from a 7-layer $\eta'$ nanoplate by forming $M_R$-type BSTU based on the (7)-plane after the glide of (a/6)<$11\bar{2}$> partials along the (9)-plane. Moreover, this $\eta'\rightarrow\eta_2$ also incorporates $\omega_T\rightarrow Pu_T$ transition (representing "O" subunit change into "R" subunit) by which mechanism can refer to recent Zhang et al. work[45]. Overall, through the "*interlayer-sliding+shuffling*" mode shown in Fig. S17b, the thickness of the $\eta_2$ nanoplate evolves in a pattern of 9-layer→11-layer→13-layer→…, with each time adding two new layers (i.e., $U_T|K$) of $\eta_2$ material. This structural evolution is perfectly confirmed by the $\eta_2$ nanoplate having 9, 11, and 13 layers shown in Fig. S15c[12].

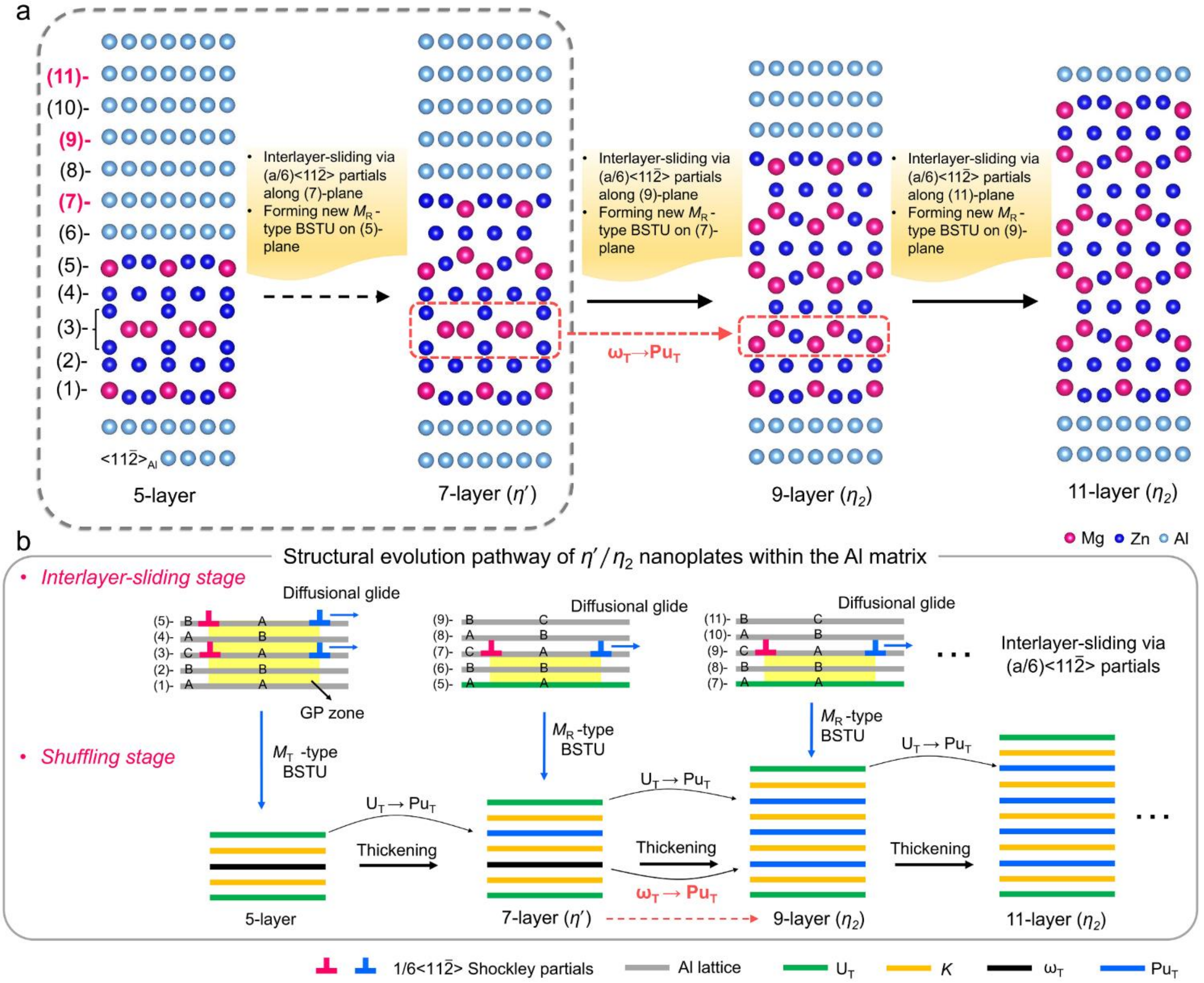

**Figure S17. a.** Schematic illustrating the structural evolution of the $\eta'$ and $\eta_2$ nanoplates within the Al matrix through forming 3-layer $M_R$-type BSTUs after a local hcp lattice is created. **b**. The structural evolution pathway of the $MgZn_2$ phase within the Al matrix.

In contrast to the direct precipitation of $MgZn_2$ nanoplate in Mg alloys, one may wonder why the $\eta_2$-$MgZn_2$ nanoplate in Al alloys requires evolution from the precursor $\eta'$. This study reveals that the 3-layer $M_R$-type BSTU required for $MgZn_2$ nucleation cannot directly form from the Al matrix during the initial nucleation stage, but adopting $\eta'$ serves as a kinetically favorable precursor phase that facilitates $\eta_2$ nanoplates form indirectly. Forming the 3-layer $M_R$-type BSTU in the Al matrix requires a single stacking fault, but as Fig. S18a-b illustrates, GP zones with diverse compositions, including monolayer Cu-GP zone, (Mg, Cu)-GP zone, and (Mg, Zn)-zone, fail to remarkably lower the $\gamma_{sf}$ of $Plane_{1/2/3}$ to enable dislocation dissociation and migration. Hence, during initial nucleation, the formation of multiple stacking faults through thicker GP zones, such as the 5-layer GP zone observed in $T_1/\Omega$ nucleation or even thicker GP zones, is kinetically preferred over individual stacking fault generation from the Al matrix. Supplementary Section 8 further demonstrates that the 7-layer $\eta'$ nucleus, formed by $M_T$- and $M_R$-type BSTUs, is kinetically favored as critical nuclei over the 5-layer $U_T|K|\omega_T|K|U_T$ nucleus for nucleation. Consequently, $MgZn_2$-$\eta_2$ precipitation cannot occur directly but evolves through the precursor $\eta'$ due to kinetic constraints, as depicted in Fig. S18c. Overall, this unique precipitation pathway of $MgZn_2$-$\eta_2$ in Al alloys results from the co-action of the fcc matrix and specific alloying solutes.

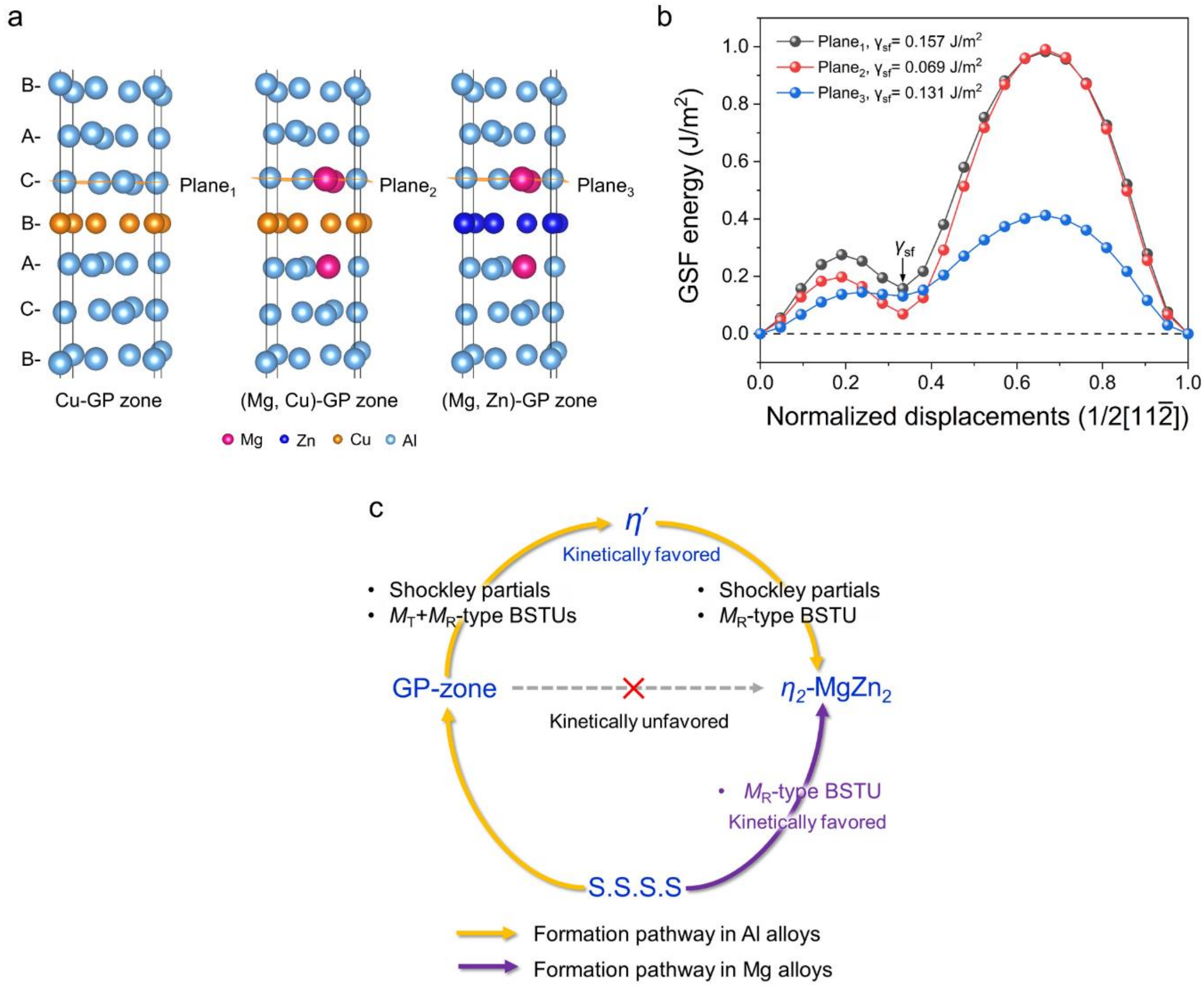

**Figure S18**. **a**. Atomic structural models for the Cu, (Mg, Cu), and (Mg, Zn)-GP zones. **b**. GSF energy curves along the $[11\bar{2}]$ direction of $Plane_{1,2,3}$, computed using the DFT calculations. **c**. Schematic showing the relation between the formation pathways of $MgZn_2$ nanoplates in Al and Mg alloys.

## 5. Supplementary Section 5: Structural evolution pathway of the Ω nanoplate within the Al matrix

As introduced in Supplementary Section 1, the addition of Mg and Ag to Al-Cu alloys dramatically facilitates the coherent precipitation of the distorted θ-$Al_2Cu$ (i.e., $θ_d$-$Al_2Cu$) *C*16 structure along $\{111\}_{Al}$ matrix planes, with Mg and Ag "segregating" exclusively to the outermost layers of matured Ω nanoplates rather than their interiors. This indicates that such co-addition does not change the alloys' thermodynamics but provides a kinetically favorable pathway for inducing the $θ_d$-$Al_2Cu$ to form coherently. Despite its distorted *C*16 structure, which differs from typical TCP structures, $θ_d$-$Al_2Cu$ exhibits high structural similarity to TCP structures, enabling coherent precipitation. Figure S19a illustrates the $θ_d$-$Al_2Cu$ structure can be formed by stacking $U_T$-like and *K*-like lattice nets alternatively from a side view. The $U_T$-like and *K*-like lattice nets (Fig. S19b, c), named based on their side and top views, respectively, maintain the same atomic density per unit-cell plane ($\rho$) as that of the Al-matrix (i.e., $\rho_{U_T\text{-like}}=\rho_{K\text{-like}}=\rho_{\text{Al-matrix}}=6$). This means that the $θ_d$-$Al_2Cu$ structure can transform directly from the Al matrix without requiring additional atoms, as demonstrated in our simulations below.

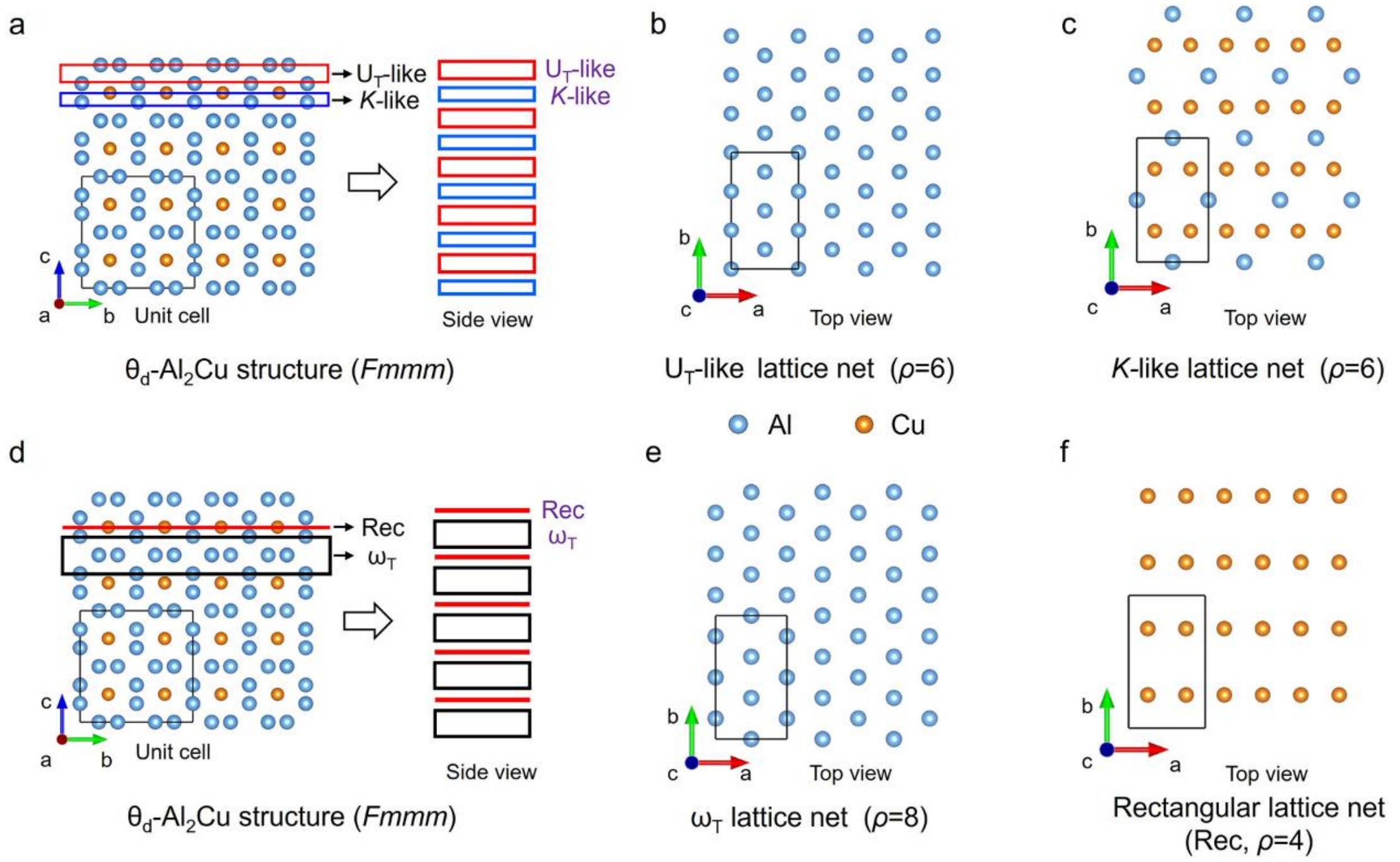

**Figure S19.** New structural definition for $θ_d$-$Al_2Cu$ structure. **a-c**. Schematic showing the $θ_d$-$Al_2Cu$ structure can be assembled by alternating $U_T$-like and *K*-like lattice nets. **d-f**. Schematic depicting the $θ_d$-$Al_2Cu$ structure can be stacked by alternating Rec and $ω_T$ lattice nets.

Apart from the abovementioned crystallographic definition, the $θ_d$-$Al_2Cu$ structure can also be described by alternating Rectangular (Rec) and $ω_T$ lattice nets (see Fig. S19d-f), where the Rec lattice net derives its name from its top view atomic packing pattern. According to this alternative definition,

Fig. S20b-c shows matured/thickened Ω nanoplates (i.e., containing the $\theta_d$-$Al_2Cu$ structure) belonging to a sandwiched structure in the form of $U_T|K|\theta_d$-$Al_2Cu|K|U_T$. For instance, a 9-layer Ω nanoplate comprises $U_T|K|\omega_T|Rec|\omega_T|Rec|\omega_T|K|U_T$ lattice nets, with the outermost $U_T(MgAg_2)|K$(Cu-riched) constantly serving as the shell of the Ω nanoplates. Here, we found that the outermost $U_T|K$ layers primarily serve as a transition region between $\theta_d$-$Al_2Cu$ and the Al matrix, playing a crucial role in template-inducing coherent precipitation of $\theta_d$-$Al_2Cu$ along $\{111\}_{Al}$ planes.

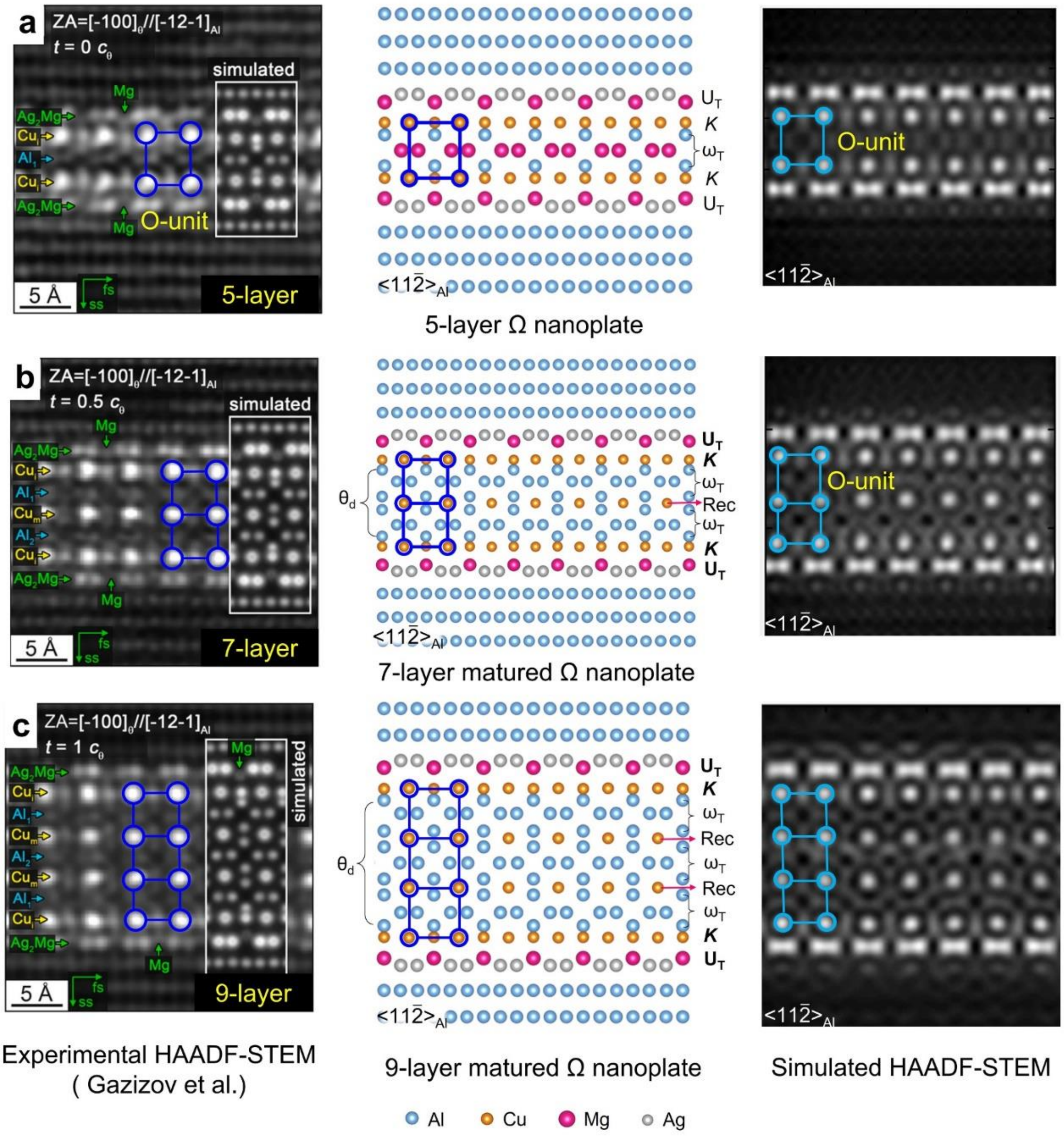

**Figure S20.** Ω nanoplate with 5, 7, and 9 layers. Experimental HAADF-STEM images from ref.[7] on the left, atomic configurations in the middle, and corresponding simulated HAADF-STEM images on the right. Mg, Cu, Al, and Ag atoms are represented by pink, yellow, light-blue, and light-white balls, respectively.

Consistent with the 5-layer $T_1$ nanoplate formation described in the main text, the formation of 5-layer Ω nanoplate (Fig. S21a) also requires initially creating a local hcp lattice through the diffusional glide of $(a/6)<11\bar{2}>$ partial dislocations along the (3, 5)-planes, followed by atomic shuffling via $M_T$- type BSTU. Supplementary Section 1 has demonstrated that the (Mg, Cu, Ag)-rich GP zone favors dislocation dissociation and migration. Regarding the subsequent thickening process evolving towards matured Ω nanoplates, distinct from $T_1$ and $\eta'/\eta_2$ nanoplates, we found that the structural transformation requires sequential participation of (a/6)<11$\bar{2}$> and (a/4)<1$\bar{1}$0> partial dislocations, where the latter triggers atomic shuffling to complete thickening. Figure S21 details the thickening process from 5-layer $U_T|K|\omega_T|K|U_T$ to 7-layer $U_T|K|\omega_T|Rec|\omega_T|K|U_T$ Ω nanoplates through these partial dislocations. Specifically, the (4)-plane transitions into a *K* lattice net, which significantly changes dislocation behavior on the neighboring (5)-plane. The γ-surface of the (5)-plane (see Fig. S21b), calculated using the model presented in Fig. S21a, illustrates that dislocation on this plane maintains as (a/2)<1$\bar{1}$0> perfect dislocation (indicated by a black arrow from A to B) without dissociating, with the detailed GSF energy curve along this path (γ-1) depicted in Fig. S21c.

Through appropriate modelings to capture unstable structures capable of generating products, we found that a thickening process requires undergoing three stages, as illustrated in Fig. S21d. In stage-1, composition adjustments occur within the (4-7)-planes: (Mg, Ag) in the (5)-plane dissolve and diffuse towards the (7)-plane, while Cu enriches the (6)-plane. In this stage, the shape of the γ-2 GSF energy curve remains largely unchanged (see Fig. S21c). However, in stage-2, if (a/6)<11$\bar{2}$> partials glide along the (7)-plane and create a local hcp lattice within the (5-7)-planes, which can be attributed to the redistributed (Mg, Ag) and Cu in the (6, 7)-planes that facilitate dislocation nucleation and migration under such chemically favorable environment, the γ-3 GSF energy curve unexpectedly shifts into negative values range and reaches a minimum at the (a/4)<1$\bar{1}$0> position. This means that, under such structural and compositional environments, a glide of (a/4)<1$\bar{1}$0> partial along the (5)-plane (i.e., stage-3) will trigger a structural transition and produce a 7-layer $U_T|K|\omega_T|Rec|\omega_T|K|U_T$ Ω nanoplate directly. Specifically, as shown in Fig. S21e, once (a/4)<1$\bar{1}$0> partials glide across the (5)-plane, Al at the (5)-plane will interact with its neighboring Al of the (4)-plane, causing these two Al atoms move outward to form $\omega_T$ at the (5)-plane and Rec at the (4)-plane via dissociative displacement ($\xi_y$, marked by green arrows); meanwhile, $\xi_y$ induces a collective displacement ($\xi_x$, marked by black arrows) of Cu atoms on the (6)-plane, creating an opening that also attract Mg from the (7)-plane to move towards it via $\xi_y$.

Notably, in practical aging scenarios, localized compositional adjustments (stage-1) are constantly occurring and are strongly coupled with stages-2 and 3. Moreover, both stages-2 and 3 also require compositional adjustments (stage-1) to drive their diffusional glide. Once at a specific moment, these localized compositional adjustments lead to sequential nucleation and migration of (a/6)<11$\bar{2}$> and (a/4)<1$\bar{1}$0> partials along the (7)-plane and (5)-plane, respectively, the thickening and lengthening of Ω nanoplates can proceed. More importantly, Yang et al.[9] have recently explicitly observed (a/6)<11$\bar{2}$> and (a/4)<1$\bar{1}$0> partials at the edges of (7)-and (5)-planes of Ω nanoplates, validating our findings. Regarding their source, Supplementary Section 7 reveals that excessive quenched-in vacancies tend to segregate towards Cu-occupied planes, possibly leading to dislocation nucleation at (5, 7)-planes.

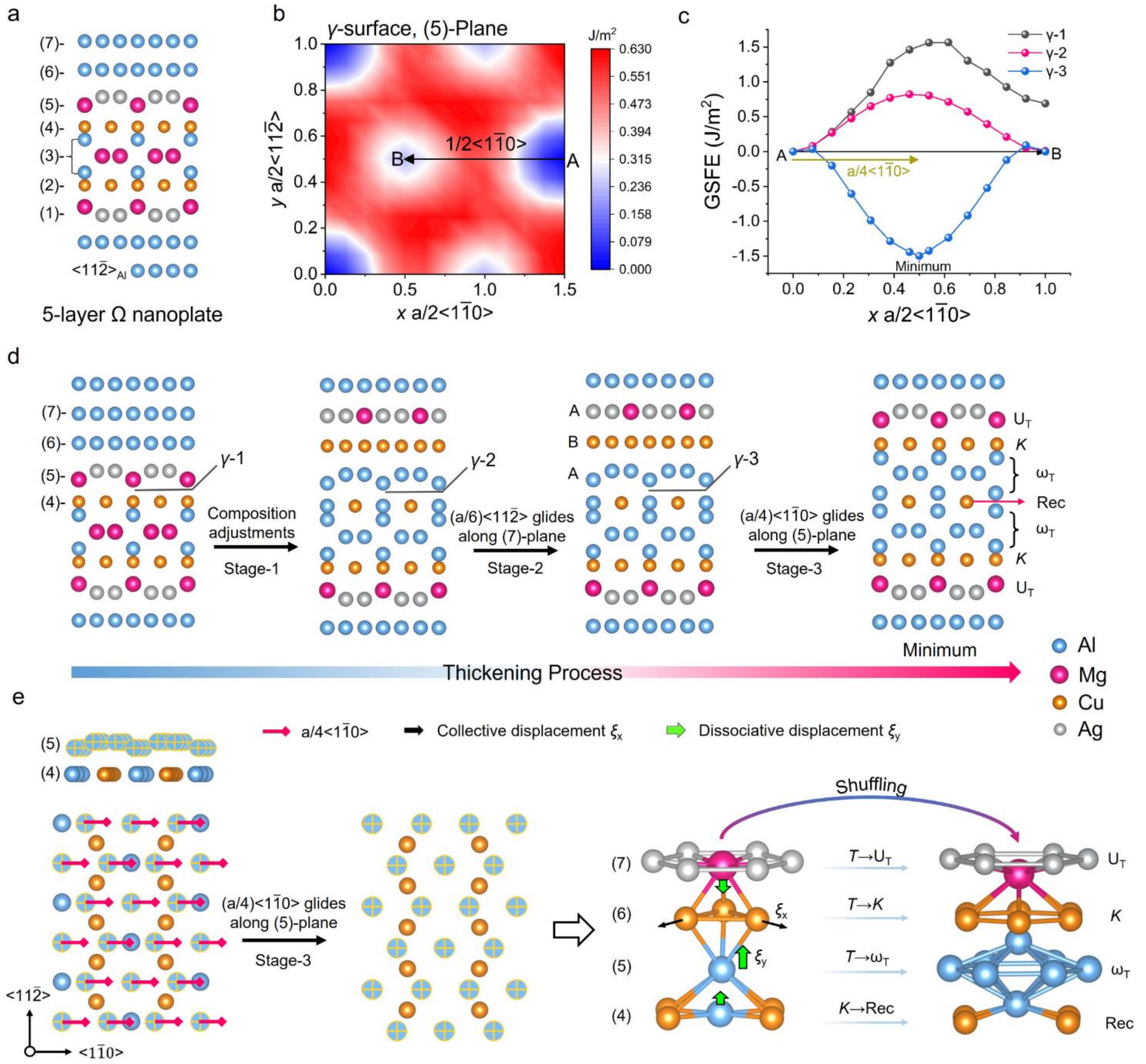

**Figure S21. a.** Atomic structural model showing a 5-layer Ω nanoplate with $U_T|K|\omega_T|K|U_T$ lattice nets. **b**. GSF energy (γ) surface for the (5)-plane, calculated based on the model in **a**. **c**. GSF energy curves along the <1$\bar{1}$0> direction in the γ-1, 2, and 3 planes with varying atomic compositions. **d**. The thickening process of the Ω nanoplate changing from a 5-layer to a 7-layer $U_T|K|\omega_T|Rec|\omega_T|K|U_T$ structure undergoes 3 stages. The γ-1, 2, and 3 planes are marked on the panel. **e**. Schematic illustrating the structural transformation through shuffle-based displacements in Stage-3. Al, Mg, Cu, and Ag are represented by light-blue, pink, yellow, and light-white balls, respectively.

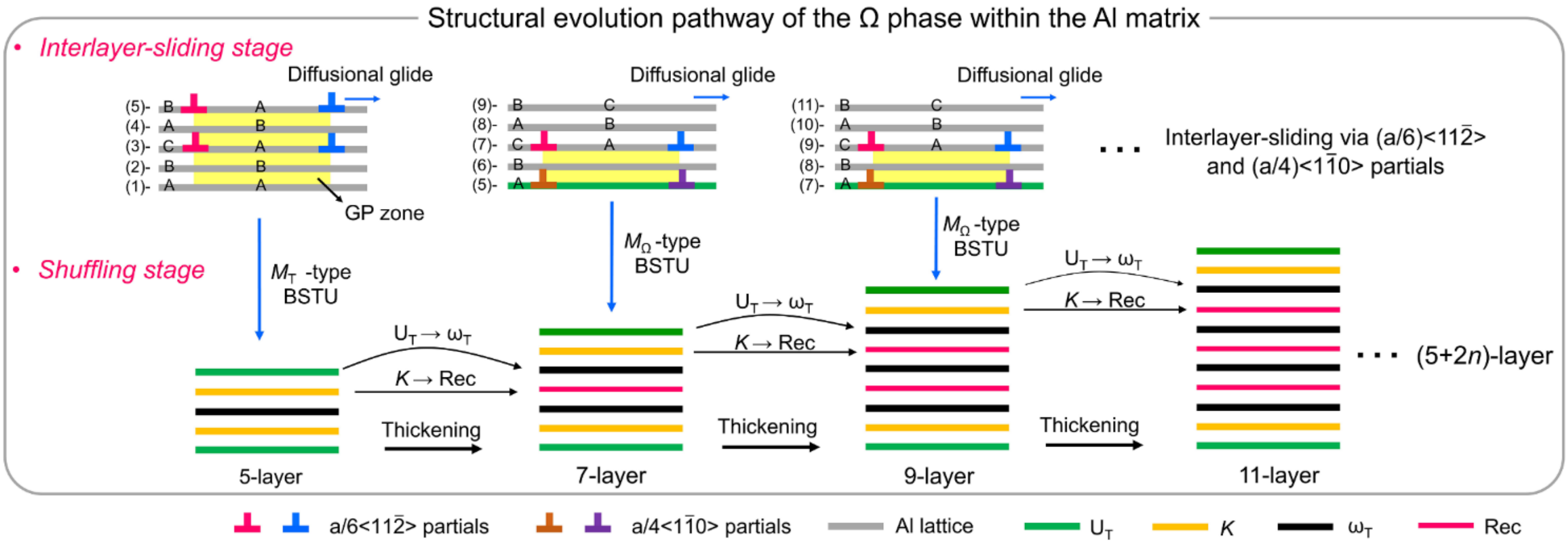

**Figure S22.** Schematic illustrating the structural evolution pathway of the Ω phase within the Al matrix.

Besides, Ω nanoplate thickening involves interfacial transitions of the outermost two layers: the (4)-plane occurs *K*→Rec, and the (5)-plane occurs $U_T$→$\omega_T$, which contrasts with TCP nanoplates where thickening involves only the outermost layer. This indicates that Ω nanoplate thickening involves four {111} planes but increases their height by two {111} planes each time. Here, we designate this 4-layer unstable structure as $M_\Omega$-type BSTU, governing both the lengthening and thickening of Ω nanoplates, as the structural evolution pathway depicted in Fig. S22. In short, Al→Ω transformations still proceed via an "*interlayer-sliding*+*shuffling*" mode, with the thickness varying according to $N$=5+2$n$, where $N$ represents the total number of layers in Ω nanoplates and $n$ specifies the number of thickening steps. Consequently, $N$ is typically an odd number instead of an even number, as confirmed by past experimental observations, including HAADF-STEM images shown in Fig. S20 and many other characterizations found in refs.[7–9]. These experimental observations consistently support our proposed structural evolution pathway.

## 6. Supplementary Section 6: Nonclassical nucleation behavior in simple→complex transitions

In Mg alloys, TCP formation exhibits nonclassical nucleation behaviors[33], where the onset of structural transformations primarily depends on the distribution of large atoms in the nucleus rather than smaller ones. Unlike classical nucleation theory (CNT), which assumes that the nucleus has the same crystal structure and composition as the final equilibrium phase, the nonclassical nucleation mechanism[46–48] allows nuclei to have composition distinct from the target precipitate phase or nucleation via intermediates with different crystal structures. This mechanism serves as a supplement of CNT[47], aiming to elucidate intricate nucleation phenomena that fall outside CNT's scope. Likewise, in Al alloys, $T_1$, $\eta'/\eta_2$, and Ω (with TCP structures in its "shell" part) phases also exhibit nonclassical nucleation behaviors during precipitation. For instance, as shown in Fig. S23, despite having a fully non-uniform composition in their nuclei, both 5-layer $T_1$ and 3-layer *M*-type nuclei can yet form. This indicates that the atomic composition in BSTUs is not fixed, allowing for a non-uniform composition within the nucleus.

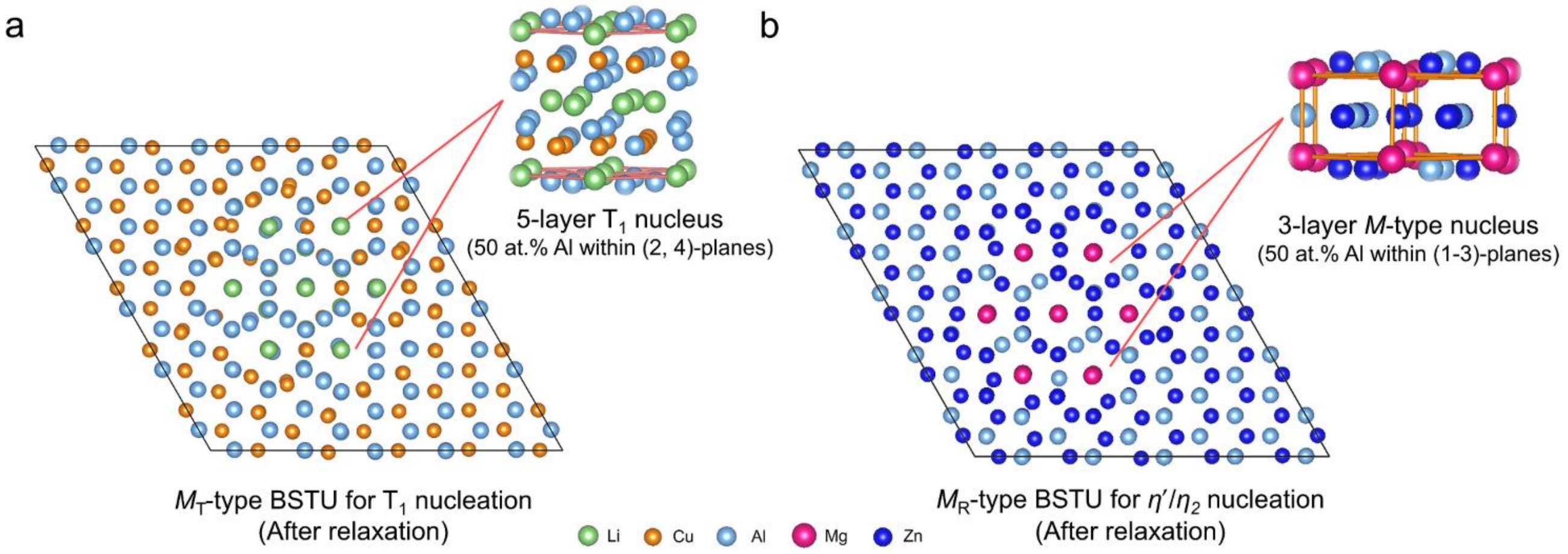

**Figure S23.** Schematic showing nonclassical nucleation features for $M_T$-type BSTU in $T_1$ nucleation and $M_R$-type BSTU in $\eta'/\eta_2$ nucleation.

During precipitation, $T_1$[36], $\eta'/\eta_2$[13,14], and Ω[7] nanoplates often exhibit intricate phenomena, mainly involving off-stoichiometry (i.e., containing excess solvent Al) or polymorphism. For instance, $\eta'/\eta_2$ nanoplates and their corresponding GP zones frequently exhibit dynamic atomic compositions[13,14], lacking a fixed stoichiometry, while $\eta_2$-$MgZn_2$ nanoplates form in various polymorphs, including *C*14, *C*15, *C*36, or combinations of these structures[49]. These intricate phenomena are also common in TCP nanoplates of Mg alloys. As demonstrated in our prior work[33], these phenomena primarily stem from nonclassical nucleation behaviors and are sensitively influenced by the interplay between thermodynamics and kinetics. Consequently, it's easy to understand that the nonclassical nucleation nature in $T_1$, $\eta'/\eta_2$, and Ω precipitation, along with kinetic factors like varying diffusion rates among distinct constituents, readily leads to off-stoichiometry in their nanoprecipitates. This, in turn, may further facilitate the emergence of many metastable polymorphs like *C*15 and *C*36 $MgZn_2$-$\eta_2$ plates due to competitive precipitation between these structures driven by the delicate interplay between thermodynamics and kinetics, as elucidated in our prior study[33]. Overall, unraveling the nature of nucleation can offer mechanistic insights into complex precipitation phenomena observed in alloys.

## 7. Supplementary Section 7: Vacancy-induced dislocation nucleation mechanism

This study elucidates that complex-structured nanoplates in the Al matrix typically form through an "*interlayer-sliding+shuffling*" mode, with partial dislocations at the edges of plates/ledges acting as the carrier for interlayer-sliding. Our identified structural evolution pathways for $T_1$, Ω, and $\eta'/\eta_2$ nanoplates indicate that these partial dislocations should occur on every other one $\{111\}_{Al}$ planes to sustain nanoplate growth. This naturally raises the question: What is the origin of these dislocations? Moreover, this issue also exists in simple→simple$_{distinct}$ transitions, such as $AlAg_2$-γ′ precipitation (fcc→hcp) in Al-Ag alloys. Although Cassada et al.[50] proposed a dislocation-jog mechanism (see Fig. S24) that can explain the initial nucleation of $T_1$ nanoplates facilitated by pre-deformation before artificial aging, this mechanism cannot provide dislocations continuously on alternating {111} planes during the subsequent thickening process. Hence, a new dislocation nucleation mechanism, independent of pre-existing dislocations and external load, needs to be considered.

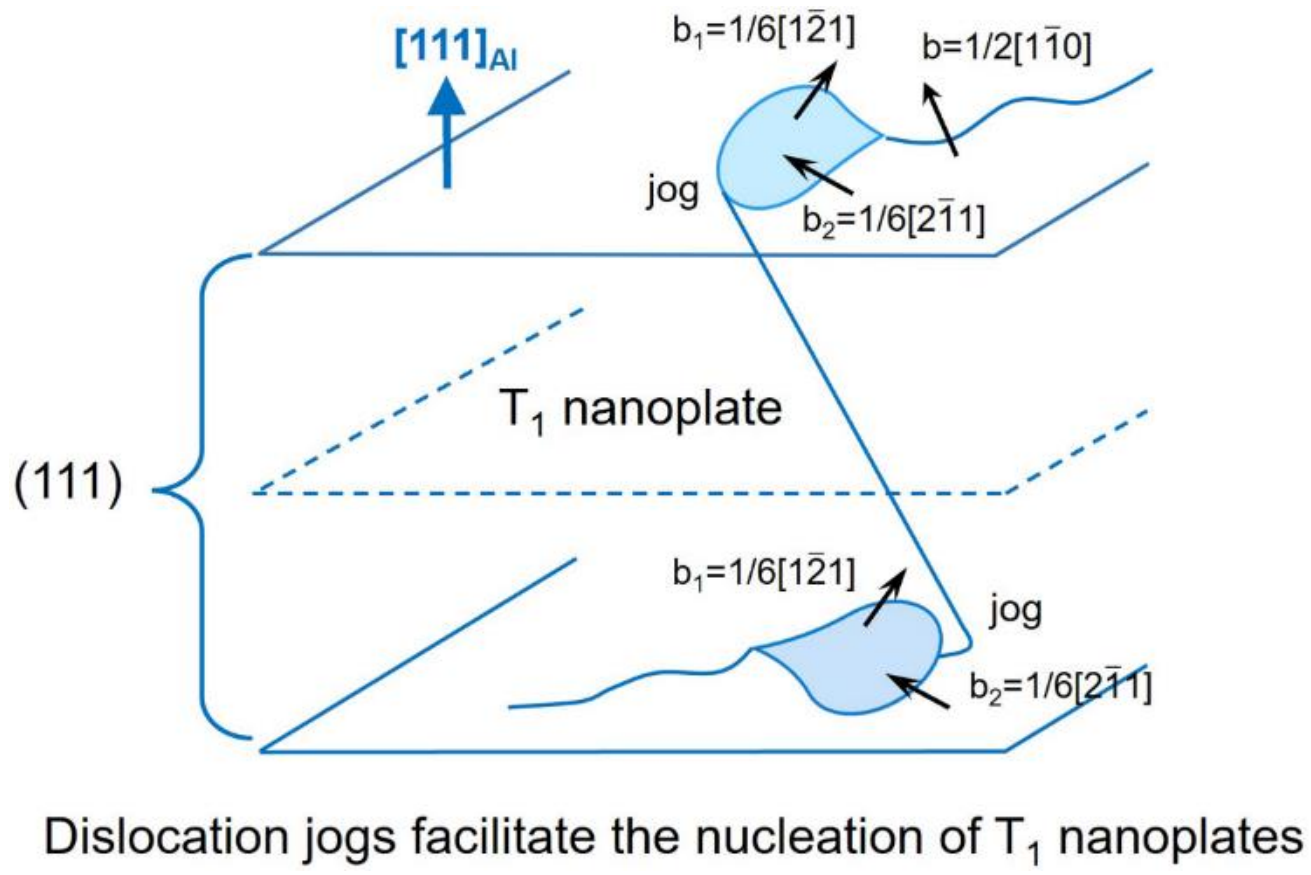

**Figure S24.** Schematic of mechanism for $T_1$ nucleation via dislocation jogs, suggested by Cassada et al.[50].

Recent evidence increasingly suggests that the nucleation of the first dislocation in a pristine crystal is associated with a diffusion-controlled process. Specifically, Zou et al.[51] observed that misfit-accommodating dislocations at the surface of Cu-Au alloy nucleate through surface diffusion and trapping, while Chen et al.[52] measured that the nucleation of surface dislocation in defect-scarce nanostructures is assisted by thermal fluctuations. Given that vacancies are superaturated after quenching and have been explicitly observed in the GP zones of $\eta'/\eta_2$ precipitates[13,14], it is reasonable to speculate that their diffusion and accumulation might play a key role in dislocation nucleation. However, direct simulations of dislocation nucleation are prohibitively challenging due to the long time scales required for thermally activated processes like vacancy diffusion beyond the accessible time scales of molecular dynamics. Consequently, we adopt DFT calculations in conjunction with current experimental findings to examine the feasibility of vacancy-induced dislocation nucleation.

The segregation energy $E_{seg}$ of a vacancy at different atomic planes of the GP zone is defined as the energy difference between a vacancy residing on the GP zone and the bulk site. This energy can be expressed as:

$$E_{seg}=\left(E_{GP}^{Vac}+E_{bulk}\right)-\left(E_{GP}+E_{bulk}^{Vac}\right) \tag{S-2}$$

where the subscript represents the structure of the host material, and the presence or absence of the superscript indicates whether or not a host atom is replaced with a vacancy. Based on this definition, a negative $E_{seg}$ means energetically favorable segregation of a vacancy towards this site, and a higher negative value indicates a stronger tendency for vacancy segregation. Figure S25a-d evaluates the $E_{seg}$ of vacancies at various atomic sites within four distinct chemical environments: Cu-GP zone, (Li, Cu)-GP zone, (Mg, Cu)-GP zone, and a Ω nanoplate thickening from 5-layer to 7-layer. In these models (left side of the panel), solutes are randomly distributed at atomic planes with specified compositions.

Our results indicate that only atomic planes with Cu distribution display negative $E_{seg}$, indicating a preferential migration of vacancies towards these planes over those enriched with larger Mg and Li atoms or pure-Al planes. Consequently, Cu-rich atomic planes readily enriched with vacancies during aging treatments. As shown in Fig. S25e, exemplified with dislocation nucleation at the (3, 5)-planes of the (Li, Cu)-GP zone, the enrichment of vacancies in the (2, 4)-planes will inevitably induce substantial interfacial strain across these planes with their adjacent (3, 5(1))-planes, which are rich in larger Li atoms. To release this interfacial strain, like dislocation nucleation at the metallic surface[51], misfit-accommodating dislocations may initially nucleate at the (3)-plane and then at the (5)-plane, as the (3)-plane contains a higher Li content than the (5)-plane. Upon nucleation, these misfit-accommodating dislocations are expected to dissociate into partial dislocations and glide diffusively toward the edge of the GP zone, driven by the favorable chemical environment, as illustrated in the main text.

As a result, excess vacancies may induce dislocation nucleation through a diffusion-controlled process. While pre-existing dislocations can facilitate initial nucleation via dislocation jogs, our suggested vacancy-induced mechanism is expected to dominate during precipitation. A recent study on θ′-$Al_2Cu$ nanoplates[53] in Al alloys has highlighted the crucial role of vacancies in facilitating the fcc→bct transformation; notably, a reduction in quenched-in vacancies has been shown to impede nanoplates' nucleation and growth rates dramatically[54]. Thus, we suggest that this vacancy-induced dislocation nucleation mechanism is the main source of dislocations for both simple→complex and simple→$simple_{distinct}$ transitions. However, further validation using advanced simulation and experimental techniques is needed to confirm this mechanism.

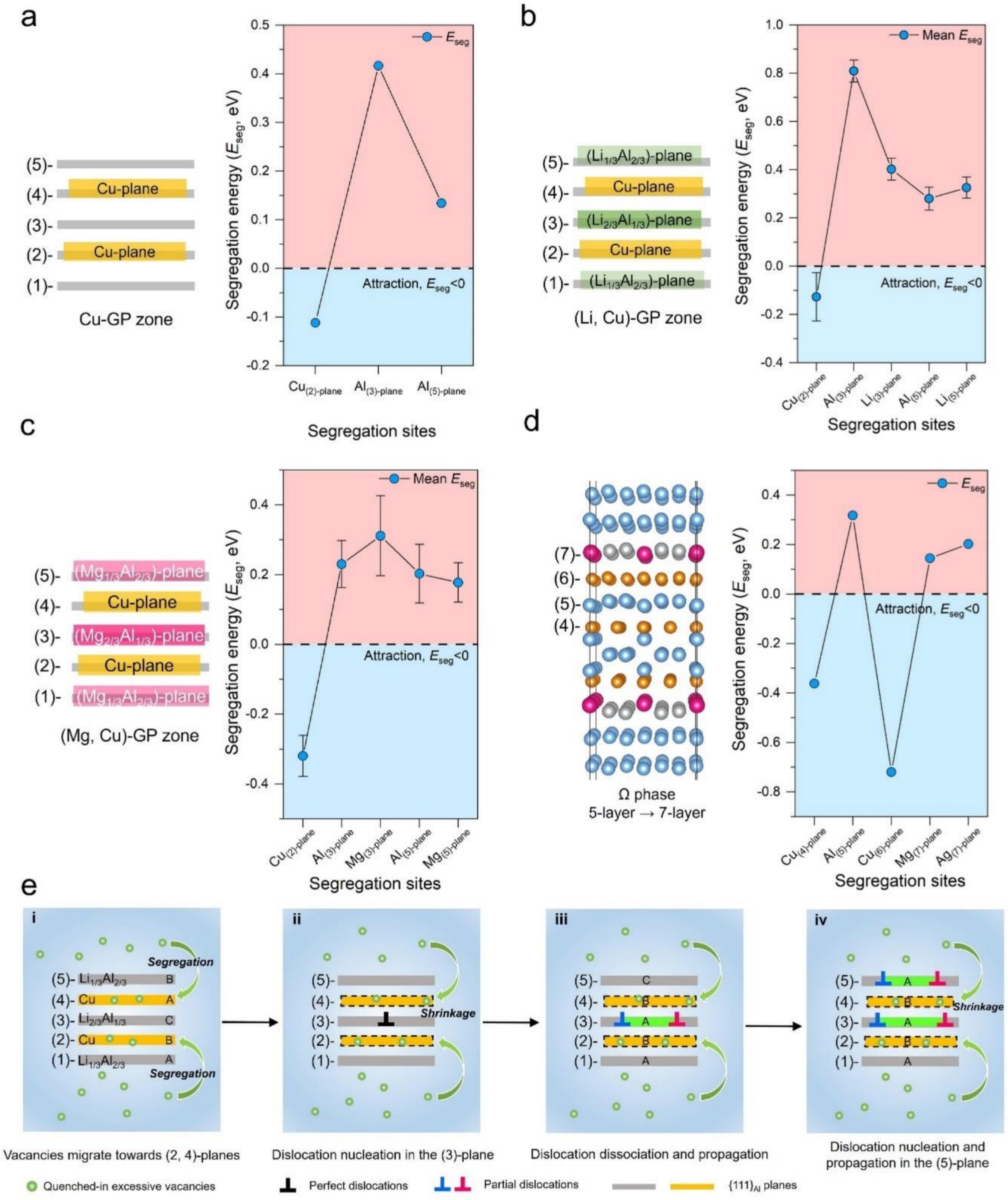

**Figure S25. a-d**. The segregation energy variation for excessive quenched-in vacancies across atomic sites in various atomic planes of GP zones. Error bars represent the standard deviation (SD) of $E_{seg}$ from 10 independent sampling configurations. **e**. Schematic illustrating the vacancy-induced dislocation nucleation mechanism.

## 8. Supplementary Section 8: Evaluations of critical nucleus during the nucleation stage

Based on the identified structural evolutional pathway for precipitate plates, we apply CNT to further evaluate the critical size $n^*$ and nucleation barriers $\Delta G^*$ for the homogeneous nucleation of Ω and $\eta'$ phases, respectively. It's important to note that although TCP/Ω nucleation exhibits nonclassical features, nonclassical nucleation theory does not replace CNT but supplements[47] it by addressing complex behaviors beyond CNT's scope. In other words, regardless of the nucleation type—classical or nonclassical—nuclei must evolve into critical nuclei to achieve stable growth.

The energy change $\Delta G$ associated with the nucleation of a precipitate with a volume $V$, surface area $S$ (including the areas of broad interface $S_b$ and rim interface $S_r$), and interfacial energies $\gamma$ (including $\gamma_b$ and $\gamma_r$) can be expressed as follows:

$$\Delta G = V\Delta G_{\text{chem}} + \left(S_b\gamma_b + S_r\gamma_r\right) + \Delta G_{\text{elastic}} \tag{S-3}$$

where $\Delta G_{\text{chem}}$ represents the nucleation driving force, while $\Delta G_{\text{elastic}}$ signifies the elastic energy within an infinite matrix containing a precipitate. Since experimentally observed Ω and $\eta'$ precipitates often keep a plate-like morphology, we assume that the nucleus is in the form of an equilateral hexagonal plate, with its radius $R$ and the height of the precipitate rim being $t$. Based on recent insights from Mg alloy research[33], we simplify Eq. (S-3) for Ω homogeneous nucleation in Al-Cu-Mg-Ag alloys as follows:

$$\Delta G_{\Omega}^{(5,\,7)\text{-layer}} = n_{\text{Cu}}\Delta\mu_{\Omega}^{(5,\,7)\text{-layer}} + \frac{2n_{\text{Cu}}V_{\Omega}^{(5,\,7)\text{-layer}}}{t_{\Omega}^{(5,\,7)\text{-layer}}}\gamma_b + 6t_{\Omega}^{(5,\,7)\text{-layer}}\sqrt{\frac{2n_{\text{Cu}}V_{\Omega}^{(5,\,7)\text{-layer}}}{3\sqrt{3}t_{\Omega}^{(5,\,7)\text{-layer}}}}\gamma_r + \Delta G_{\text{elastic}}^{(5,7)\text{-layer}} \tag{S-4}$$

Here, $n_{\text{Cu}}$ signifies the number of Cu atoms within the nucleus, $\Delta\mu_{\Omega}^{(5,\,7)\text{-layer}}$ represents the reduction in chemical potential associated with the formation of (5, 7)-layer Ω nucleus in the system, while $n_{\text{Cu}}V_{\Omega}^{(5,\,7)\text{-layer}}$ denotes the volume of (5, 7)-layer Ω nanoplates. Based on the stoichiometry of (5, 7)-layer Ω nanoplates, the detailed expression of $\Delta\mu_{\Omega}^{(5,\,7)\text{-layer}}$ and $V_{\Omega}^{(5,\,7)\text{-layer}}$ are:

$$\Delta\mu_{\Omega}^{5\text{-layer}} = \Delta\mu_{\text{Cu}} + \tfrac{2}{3}\Delta\mu_{\text{Mg}} + \tfrac{1}{3}\Delta\mu_{\text{Al}} + \tfrac{2}{3}\Delta\mu_{\text{Ag}},\quad V_{\Omega}^{5\text{-layer}} = V_{\text{Cu}} + \tfrac{2}{3}V_{\text{Mg}} + \tfrac{1}{3}V_{\text{Al}} + \tfrac{2}{3}V_{\text{Ag}};$$

$$\Delta\mu_{\Omega}^{7\text{-layer}} = \Delta\mu_{\text{Cu}} + \tfrac{1}{4}\Delta\mu_{\text{Mg}} + \Delta\mu_{\text{Al}} + \tfrac{1}{2}\Delta\mu_{\text{Ag}},\quad V_{\Omega}^{7\text{-layer}} = V_{\text{Cu}} + \tfrac{1}{4}V_{\text{Mg}} + V_{\text{Al}} + \tfrac{1}{2}V_{\text{Ag}}. \tag{S-5}$$

where $\Delta\mu_{\text{x}}$(X=Cu, Mg, Al, Ag) is the chemical potential difference between X in the (5, 7)-layer Ω nanoplates and the GP zone, and $V_{\text{X}}$ (X=Cu, Mg, Al, Ag) represents the Voronoi volume of a single X atom obtained from Ovito[55] software. To simplify the calculations, the nonclassical nucleation features were not considered, with the focus on normal stoichiometry scenarios. The chemical potentials of Cu, Mg, Al, and Ag in the Ω nanoplates were determined by solving the following equations based on three reference (i.e., co-existing) phases, Al, $Al_2CuMg$ (S), and Ω (θ-$Al_2Cu$) in experiment[2]:

$$\mu_{Al} = E_{FCC\text{-}Al},\ 2\mu_{Al}+\mu_{Cu}=E_{Al_2Cu(\theta)},\ 2\mu_{Al}+\mu_{Cu}+\mu_{Mg}=E_S,\ \mu_{Ag}=E_{GP} \tag{S-6}$$

For interfacial energies of coherent broad interface $\gamma_b$ and the rim interface $\gamma_r$, we construct supercells of Ω and $\alpha$-Al to calculate these two quantities. The energy of formation per atom relative to the energies of $\alpha$-Al and Ω phases can be expressed as follows:

$$\Delta E_f = \delta E_{cs} + \frac{S_b\gamma_b}{N} \tag{S-7}$$

$$\Delta E_f = \delta E_{cs} + \frac{S_b\gamma_b + S_r\gamma_r}{N} \tag{S-8}$$

where $\Delta E_f$ is the energy of supercell formation (eV/atom) relative to the bulk energies of $\alpha$-Al and Ω phases. $\delta E_{cs}$ denotes the coherency strain per atom caused by the lattice mismatch between $\alpha$-Al and Ω phases. $N$ represents the total number of atoms in the supercells, and $\gamma_{b(r)}$ and $S_{b(r)}$ are the interfacial energy and area, respectively. The interfacial energy can be obtained from the slope of $\Delta E_f$ (eV/atom) *vs.* $1/N$ as shown in Fig. S26a. Here, to rigorously evaluate $\gamma_r$, we initially calculated $\gamma_b$ using Eq. (S-7), which was then used in Eq. (S-8) to determine $\gamma_r$. Fig. S26a indicates that the broad interface of $MgAg_2$ composition is energetically favorable compared to $MgAl_2$- and $MgCu_2$-terminations, consistent with numerous experimental observations[2,7–9] for Ω nanoplates. Therefore, we derived $\gamma_b$ and $\gamma_r$ values of 0.143 J/m$^2$ and 1.644 J/m$^2$, respectively. The high ratio of $\gamma_r/\gamma_b$ (~11.5) aligns with the characteristic of Ω nanoplates having a large aspect ratio.

With the assumption of homogeneous and isotropic elasticity, for a plate-shaped nucleus with a radius $R$ and an aspect ratio ($a=2R/t$), $\Delta G_{elastic}$ can be analytically calculated from the Eshelby's solution[56]:

$$\Delta G_{elastic}=\mu\Big[\left(\varepsilon_{11}^{*2}+\varepsilon_{22}^{*2}\right)\left(\frac{v}{1-v}-\frac{13}{32(1-v)}\frac{\pi}{a}+1\right)+\varepsilon_{33}^{*2}\frac{1}{4(1-v)}\frac{\pi}{a}+2\varepsilon_{11}^{*}\varepsilon_{22}^{*}\left(\frac{v}{1-v}-\frac{16v-1}{32(1-v)}\frac{\pi}{a}\right)+\left(2\varepsilon_{11}^{*}\varepsilon_{33}^{*}+2\varepsilon_{22}^{*}\varepsilon_{33}^{*}\right)\frac{2v+1}{8(1-v)}\frac{\pi}{a}+\left(2\varepsilon_{23}^{*2}+\varepsilon_{31}^{*2}\right)\frac{2-v}{4(1-v)}\frac{\pi}{a}+2\varepsilon_{12}^{*2}\left(1-\frac{7-8v}{16(1-v)}\frac{\pi}{a}\right)\Big] \tag{S-9}$$

where $\varepsilon_{ij}^{*}$ are components of the stress-free strain tensor associated with lattice mismatch, $\mu$ is the shear modulus, and $v$ is the Poisson ratio. A Poisson's ratio of $v$=0.33 was used for Al. Despite the relatively large difference in shear modulus $\mu$ between Al (26 GPa) and $Al_2Cu$ (48 GPa), the elastic inhomogeneity was found to have a negligible effect on the critical nucleus size after the test (detailed results are not shown here). This can be attributed to the $\Delta G_{elastic}$ of the plate-like nucleus being inherently small compared to other morphologies, owing to slight local lattice distortions and elastic relaxation effects during nucleation. Similar results have been found in the nucleation of other plate-like precipitates, such as the θ′-$Al_2Cu$ phase[57] in Al alloys and the $Mg_2Ca$ phase[33] in Mg alloys.

For the critical nucleus structures of Ω nanoplates, we examined the 5-layer $U_T|K|\omega_T|K|U_T$ and 7-layer $U_T|K|\omega_T|Rec|\omega_T|K|U_T$ nuclei. According to our recent study[33], plotting $\Delta G$ against $n$ (the number

of specific atoms in the nucleus) facilitates us to compare critical sizes of nuclei with varying thicknesses. In this way, Fig. S26c shows that the 5-layer U$_T$|$K$|ω$_T$|$K$|U$_T$ nucleus has the lowest critical nucleation barrier and smallest size ($\Delta G^*_{5\text{-layer}}$=667 $k_BT$, with $n^*_{5\text{-layer}}$=58 Cu atoms, $R^*_{5\text{-layer}}\approx$1 nm) compared to that of the 7-layer U$_T$|$K$|ω$_T$|Rec|ω$_T$|$K$|U$_T$ nucleus. Consequently, the 5-layer U$_T$|$K$|ω$_T$|$K$|U$_T$ nucleus is predicted to be the most energetically and kinetically favorable for Ω nucleation, consistent with experimental observations[7–9] showing it as the thinnest observed Ω nanoplate.

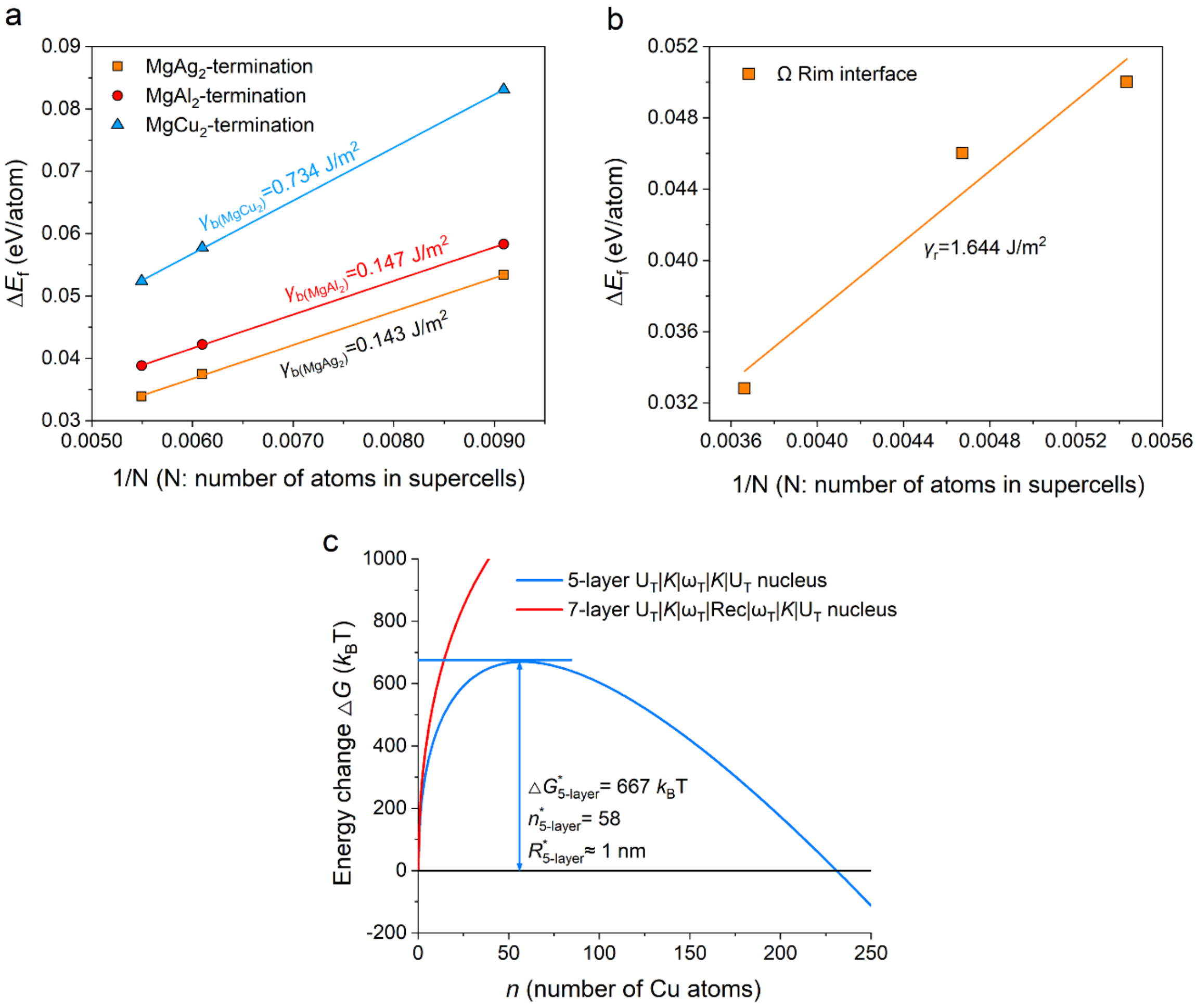

**Figure S26. a-b**. First-principles formation energies of N-atom supercells as a function of 1/N for the $\gamma_b$ and $\gamma_r$ interface. The interfacial energies $\gamma_{b(r)}$ were extracted from the slope by Eqs. (S7-S8). **c**. Theoretical predictions of critical size $n$* and nucleation barriers $\Delta G^*$ in the homogeneous nucleation of Ω nanoplates. The unit of energy change $\Delta G$ is $k_B$T, where $k_B$ is the Boltzmann constant and T=500 K.

Likewise, the total energy change associated with $\eta'$ homogeneous nucleation can be written as:

$$\Delta G_{\eta'}^{(5,\,7)\text{-layer}}=n_{\mathrm{Mg}}\Delta\mu_{\eta'}^{(5,\,7)\text{-layer}}+\frac{2n_{\mathrm{Mg}}V_{\eta'}^{(5,\,7)\text{-layer}}}{t_{\eta'}^{(5,\,7)\text{-layer}}}\gamma_{\mathrm{b}}+6t_{\eta'}^{(5,\,7)\text{-layer}}\sqrt{\frac{2n_{\mathrm{Mg}}V_{\eta'}^{(5,\,7)\text{-layer}}}{3\sqrt{3}t_{\eta'}^{(5,\,7)\text{-layer}}}}\gamma_{\mathrm{r}}+\Delta G_{\mathrm{elastic}}^{(5,\,7)\text{-layer}} \quad \text{(S-10)}$$

where $n_{\mathrm{Mg}}$ is the number of Mg atoms in the $\eta'$ nucleus, $\Delta\mu_{\eta'}^{(5,\,7)\text{-layer}}$ represents the reduction in chemical potential associated with the formation of (5, 7)-layer $\eta'$ nucleus in the system, while $n_{Mg}V_{\eta'}^{(5,\,7)\text{-layer}}$ denotes the volume of (5, 7)-layer $\eta'$ nanoplates. Based on the stoichiometry of (5, 7)-layer $\eta'$ nanoplates, the detailed expression of $\Delta\mu_{\eta'}^{(5,\,7)\text{-layer}}$ and $V_{\eta'}^{(5,\,7)\text{-layer}}$ are:

$$\Delta\mu_{\eta'}^{5\text{-layer}}=\Delta\mu_{Mg}+3\Delta\mu_{Zn},\;\; V_{\eta'}^{5\text{-layer}}=V_{Mg}+3V_{Zn}$$
$$\Delta\mu_{\eta'}^{7\text{-layer}}=\Delta\mu_{Mg}+\frac{8}{3}\Delta\mu_{Zn},\;\; V_{\eta'}^{7\text{-layer}}=V_{Mg}+\frac{8}{3}V_{Zn} \quad \text{(S-11)}$$

where $\Delta\mu_{X}$(X=Mg, Zn) is the chemical potential difference between X in the (5, 7)-layer $\eta'$ nanoplates and the (Mg, Zn)-GP zone, and $V_X$ (X= Mg, Zn) represents the Voronoi volume of a single X atom extracted from Ovito[55] software.

Different from the Ω nanoplate, which utilizes reference phases to determine the chemical potential of its constituents, the $Mg_6Zn_{13}$-$\eta'$ nanoplates lack a co-existing phase. Nevertheless, $\mu_{\eta'}^{Mg(Zn)}$ is expected to be lower than $\mu_{GP}^{Mg(Zn)}$ and satisfyies the condition of $6\mu_{\eta'}^{Mg}+13\mu_{\eta'}^{Zn}=E_{\eta'}$, resulting in $\mu_{\eta'}^{Zn}$ attain its lower range value ($\mu_{Zn}^{lower}$) when $\mu_{\eta'}^{Mg}=\mu_{GP}^{Mg}$. Thus, a range of $\frac{E_{\eta'}\text{-}6\mu_{GP}^{Mg}}{13}=\mu_{Zn}^{lower}<\mu_{\eta'}^{Zn}<\mu_{Zn}^{upper}=\mu_{GP}^{Zn}$ can be established. Subsequently, within this constrained range, Fig. S27a-b evaluated the $\gamma_b$ range for $MgAl_2$- and $MgZn_2$-terminated scenarios, with specific values marked on the panel. Based on the determined $\gamma_b$ range, Fig. S27c constructed an interfacial phase diagram to ascertain the stability of either termination type within a specified range. The results indicate that the $MgAl_2$-termination is energetically favored when $\mu_{Zn}^{lower}<\mu_{\eta'}^{Zn}<\mu_{Zn}^{o}$=-1.215 eV, whereas the $MgZn_2$-termination is preferred within the range of $\mu_{Zn}^{o}<\mu_{\eta'}^{Zn}<\mu_{Zn}^{upper}$. Given previous experimental findings[12–14] observing the enrichment of Zn along the broad interface of the GP zone, it is expected that $\mu_{\eta'}^{Zn}$ should fall within the latter range. The calculated $\gamma_r$ range, as shown in Fig. S27d, falls between 0.482 J/m$^2$ and 0.491 J/m$^2$, based on Eq. (S-8). Additionally, the low $\gamma_r/\gamma_b$ ratio (1.70~7.76) derived reflects the nature of $\eta'$ nanoplates having a low aspect ratio[12–14], indicating the validity of the approach employed.

By taking the above values into Eq.(S-10), Fig. S27e plots $\Delta G$ variations for both 5-layer $U_T|K|\omega_T|K|U_T$ and 7-layer $U_T|K|\omega_T|K|Pu_T|K|U_T$ nuclei against $n$ (number of Mg atoms in the nucleus) within the range of $\mu_{Zn}^{o}<\mu_{\eta'}^{Zn}<\mu_{Zn}^{upper}$. In this range, although 5-layer nuclei generally exhibit lower $\Delta G^*$, the smaller $n^*$ of 7-layer nuclei suggest they are kinetically favored as critical nuclei for $\eta'$ nucleation. Moreover, $\Delta G_{7\text{-layer}}^*$ is only slightly higher than that of the 5-layer nucleus by 1~2 $k_B$T (see Table S6). Considering Mg and Zn are fast-diffusing solutes in Al alloys[58], kinetic factors readily dominate, resulting in 7-layer nuclei being the critical state in reality. Hence, GP zones/$\eta'$ form in the matrix primarily in a 7-layer form rather than a 5-layer one.

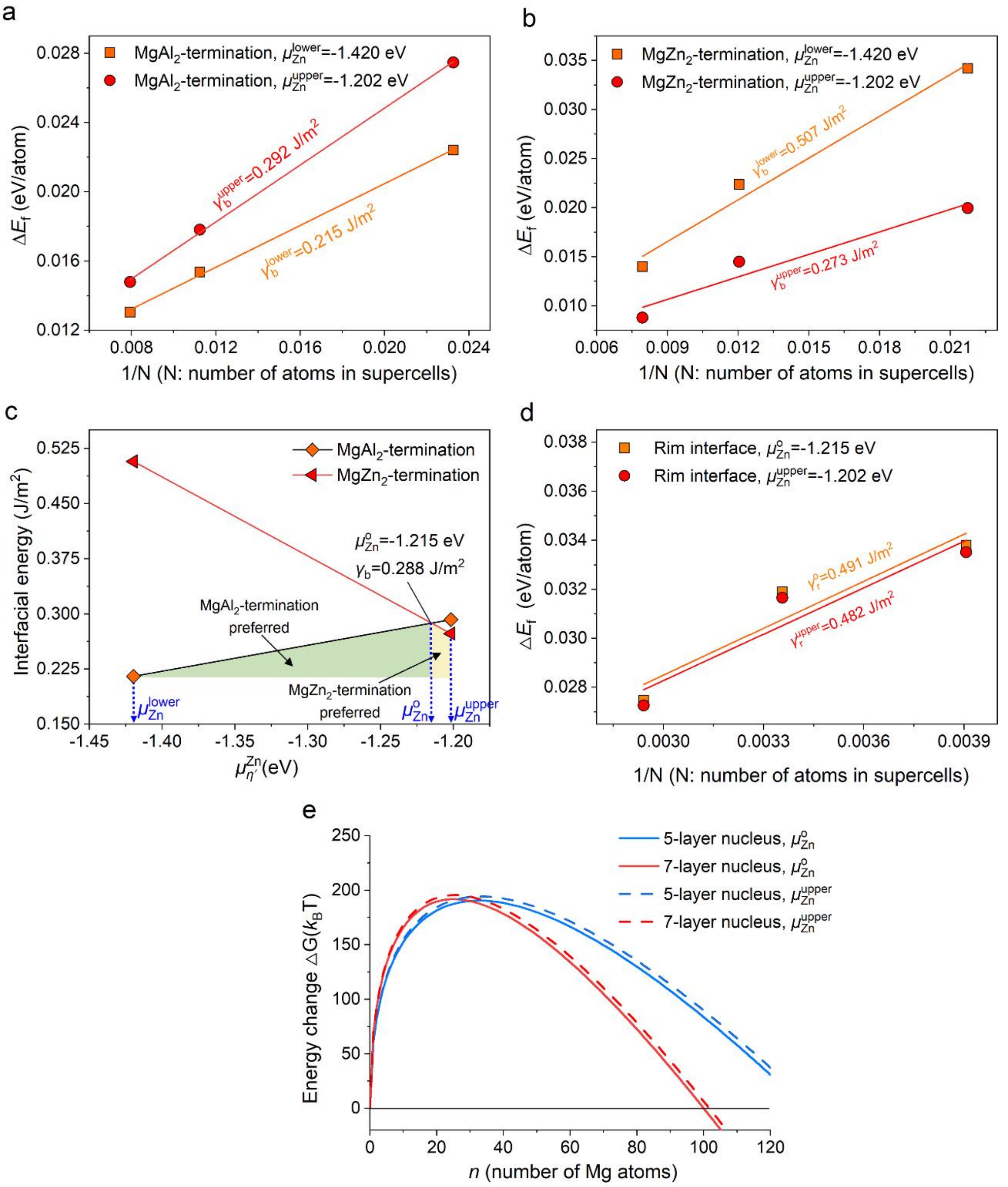

**Figure S27. a-b**. First-principles formation energies of N-atom supercells as a function of 1/N for the $\gamma_b$ interface under the $\mu_{Zn}^{lower}$ and $\mu_{Zn}^{upper}$ values. **c**. The calculated interfacial phase diagram to determine the most energy-favorable interface composition under a given range of $\mu_{\eta'}^{Zn}$. **d**. First-principles formation energies of N-atom supercells as a function of 1/N for the $\gamma_r$ interface under the $\mu_{Zn}^{o}$ and $\mu_{Zn}^{upper}$ values. **e**. Theoretical predictions of critical size $n^*$ and nucleation barriers $\Delta G^*$ in the homogeneous nucleation of $\eta'$ nanoplates within the range of $\mu_{Zn}^{o}<\mu_{\eta'}^{Zn}<\mu_{Zn}^{upper}$. The unit of energy change $\Delta G$ is $k_B$T, where $k_B$ is the Boltzmann constant and T=500 K.

**Table S6.** Summary of $n^*$ and $\Delta G^*$ for 5-layer and 7-layer nuclei under the $\mu_{Zn}^{o}$ and $\mu_{Zn}^{upper}$ values.

| Nucleus | $n^*$ | $\Delta G^*$ ($k_B$T) | $R^*$ (nm) |
|---|---|---|---|
| 5-layer nucleus $\mu_{Zn}^{o}$ | 33 | 190 | ~0.760 |
| 7-layer nucleus $\mu_{Zn}^{o}$ | 25 | 192 | ~0.500 |
| 5-layer nucleus $\mu_{Zn}^{upper}$ | 33 | 194 | ~0.760 |
| 7-layer nucleus $\mu_{Zn}^{upper}$ | 25 | 195 | ~0.500 |